   \documentstyle[12pt]{article}
\if@twoside\oddsidemargin25mm
\else\oddsidemargin25mm\evensidemargin25mm\marginparwidth25mm\fi%
\begin{document}
\newtheorem{definizione}{Definition}[section]
\newtheorem{teorema}{Theorem}[section]
\newcommand{\bth}{\begin{teorema}}
\newcommand{\eth}{\end{teorema}}
\newtheorem{lemma}{Lemma}[section]
\newcommand{\blem}{\begin{lemma}}
\newcommand{\elem}{\end{lemma}}
\newcommand{\brr}{\begin{array}}
\newcommand{\err}{\end{array}}
\newtheorem{corollario}{Corollary}[section]
\newcommand{\bcorol}{\begin{corollario}}
\newcommand{\ecorol}{\end{corollario}}
\newcommand{\bd}{\begin{definizione}}
\newcommand{\ed}{\end{definizione}}
\def\T{T}
\def\inbar{\vrule height1.5ex width.4pt depth0pt}
\def\bfzero{\relax{\rm I\kern-.18em 0}}
\def\bfone{\relax{\rm 1\kern-.35em 1}}
\def\IR{\relax{\rm I\kern-.18em R}}
\def\IC{\relax\,\hbox{$\inbar\kern-.3em{\rm C}$}}
\def\IH{\relax{\rm I\kern-.18em H}}
\def\ID{\relax{\rm I\kern-.18em D}}
\def\IF{\relax{\rm I\kern-.18em F}}
\def\II{\relax{\rm I\kern-.17em I}}
\def\IN{\relax{\rm I\kern-.18em N}}
\def\IP{\relax{\rm I\kern-.18em P}}
\def\IQ{\relax\,\hbox{$\inbar\kern-.3em{\rm Q}$}}
\def\IG{\relax\,\hbox{$\inbar\kern-.3em{\rm G}$}}
\font\cmss=cmss10 \font\cmsss=cmss10 at 7pt
\def\ZZ{\relax\ifmmode\mathchoice
{\hbox{\cmss Z\kern-.4em Z}}{\hbox{\cmss Z\kern-.4em Z}}
{\lower.9pt\hbox{\cmsss Z\kern-.4em Z}}
{\lower1.2pt\hbox{\cmsss Z\kern-.4em Z}}\else{\cmss Z\kern-.4em
Z}\fi}
\def\half{{1 \over 2}}
\def\cA{{\cal A}} \def\cB{{\cal B}}
\def\cC{{\cal C}} \def\cD{{\cal D}}
\def\cF{{\cal F}} \def\cG{{\cal G}}
\def\cH{{\cal H}} \def\cI{{\cal I}}
\def\cJ{{\cal J}} \def\cK{{\cal K}}
\def\cL{{\cal L}} \def\cM{{\cal M}}
\def\cN{{\cal N}} \def\cO{{\cal O}}
\def\cP{{\cal P}} \def\cQ{{\cal Q}}
\def\cR{{\cal R}} \def\cV{{\cal V}}\def\cW{{\cal W}}
\def\twomat#1#2#3#4{\left(\begin{array}{cc}
 {#1}&{#2}\\ {#3}&{#4}\\
\end{array}
\right)}
\def\twovec#1#2{\left(\begin{array}{c}
{#1}\\ {#2}\\
\end{array}
\right)}
\newcommand{\nn}{\nonumber}
\newcommand{\La}{{\Lambda}}
\newcommand{\Si}{{\Sigma}}
\def\a{\alpha} \def\b{\beta} \def\d{\delta}
\def\e{\epsilon} \def\c{\gamma}
\def\G{\Gamma} \def\l{\lambda}
\def\L{\Lambda} \def\s{\sigma}
%
\newcommand{\ft}[2]{{\textstyle\frac{#1}{#2}}}
\newcommand{\QED}{{\hspace*{\fill}\rule{2mm}{2mm}\linebreak}}
\def\dop{{\rm d}\hskip -1pt}
\def\Psi{{\psi}}
%
%
%
\def\crr{\crcr\noalign{\vskip {8.3333pt}}}
\def\tilde{\widetilde}
\def\bar{\overline}
\def\us#1{\underline{#1}}
\let\shat=\hat
\def\hat{\widehat}
\def\hyp{\vrule height 2.3pt width 2.5pt depth -1.5pt}
\def\square{\mbox{.08}{.08}}
\def\Coeff#1#2{{#1\over #2}}
\def\Coe#1.#2.{{#1\over #2}}
\def\coeff#1#2{\relax{\textstyle {#1 \over #2}}\displaystyle}
\def\coe#1.#2.{\relax{\textstyle {#1 \over #2}}\displaystyle}
\def\half{{1 \over 2}}
\def\shalf{\relax{\textstyle {1 \over 2}}\displaystyle}
\def\dag#1{#1\!\!\!/\,\,\,}
\def\to{\rightarrow}
\def\notin{\hbox{{$\in$}\kern-.51em\hbox{/}}}
\def\shdot{\!\cdot\!}
\def\ket#1{\,\big|\,#1\,\big>\,}
\def\bra#1{\,\big<\,#1\,\big|\,}
\def\equaltop#1{\mathrel{\mathop=^{#1}}}
\def\Trbel#1{\mathop{{\rm Tr}}_{#1}}
\def\inserteq#1{\noalign{\vskip-.2truecm\hbox{#1\hfil}
\vskip-.2cm}}
\def\attac#1{\Bigl\vert
{\phantom{X}\atop{{\rm\scriptstyle #1}}\phantom{X}}}
\def\exx#1{e^{{\displaystyle #1}}}
\def\del{\partial}
\def\delbar{\bar\partial}
\def\nex#1{$N\!=\!#1$}
\def\dex#1{$d\!=\!#1$}
\def\cex#1{$c\!=\!#1$}
\def\eg{{\it e.g.}} \def\ie{{\it i.e.}}
%
\def\cS{{\cal K}}
\def\IE{\relax{{\rm I\kern-.18em E}}}
\def\cE{{\cal E}}
\def\rt{{\cR^{(3)}}}
\def\IGam{\relax{{\rm I}\kern-.18em \Gamma}}
\def\IGa{\IA}
\def\LG{Lan\-dau-Ginz\-burg\ }
\def\cV{{\cal V}}
\def\Rt{{\cal R}^{(3)}}
\def\wabc{W_{abc}}
\def\WABC{W_{\a\b\c}}
\def\W{{\cal W}}
\def\tft#1{\langle\langle\,#1\,\rangle\rangle}
\def\IA{\relax{\hbox{{\rm A}\kern-.82em {\rm A}}}}
\let\picfuc=\fp
\def\hata{{\shat\a}}
\def\hatb{{\shat\b}}
\def\hatA{{\shat A}}
\def\hatB{{\shat B}}
\def\bv{{\bf V}}
\def\spg{special geometry}
\def\sc{SCFT}
\def\leel{low energy effective Lagrangian}
\def\pf{Picard--Fuchs}
\def\pfS{Picard--Fuchs system}
\def\el{effective Lagrangian}
\def\Fb{\overline{F}}
\def\nablab{\overline{\nabla}}
\def\Ub{\overline{U}}
\def\Db{\overline{D}}
\def\zb{\overline{z}}
\def\eb{\overline{e}}
\def\fb{\overline{f}}
\def\tb{\overline{t}}
\def\Xb{\overline{X}}
\def\Vb{\overline{V}}
\def\Cb{\overline{C}}
\def\Sb{\overline{S}}
\def\delb{\overline{\del}}
\def\Gammab{\overline{\Gamma}}
\def\Ab{\overline{A}}
\def\Anh{A^{\rm nh}}
\def\alphab{\bar{\alpha}}
\def\cy{Calabi--Yau}
\def\cabg{C_{\alpha\beta\gamma}}
\def\B{\Sigma}
\def\Bh{\hat \Sigma}
\def\Kh{\hat{K}}
\def\Knh{{\cal K}}
\def\A{\Lambda}
\def\Ah{\hat \Lambda}
\def\R{\hat{R}}
\def\V{{V}}
\def\Gammah{\hat{\Gamma}}
\def\twot{$(2,2)$}
\def\K{K\"ahler}
\def\rat{({\theta_2 \over \theta_1})}
\def\lv{{\bf \omega}}
\def\w{w}
\def\CP{C\!P}
\def\o#1#2{{{#1}\over{#2}}}
\newcommand{\js}{{j^\star}}
\newcommand{\im}{{\rm Im\ }}
\def\bfzero{\relax{\rm I\kern-.18em 0}}
\def\inbar{\vrule height1.5ex width.4pt depth0pt}
\def\IC{\relax\,\hbox{$\inbar\kern-.3em{\rm C}$}}
\def\ID{\relax{\rm I\kern-.18em D}}
\def\IF{\relax{\rm I\kern-.18em F}}
\def\IH{\relax{\rm I\kern-.18em H}}
\def\II{\relax{\rm I\kern-.17em I}}
\def\IN{\relax{\rm I\kern-.18em N}}
\def\IP{\relax{\rm I\kern-.18em P}}
\def\IQ{\relax\,\hbox{$\inbar\kern-.3em{\rm Q}$}}
\def\IR{\relax{\rm I\kern-.18em R}}
\def\IG{\relax\,\hbox{$\inbar\kern-.3em{\rm G}$}}
\font\cmss=cmss10 \font\cmsss=cmss10 at 7pt
\def\ZZ{\relax\ifmmode\mathchoice
{\hbox{\cmss Z\kern-.4em Z}}{\hbox{\cmss Z\kern-.4em Z}}
{\lower.9pt\hbox{\cmsss Z\kern-.4em Z}}
{\lower1.2pt\hbox{\cmsss Z\kern-.4em Z}}\else{\cmss Z\kern-.4em
Z}\fi}
\def\a{\alpha} \def\b{\beta} \def\d{\delta}
\def\e{\epsilon} \def\c{\gamma}
\def\G{\Gamma} \def\l{\lambda}
\def\L{\Lambda} \def\s{\sigma}
\def\cA{{\cal A}} \def\cB{{\cal B}}
\def\cC{{\cal C}} \def\cD{{\cal D}}
\def\cF{{\cal F}} \def\cG{{\cal G}}
\def\cH{{\cal H}} \def\cI{{\cal I}}
\def\cJ{{\cal J}} \def\cK{{\cal K}}
\def\cL{{\cal L}} \def\cM{{\cal M}}
\def\cN{{\cal N}} \def\cO{{\cal O}} \def\cU{{\cal U}}
\def\cP{{\cal P}} \def\cQ{{\cal Q}} \def\cS{{\cal S}}
\def\cR{{\cal R}} \def\cV{{\cal V}}\def\cW{{\cal W}}
%
%
%
\def\crr{\crcr\noalign{\vskip {8.3333pt}}}
\def\tilde{\widetilde}
\def\bar{\overline}
\def\us#1{\underline{#1}}
\let\shat=\hat
\def\hat{\widehat}
\def\hyp{\vrule height 2.3pt width 2.5pt depth -1.5pt}
\def\square{\mbox{.08}{.08}}
\def\Coeff#1#2{{#1\over #2}}
\def\Coe#1.#2.{{#1\over #2}}
\def\coeff#1#2{\relax{\textstyle {#1 \over #2}}\displaystyle}
\def\coe#1.#2.{\relax{\textstyle {#1 \over #2}}\displaystyle}
\def\shalf{\relax{\textstyle {1 \over 2}}\displaystyle}
\def\dag#1{#1\!\!\!/\,\,\,}
\def\to{\rightarrow}
\def\notin{\hbox{{$\in$}\kern-.51em\hbox{/}}}
\def\shdot{\!\cdot\!}
\def\ket#1{\,\big|\,#1\,\big>\,}
\def\bra#1{\,\big<\,#1\,\big|\,}
\def\equaltop#1{\mathrel{\mathop=^{#1}}}
\def\Trbel#1{\mathop{{\rm Tr}}_{#1}}
\def\inserteq#1{\noalign{\vskip-.2truecm\hbox{#1\hfil}
\vskip-.2cm}}
\def\attac#1{\Bigl\vert
{\phantom{X}\atop{{\rm\scriptstyle #1}}\phantom{X}}}
\def\exx#1{e^{{\displaystyle #1}}}
\def\del{\partial}
\def\delbar{\bar\partial}
\def\nex#1{$N\!=\!#1$}
\def\dex#1{$d\!=\!#1$}
\def\cex#1{$c\!=\!#1$}
\def\eg{{\it e.g.}} \def\ie{{\it i.e.}}
%
\newcommand{\be}{\begin{equation}}
\newcommand{\ee}{\end{equation}}
\newcommand{\ba}{\begin{eqnarray}}
\newcommand{\ea}{\end{eqnarray}}

\begin{titlepage}
\hskip 5.5cm
\vbox{
}
\hskip 2.5cm
\vbox{
\hbox{POLFIS-TH.03/96} 
\hbox{UCLA/96/TEP/9}
\hbox{hep-th/9605032}}
\vfill
\begin{center}
{{\LARGE   N=2 Supergravity
and N=2 Super Yang-Mills Theory on General Scalar Manifolds:
\vskip 1.5mm
}
{\Large {\it
  Symplectic Covariance, Gaugings and
the Momentum Map.$^*$}  } }\\
\vskip 1.5cm
{  {\bf L. Andrianopoli$^1$, M. Bertolini$^2$ $^\dagger$, A.
Ceresole$^2$, R. D'Auria$^2$,
\vskip 1.5mm
S. Ferrara$^3$, P. Fr\'e$^4$ $^{\dagger}$ and T. Magri$^4$ }} \\
\vskip 0.5cm
{\small
$^1$ Dipartimento di Fisica, Universit\'a di Genova, via Dodecaneso 33,
I-16146 Genova, Italy\\
\vspace{6pt}
$^2$ Dipartimento di Fisica, Politecnico di Torino,\\
 Corso Duca degli Abruzzi 24, I-10129 Torino, Italy\\
\vspace{6pt}
$^3$ CERN, CH 1211 Geneva 23, Switzerland\\
and Department of Physics, University of California,  Los Angeles, CA 90024,
 USA\\
\vspace{6pt}
$^4$ Dipartimento di Fisica Teorica, Universit\`a di Torino, via
P. Giuria, 1
I-10125 Torino, Italy\\
}
\end{center}
\vfill
\begin{center} {\bf Abstract}
\end{center}
{
\small
The general form of $N\!=\!2$ supergravity  coupled to an arbitrary
number of vector multiplets and hypermultiplets, with a generic
gauging of the scalar manifold isometries is given. This extends
the results already available in the literature in that we use a
coordinate independent and manifestly symplectic covariant formalism
which allows to cover theories difficult to formulate within superspace
or tensor calculus approach.
We  provide the  complete lagrangian and supersymmetry variations
with all fermionic terms, and the form of the scalar potential for
arbitrary quaternionic manifolds and special geometry, not necessarily
in special coordinates. Lagrangians for rigid theories are also written
in this general setting and the connection with local theories elucidated.
The derivation of these results using geometrical techniques is briefly
summarized.
}
\vspace{2mm} \vfill \hrule width 3.cm
{\footnotesize
\noindent $^\dagger$ Fellow by Ansaldo Ricerche srl C.so Perrone 24,
I-16152 Genova. \\
\noindent
$^*$ Supported in part by DOE grant
DE-FGO3-91ER40662 Task C., EEC Science Program SC1*CT92-0789,
NSF grant no. PHY94-07194, and INFN.
}
\end{titlepage}

\section{Introduction}
Impressive results over the last year on non perturbative properties of $N=2$
supersymmetric Yang-Mills theories\cite{SW12,kltold} 
and their extension to string theory\cite{CDF}-\cite{Wdy}
through the notion of string-string duality\cite{Duff,SS},
 have used the deep
underlying mathematical structure of these theories and its relation to
algebraic geometry 
\cite{stromco}-
\cite{cynoi}.

In the case of $N=2$ vector multiplets, describing the effective
interactions in the Abelian (Coulomb) phase of a spontaneously
broken gauge theory, Seiberg and Witten \cite{SW12}
have shown that positivity of
the metric on the underlying moduli space identifies the geometrical
data of the effective $N=2$ rigid theory with the periods of
a particular torus.
\par
In the coupling to gravity it was conjectured by some of the present
authors \cite{CDF,CDFVP}
and later confirmed by heterotic-Type II duality
\cite{FHSV,kava,cynoi,KLM},  that the very same argument based on positivity
of the vector multiplet kinetic
metric identifies the corresponding geometrical data of the
effective $N=2$ supergravity with the periods of Calabi-Yau
threefolds.
\par
On the other hand, when matter is added, the underlying geometrical
structure is much richer, since  $N=2$  matter hypermultiplets are
associated with quaternionic geometry\cite{bagwit,hklr,gal}, 
and charged hypermultiplets are
naturally associated with the gauging of triholomorphic isometries of
these quaternionic manifolds \cite{DFF,ans123}.

\par
It is the aim of this paper to complete the general form of  the
$N=2$ supergravity lagrangian coupled to an arbitrary number of
vector multiplets and hypermultiplets  in presence of a general gauging 
of the isometries of both the vector multiplets and hypermultiplets 
scalar manifolds.
Actually this extends  
results already obtained years ago by some of us \cite{DFF},
that in turn extended previous work by Bagger and Witten  on ungauged
general quaternionic manifolds coupled to $N=2$ supergravity\cite{bagwit},
by de Wit, Lauwers and Van Proeyen on gauged special geometry and gauged
quaternionic manifolds obtained by quaternionic quotient in the
tensor calculus framework \cite{dWLVP},
and by Castellani, D'Auria and Ferrara on covariant formulation of
special geometry for matter coupled supergravity \cite{CaDF}.
\par
This paper firstly provides in a geometrical setting the full lagrangian
with all the fermionic terms and the supersymmetry variations. Secondly,
it uses a coordinate independent and manifestly
symplectic covariant formalism which in particular does not require
the use of a prepotential function $F(X)$.
Whether a prepotential $F(X)$ exists or not depends on the choice
of a symplectic gauge\cite{CDFVP}. Moreover, some physically interesting
cases are precisely instances where $F(X)$ does not exist\cite{CDFVP}. 
\par
Of particular relevance is the fact that we exhibit a scalar potential for
arbitrary quaternionic geometries
and for special geometry not necessarily in special coordiantes. This
allows us to go beyond what is obtainable with the tensor calculus (or
superspace) approach. 
Among many applications, our results  allow the study of general conditions
for spontaneous supersymmetry breaking in a manner analogous to what was done
for  $N=1$ matter coupled supergravity \cite{CFGVP}.
Many examples of supersymmetry breaking studied in the past are then
reproduced in a unified framework.
\par
Recently the power of using simple geometrical formulae for the
scalar potential was exploited while studying the breaking of half
supersymmetries in a particular simple model, using a symplectic basis
where $F(X)$ is not defined\cite{FGP1}. The method has potential
applications in string theory to study  non perturbative phenomena
such as conifold transitions \cite{GMS}, $p$-forms  condensation 
\cite{pols} and  Fayet-Iliopoulos terms \cite{FGP1,APT}.

$N=2$ supergravity  displays a high degree of
complexity in its structure, based however on the simplicity of
few principles. The supersymmetric Lagrangian and the transformation
rules are indeed quite involved but all the couplings, the mass matrices
and the vacuum energy  are completely fixed and
organized in terms of three geometrical data:
\begin{enumerate}
\item {\it The choice of a special K\"ahler manifold ${\cal SM}$
describing the self-interactions of the vector multiplets}
\item {\it The choice of a quaternionic manifold ${\cal HM}$
describing the self-interaction of the hypermultiplets}
\item{{\it The choice of a gauge group $\cal G$, that in the non
abelian case  must be a subgroup of the isometry
group of the scalar manifold ${\cal M}_{scalar} \, \equiv \,
 {\cal SM} \otimes {\cal HM}$ with a block diagonal immersion
 in the symplectic group $Sp(2{\bar n}+2,\IR ) $ of
 electric--magnetic duality rotations} (see eq.~\ref{classym}).}
\end{enumerate}
For this reason we devote the first and largest part of the paper
(sections 2-7) to review and discuss, in a way independent from
supersymmetric Lagrangians and supersymmetry algebras,
the geometrical ingredients
of the construction that we listed above. This part of the paper can be
read as an independent essay and should be quite accessible to mathematicians
as well as to readers  who have no background or interest in supersymmetry.
\par
The second part of the paper (sections 8-9) presents instead the
Lagrangian and supersymmetry transformation rules for both $N=2$
supergravity and $N=2$ matter coupled rigid Yang--Mills theory that
is retrieved from supergravity in the infinite Planck mass limit
$\mu \, \to \, \infty$. The theory is presented in a completely
explicit component formalism, and no formulae employ  or require
the use of superfields, superspace or conformal tensor calculus.
All items entering such formulae are rather geometrical
objects whose nature and properties were described and explained
in previous sections.
\par
The reader interested in applications of $N=2$ supergravity or Yang--Mills
theory can directly jump  to sections 8-9, that are
self--contained, and insert, in the ready-to-use formulae the
specific geometrical data corresponding to the problem  considered.
References to formulae in previous sections are given to fix
normalizations.
\par
The derivation of the results presented in sections 8-9 was obtained
by means of the geometric (``rheonomic'') approach (for a general review
see the book by some of us \cite{CaDFb}). The details of the
derivation are given in the Appendices for the interested reader,
while the results are presented
in the main text. It is indeed one of the main advantages of the
geometrical approach to supersymmetry that the final outcome of
the construction is directly written in space--time component
formalism.
\par
As emphasized our results are general and apply to generic choice
of the scalar manifold. As an illustration of our formulae
in the appendix we specialize them to the case of the manifolds
~\ref{grancaso}.
More specifically, our paper is organized as follows:
\begin{enumerate}
\item {{\bf Section 2} reviews duality
rotations and symplectic covariance in field theory.}
\item {{\bf Section 3} describes the symplectic embedding of the
homogeneous spaces,in particular the special symmetric spaces which
appear at tree level in heterotic string theory.}
\item{{\bf Section 4}
reviews Special K\"ahler geometry, both for rigid
and local supersymmetry.}
\item{{\bf Section 5}
describes the geometry of hypermultiplets, their associated
quaternionic and hyperK\"ahler manifolds in local and rigid
supersymmetry.}
\item{{\bf Section 6}
faces the gauging of special and quaternionic manifolds.}
\item{{\bf Section 7}
deals with the so called momentum map on Special K\"ahler
and quaternionic manifolds giving rise to the introduction of
prepotential functions which enter in the construction of the scalar
potential.}
\item{{\bf Section 8}
reports the full $N=2$ Lagrangian in a symplectic
covariant form}
\item{{\bf Section 9} contains the rigid limit and reports
the general form of a matter coupled $N=2$ super Yang--Mills theory
on a generic rigid special manifold  and a generic rigid hyperK\"ahler
manifold.}
\item{{\bf Appendices A, B} give  a detailed derivation of the
Lagrangian and transformation rules using the geometrical
 approach.}
\item{{\bf Appendix C} deals with the relevant
formulas for $N=2$ supergravity based on the manifolds
\begin{eqnarray}
 \mbox{special manifold} &=& ST[2,n] \, \equiv \,
 {{SU(1,1)}\over{U(1)}} \, \otimes \, {{SO(2,n)}\over{SO(2)\times
 SO(n)}} \nonumber\\
\mbox{quaternionic manifold} &=& \, HQ[m]  \, \equiv  \,
{{SO(4,m)}\over{SO(4)\times SO(m)}}
\label{grancaso}
\end{eqnarray}
This is done as an exemplification of the general formulae for the
potential, mass matrices and kinetic period matrices and for its intrinsic
interest in applications to tree level string theory}
\item{{\bf Appendix D} contains a list of  conventions and normalizations
that we have employed.}

\end{enumerate}

An expanded version of this paper, with particular attention to the
geometrical properties of the scalar manifolds,  the rigidly
supersymmetric version and further related issues is given
in \cite{corto}.
\section{Duality Rotations and Symplectic Covariance}
\label{LL1}
\setcounter{equation}{0}
In this section, both for completeness and in order to fix
our conventions and notations, we review the general structure of an
abelian theory of vectors and scalars displaying covariance
under a group of duality rotations.
The basic reference is the 1981 paper by Gaillard and Zumino
\cite{gaizum}. A general presentation in $D=2p$ dimensions was
recently given in \cite{pietrolectures}. Here we fix
 $D=4$.
\par
We consider a theory of  $\bar n$ gauge fields $A^\Lambda_{\mu}$,
in a $D=4$  space--time with Lorentz signature.
They  correspond to a set of  $\bar n$
differential $1$--forms
\begin{equation}
A^\Lambda ~ \equiv ~
A^\Lambda_{\mu} \, dx^{\mu} \quad \quad
 \left ( \Lambda = 1,
\dots , {\bar n} \right )
\end{equation}
The corresponding field strengths and their Hodge duals are defined
by
\begin{eqnarray}
{ F}^\Lambda & \equiv & d \, A^\Lambda \,  \equiv   \,
   {\cal F}^\Lambda_{\mu  \nu} \,
dx^{\mu } \, \wedge \, dx^{\nu} \nonumber\\
{\cal F}^\Lambda_{\mu \nu} & \equiv & {{1}\over{2 }} \,\left ( \partial_{\mu }
A^\Lambda_{\nu} \, - \, \partial_{\nu }
A^\Lambda_{\mu} \right )
\nonumber\\ ^{\star}{ F}^{\Lambda} & \equiv &\, {\tilde
{\cal F}}^\Lambda_{\mu \nu} \, dx^{\mu } \, \wedge \,
 dx^{\nu} \nonumber\\
{\tilde {\cal
F}}^\Lambda_{\mu \nu} & \equiv &{{1}\over{2}}
\varepsilon_{\mu  \nu \rho\sigma}\, {\cal F}^{\Lambda
\vert \rho \sigma}
\label{campfort}
\end{eqnarray}
Defining the space--time integration volume as 
\begin{equation}
\mbox{d}^4 x \, \equiv \, -{{1}\over{4!}} \, \varepsilon_{\mu_1\dots
\mu_4} \, dx^{\mu_1} \, \wedge \, \dots \, \wedge dx^{\mu_{4}}\ ,
\label{volume}
\end{equation}
we obtain
\begin{equation}
F^\Lambda \, \wedge \, F^\Sigma \,  =
  \varepsilon^{\mu  \nu \rho \sigma }\,
{\cal F}^\Lambda_{\mu\nu} \, {\cal
F}^\Sigma_{\rho\sigma}\,d^4 x \qquad ; \qquad
F^\Lambda \, \wedge
\, ^{\star}F^{\Sigma}   =   - 2 \,
{\cal F}^\Lambda_{\mu\nu} \, {\cal F}^{\Sigma \vert
\mu\nu} d^4 x\ .
\label{cinetici}
\end{equation}
In addition
to the gauge fields let us also introduce a set of real scalar
fields $\phi^I$ ( $I=1,\dots , {\bar m}$) spanning an ${\bar
m}$--dimensional manifold ${\cal M}_{scalar}$ \footnotemark
\footnotetext{Whether the $\phi^I$ can be arranged into complex
fields is not relevant at this level of the discussion. } endowed
with a metric $g_{IJ}(\phi)$. Utilizing the above field
content we can write the following action functional:
\begin{equation}
{\cal S}\, = {1\over 2}\,  \int \, \left \{ \left [ \,
  \gamma_{\Lambda\Sigma}(\phi) \, F^\Lambda \, \wedge
\,\star F^{\Sigma} \, +  \,
\theta_{\Lambda\Sigma}(\phi) \, F^\Lambda \, \wedge \, F^{\Sigma}  \,
\right  ]
     \,  + \,
g_{IJ}(\phi) \, \partial_\mu \phi^I \, \partial^\mu \phi^J   \,
\mbox{d}^4 x \,  \right \}\ ,
\label{gaiazuma}
\end{equation}
where the scalar fields
dependent ${\bar n} \times {\bar n}$ matrix
$\gamma_{\Lambda\Sigma}(\phi)$  generalizes the inverse of the
squared coupling constant $\o{1}{g^2}$ appearing in ordinary
 gauge theories. The field dependent matrix
$\theta_{\Lambda\Sigma}(\phi)$ is instead a generalization of the
$theta$--angle of quantum chromodynamics. Both $\gamma$ and $\theta$
are symmetric matrices.
Introducing a formal operator $j$ that maps a field
strength into its Hodge dual
\begin{equation} \left ( j \, {\cal
F}^\Lambda \right )_{\mu \nu} \, \equiv \,
{{1}\over{ 2 }} \, \epsilon_{\mu\nu \rho \sigma} \,
{\cal F}^{\Lambda \vert\rho\sigma}
\label{opjei}
\end{equation}
and a formal scalar product
\begin{equation} \left (
G \, , \, K \right ) \equiv G^T K \, \equiv \,
\sum_{\Lambda=1}^{\bar n} G^{\Lambda}_{\mu\nu} K^{\Lambda
\vert \mu\nu }
\label{formprod}
\end{equation} the total
Lagrangian of eq.~\ref{gaiazuma} can be rewritten as
\begin{equation}
\cL^{(tot)}\,  = \,  {\cal F}^T \, \left ( -\gamma
\otimes \bfone + \theta \otimes j \right ) {\cal F} \, + \,
\o{1}{2} \, g_{IJ}(\phi) \, \partial_\mu \phi^I \,
\partial^\mu \phi^J
\label{gaiazumadue}
\end{equation}
The operator $j$ satisfies $j^2 \, = \, - \, \bfone$ so that its
eigenvalues are $\pm {\rm i}$.
Introducing self--dual and antiself--dual combinations
\begin{eqnarray}
  {\cal F}^{\pm} &=& {1\over 2}\left({\cal F}\, \pm {\rm i} \, j
{\cal F}\right) \nonumber \\
j \, {\cal F}^{\pm}& =& \mp \mbox{i} {\cal F}^{\pm}
\label{selfduals}
\end{eqnarray}
and the field--dependent symmetric matrices
\begin{eqnarray}
  {\cal N} & = & \theta \, - \,
\mbox{i} \gamma \nonumber\\
{\bar {\cal N}} & = & \theta + \mbox{i} \gamma\ ,
\label{scripten}
\end{eqnarray}
the
vector part of the Lagrangian ~\ref{gaiazumadue} can be rewritten as
\begin{equation}
{\cal L}_{vec} \,  = \, {\mbox{i}} \, \left [{\cal F}^{-T} {\bar {\cal N}}
{\cal F}^{-}-
{\cal F}^{+T} {\cal N} {\cal F}^{+} \right]
\label{lagrapm}
\end{equation}
Introducing the new tensors
\begin{equation}
  {\tilde{\cal
G}}^\Lambda_{\mu\nu} \, \equiv \,  {1 \over 2 } { {\partial {\cal
L}}\over{\partial {\cal F}^\Lambda_{\mu\nu}}}\leftrightarrow  {\cal
G}^{\mp \Lambda}_{\mu\nu} \, \equiv \,  \mp{{\rm i} \over 2 } { {\partial {\cal
L}}\over{\partial {\cal F}^{\mp \Lambda}_{\mu\nu}}}
\label{gtensor}
\end{equation}
which, in matrix notation, corresponds to
\begin{equation}
j \, {\cal G} \, \equiv \,{1 \over 2 }  \,
{{\partial {\cal L}}\over{\partial {\cal F}^T}} \,  = \, -
\, \left ( \gamma\otimes\bfone -\theta\otimes j \right )
\, {\cal F}
\label{ggmatnot}
\end{equation}
the Bianchi identities and field
equations associated with the Lagrangian ~\ref{gaiazuma} can be
written as 
\ba 
\partial^{\mu }{\tilde {\cal F}}^{\Lambda}_{\mu\nu} &=& 0 \\
\partial^{\mu }{\tilde {\cal G}}^{\Lambda}_{\mu\nu} &=& 0
\label{biafieq}
\ea
or equivalently
\ba
 \partial^{\mu }
Im{\cal F}^{\pm \Lambda}_{\mu\nu} &=& 0 \\
\partial^{\mu } Im{\cal G}^{\pm \Lambda}_{\mu\nu} &=&
0 \ .
\label{biafieqpm}
\ea
This suggests that we introduce the $2{\bar
n}$ column vector 
\begin{equation}
{\bf V} \, \equiv \, \left ( \matrix
{ j \, {\cal F}\cr
j \, {\cal G}\cr}\right )
\label{sympvec}
\end{equation}
and that we consider general linear transformations on such a vector
\begin{equation}
\left ( \matrix
{ j \, {\cal F}\cr
j \, {\cal G}\cr}\right )^\prime \, =\,
\left (\matrix{ A & B \cr C & D \cr} \right )
\left ( \matrix
{ j \, {\cal F}\cr
j \, {\cal G}\cr}\right )
\label{dualrot}
\end{equation}
For any matrix $\left (\matrix{ A & B \cr C & D \cr} \right ) \, \in
\, GL(2{\bar n},\IR )$ the new vector ${\bf V}^\prime$ of {\it
magnetic and electric} field--strengths satisfies the same
equations ~\ref{biafieq} as the old one. In a condensed notation
we can write
\begin{equation}
\partial \, {\bf V}\, = \, 0 \quad \Longleftrightarrow \quad
\partial \, {\bf V}^\prime \, = \, 0
\label{dualdue}
\end{equation}
Separating the self--dual and anti--self--dual parts
\begin{equation}
{\cal F}=\left ({\cal F}^+ +{\cal F}^- \right ) \qquad ;
\qquad
{\cal G}=\left ({\cal G}^+ +{\cal G}^- \right )
\label{divorzio}
\end{equation}
and taking into account that   we have
\begin{equation}
{\cal G}^+ \, = \, {\cal N}{\cal F}^+  \quad
{\cal G}^- \, = \, {\bar {\cal N}}{\cal F}^-
\label{gigiuno}
\end{equation}
the duality
rotation of eq.~\ref{dualrot} can be rewritten as
\begin{equation}
\left ( \matrix
{   {\cal F}^+ \cr
 {\cal G}^+\cr}\right )^\prime  \, = \,
\left (\matrix{ A & B \cr C & D \cr} \right )
\left ( \matrix
{   {\cal F}^+\cr
{\cal N} {\cal F}^+\cr}\right ) \qquad ; \qquad
\left ( \matrix
{   {\cal F}^- \cr
 {\cal G}^-\cr}\right )^\prime \, = \,
\left (\matrix{ A & B \cr C & D \cr} \right )
\left ( \matrix
{   {\cal F}^-\cr
{\bar {\cal N}} {\cal F}^-\cr}\right )
\label{trasform}
\end{equation}
The problem is that the transformation rule
~\ref{trasform} of ${\cal G}^\pm$ must be consistent with the definition
of the latter
as variation of the Lagrangian with respect
to ${\cal F}^\pm$ (see eq.~\ref{gtensor}). This request
restricts the form of the matrix
$\Lambda =\left (\matrix{ A & B \cr C & D \cr} \right )$.
As we are going to show, $\Lambda$ must belong
to the symplectic subgroup  of the general linear group
\begin{equation}
\Lambda \equiv \left (\matrix{ A & B \cr C & D \cr} \right ) \,  \in \,
Sp(2\bar n,\IR)
\,\subset
\, GL(2\bar n ,\IR )
\label{distinguo}
\end{equation}
the   subgroup $Sp(2\bar n,\IR)$ being defined as the set of $2\bar n \times
2\bar n$ matrices that satisfy the condition
\begin{equation}
\Lambda \in Sp(2\bar n,\IR) ~ \longrightarrow ~
   \Lambda^T  \,
\left (\matrix{ {\bf 0}_{} &
\bfone_{} \cr
-\bfone_{} & {\bf 0}_{}
\cr }\right )
 \, \Lambda \,  = \,
   \left (\matrix{ {\bf 0}_{} &
\bfone_{} \cr
-\bfone_{} & {\bf 0}_{}
\cr }\right )
\label{ortosymp}
\end{equation}
that is, using $ n \otimes n$ block components
\begin{equation}
A^T C-C^T A=
B^T D - D^T B  =0 \quad\quad
A^T D - C^T B =1   \label {ortocomp}
\end{equation}

To prove the statement we just made, we calculate the transformed
Lagrangian ${\cal L}^\prime$ and then we compare its variation
${\o{\partial {\cal L}^\prime}{\partial {\cal F}^{\prime T}}}$
with  ${\cal G}^{\pm\prime}$ as it follows from the postulated
transformation rule ~\ref{trasform}. To perform such a calculation
we rely on the following basic idea. While the
duality rotation~\ref{trasform} is performed on the field strengths
and on their duals, also the scalar fields are transformed by the action
of some diffeomorphism ${\xi }\,  \in \,
{\rm Diff}\left ( {\cal M}_{scalar}\right )$ of the scalar manifold
and, as a consequence of that, also the matrix ${\cal N}$ changes.
In other words given the scalar manifold ${\cal M}_{scalar}$ we
assume that   there exists a
homomorphism of the  form 
\begin{equation}
\iota _{\delta} : \,  {\rm Diff}\left ( {\cal M}_{scalar}\right )
\, \longrightarrow \, GL(2\bar n,\IR)
\label{immersione}
\end{equation}
so that
\begin{eqnarray}
\forall &  \xi   &\in \, {\rm Diff}\left ( {\cal M}_{scalar}\right ) \, :
\, \phi^I \, \stackrel{\xi}{\longrightarrow} \,  \phi^{I\prime}
\nonumber\\
\exists  & \iota _{\delta}(\xi) & = \left (\matrix{ A_\xi & B_\xi \cr
C_\xi & D_\xi \cr }\right ) \, \in \,  GL(2\bar n,\IR)
\label{apnea}
\end{eqnarray}
(In the sequel the subfix $\xi$ will be  omitted when no confusion
can arise and be reinstalled when necessary for clarity. )
\par
Using such a homomorphism
we can define the simultaneous action of $\xi$ on
all the fields of our theory by setting
\begin{equation}
\xi \, : \, \cases{   \phi \,
\longrightarrow \, \xi (\phi) \cr
{\bf V} \,
\longrightarrow \, \iota _{\delta}(\xi) \, {\bf V} \cr
{\cal N}(\phi) \, \longrightarrow \, {\cal N}^\prime (\xi (\phi)) \cr }
\end{equation}
where the notation~\ref{sympvec} has been utilized.
In the gauge sector the transformed Lagrangian is
\begin{equation}
  {\cal L}^{\prime}_{vec}  \, =  \,
   {\rm i} \, \Bigl [{\cal F}^{-T}
\, \bigl ( A + B {\bar {\cal N}} \bigr )^T {\bar {\cal N}}^\prime
( A + B {\bar {\cal N}} \bigr ) {\cal F}^{-} \, -  \, {\cal F}^{+T}
\, \bigl ( A + B {\cal N} \bigr )^T {\cal N}^\prime
( A + B {\cal N} \bigr ) {\cal F}^{+}
\Bigr ]
\label{elleprima}
\end{equation}
Consistency with
the definition of ${\cal G}^+$ requires  that
\begin{equation}
{\cal N}^\prime \, \equiv \, {\cal N}^\prime (\xi(\phi)) \, = \,
 \left ( C  + D  {\cal N}(\phi) \right )   \left ( A  +
B  {\cal N}(\phi)\right )^{-1}
\label{Ntrasform}
\end{equation}
while consistency with the definition of ${\cal G}^-$ imposes
the transformation rule
\begin{equation}
{\bar {\cal N}}^\prime  \, \equiv \, {\bar {\cal N}}^\prime (\xi(\phi)) \,
= \,
\left ( C  + D  {\bar {\cal N}}(\phi ) \right )   \left ( A  +
B  {\bar {\cal N}}(\phi)\right )^{-1}
\label{Nbtrasform}
\end{equation}
 It is from the transformation rules~\ref{Ntrasform} and ~\ref{Nbtrasform}
that we derive a restriction on the form of the
duality rotation matrix $\Lambda \equiv \iota_\delta(\xi)$.
Indeed by requiring that the transformed matrix ${\cal N}^\prime$ be again
symmetric one easily finds that ${\Lambda}$ must obey eq. \ref {ortosymp},
namely $\Lambda \in Sp(2\bar n ,\IR)$.
Consequently the
homomorphism of eq.~\ref{immersione} specializes
as
 \begin{equation}
\iota _{\delta} : \,  {\rm Diff}\left ( {\cal M}_{scalar}\right )
\, \longrightarrow \, Sp(2\bar n,\IR)
\label{spaccoindue}
\end{equation}
Clearly, since   $Sp(2\bar n,\IR)$   is
a finite dimensional Lie group, while
${\rm Diff}\left ( {\cal M}_{scalar}\right )$ is
infinite--dimensional, the homomorphism  $\iota _{\delta}$ can
never be an isomorphism. Defining the Torelli group of the
scalar manifold as
\begin{equation}
{\rm Diff}\left ( {\cal M}_{scalar}\right ) \, \supset \,
\mbox{Tor} \left ({\cal M}_{scalar} \right ) \, \equiv \,
\mbox{ker} \, \iota_\delta
\label{torellus}
\end{equation}
we always have
\begin{equation}
\mbox{dim} \, \mbox{Tor} \left ({\cal M}_{scalar} \right ) \, = \,
\infty
\label{infitor}
\end{equation}
The reason why we have given the name of Torelli to the group defined
by eq.~\ref{torellus} is because of its similarity with the
Torelli group that occurs in algebraic geometry.
\par
What should
be clear from the above discussion is that a family of Lagrangians
as in eq.~\ref{gaiazuma} will admit a group of
duality--rotations/field--redefinitions that will map elements
of the family into each other, as long as a {\it kinetic matrix}
${\cal N}_{\Lambda\Sigma}$ can be constructed  that transforms as
in eq.~\ref{Ntrasform}. A way to obtain such an object is to identify
it with the {\it period matrix} occurring in problems of algebraic
geometry. At the level of the present discussion, however, this
identification is by no means essential: any construction of
${\cal N}_{\Lambda\Sigma}$ with the appropriate transformation
properties is acceptable.
Note also that so far we have used the
words {\it duality--rotations/field--redefinitions} and not the word
duality symmetry. Indeed the diffeomorphisms of the scalar manifold
we have considered were quite general and, as such had no pretension
to be symmetries of the action, or of the theory. Indeed the question
we have answered is the following: what are the appropriate
transformation properties of the tensor gauge fields and of the generalized
coupling constants under diffeomorphisms of the scalar manifold?
The next question is obviously that of duality symmetries.
\par
As it is the case with the difference between general covariance and
isometries in the context of general relativity, duality symmetries
correspond to the subset of duality transformations for which
we obtain an invariance in form of the theory.
In this respect, however, we have to stress that what is invariant
in form cannot be the Lagrangian but only the set of field equations
plus Bianchi identities.
Indeed, while any $\Lambda \in Sp(2\bar n,\IR)$ can, in principle,
be an invariance
in form of eqs.~\ref{biafieqpm}, the same is  not true for the
Lagrangian.  One can easily find that the vector kinetic part
of this latter transforms as follows:
\begin{eqnarray}
\mbox{Im} {\cal F}^{-\Lambda}{\bar{\cal N}}_{\Lambda\Sigma}
{\cal F}^{-\Sigma}
& \rightarrow & \mbox{Im} {\tilde{\cal F}}^{-\Lambda}\,
{\tilde{\cal G}}^{-}_{\Sigma}\nonumber\\
& \null &= \mbox{Im} \Bigl ( {\cal F}^{-\Lambda} {\cal G}^{-}_{\Lambda} +
2 {\cal F}^{-\Lambda}\, \bigl ( C^{T} B
\bigr )_{\Lambda}^{\phantom{\Lambda}\Sigma}  \,
{\cal G}^{-}_{\Sigma} \nonumber\\
& \null & + {\cal F}^{-\Lambda}\, \bigl ( C^{T} A
\bigr )_{\Lambda \Sigma} {\cal F}^{-\Sigma} +
{\cal G}^{-}_{\Lambda} \, \bigl ( D^{T} B
\bigr )^{\Lambda \Sigma} \,  {\cal G}^{-}_{\Sigma}
\Bigr )
\label{lagtrasf}
\end{eqnarray}
whence we conclude that proper symmetries of the Lagrangian are
to be looked for only among matrices with $ C=B=0 $.
If $ C \ne  0 $ and  $B=0$, the Lagrangian varies through the
addition of a topological density (see below eq.~\ref{topogatto}).
  Elements of $Sp(2{\bar n}, \IR)$ with
$B \ne  0$, cannot be symmetries of the classical action under any
circumstance.
\par
The scalar part of the Lagrangian, on the other hand, is invariant
under all those diffeomorphisms of the scalar manifolds that
are
{\it isometries} of the scalar metric
$g_{IJ}$. Naming
$\xi^\star : \, T{\cal M}_{scalar} \, \rightarrow \, T{\cal M}_{scalar}$
the push--forward of $\xi$, this means that
\begin{eqnarray}
&\forall\,  X,Y \, \in \, T{\cal M}_{scalar}& \nonumber\\
& g\left ( X, Y \right
)\, = \, g \left ( \xi^\star X, \xi^\star Y \right )&
\label{isom}
\end{eqnarray}
and $\xi$ is an exact global symmetry of the scalar  part of the
Lagrangian in eq.~\ref{gaiazuma}. In view of our previous discussion
these symmetries of the scalar sector are not guaranteed to admit an
extension to symmetries of the complete action. Yet we can insist
that they extend to symmetries of the field
equations plus Bianchi identities, namely  to  duality symmetries
in the sense defined above. This requires
that the group of isometries of the scalar metric
${\cal I} ({\cal M}_{scalar})$ be suitably embedded into
the duality group $Sp(2\bar n,\IR)$ and that the kinetic matrix ${\cal
N}_{\Lambda\Sigma}$ satisfies the covariance law:
\begin{equation}
{\cal N}\left ( \xi (\phi)\right ) \, = \,
\left ( C_\xi + D_\xi {\cal N}(\phi) \right )
\left ( A_\xi + B_\xi {\cal N}( \phi )\right )^{-1}\ .
\label{covarianza}
\end{equation}
\section{Symplectic embeddings of homogenous spaces}
\label{LL2}
\setcounter{equation}{0}
\par
A general construction of the kinetic coupling matrix $ \cal N$ 
 can be derived in the
case where the scalar manifold is taken to be a homogeneous space
${\cal G}/{\cal H}$.
This is what happens in all
extended supergravities for $N \ge 3$ and also in specific
instances of N=2 theories. For this reason we shortly review
the construction of the {\it kinetic
period matrix} ${\cal N}$ in the case of homogeneous spaces. Although
the basic construction was introduced in the literature by Gaillard
and Zumino in 1981 \cite{gaizum} and was reviewed by some of us in
\cite{CaDFb}, a derivation of the basic
formulae that matches completely
with the modern notations of N=2 and N=4 theories, such as they
emerge in string compactifications and in the discussion of S--duality,
is not available, to our knowledge,  in the existing literature.
To make the present paper self contained we consider therefore
essential to review such a construction in modern gear.
\par
The relevant homomorphism $\iota_\delta$ (see
eq.~\ref{spaccoindue}) becomes:
\begin{equation}
\iota_\delta : \, \mbox{Diff}\left ({{\cal G}\over{\cal
H}} \right ) \, \longrightarrow \, Sp(2\bar n, \IR)
\label{embeddif}
\end{equation}
In particular, focusing on the isometry group of the canonical metric
defined on ${{\cal G}\over{\cal H}}$\footnotemark
\footnotetext{Actually, in order to be true, the
equation ${\cal I}({\o{\cal G}{\cal H}})={\cal G}$ requires
that that the normaliser of ${\cal H}$ in ${\cal G}$ be the
identity group, a condition that is verified in all the relevant examples}:
$ {\cal I} \left ({{\cal G}\over{\cal H}}\right ) \, = \, {\cal G}$
we must consider the embedding:
\begin{equation}
\iota_\delta : \,  {\cal G}  \, \longrightarrow \, Sp(2\bar n, \IR)
\label{embediso}
\end{equation}
That in eq.~\ref{embeddif} is a homomorphism of finite dimensional
Lie groups and as such it constitutes a problem that can be solved
in explicit form. What we just need to know is the dimension of the
symplectic group, namely the number $\bar n$ of gauge fields appearing
in the theory. Without supersymmetry the dimension $m$ of the scalar
manifold (namely the possible choices of ${{\cal G}\over{\cal H}}$) and
the number of vectors $\bar n$ are unrelated so that the
possibilities covered by eq.~\ref{embediso} are infinitely many.
In supersymmetric theories, instead, the two numbers $m$ and $\bar n$
are related, so that there are finitely many cases to be studied
corresponding to the possible  embeddings of given groups
${\cal G}$ into a symplectic group $Sp(2\bar n, \IR)$ of fixed dimension
$\bar n$. Actually taking into account further conditions on
the holonomy of the scalar manifold that are also imposed by
supersymmetry, the solution for the symplectic embedding problem
is unique for all extended supergravities with $N \ge 3$ as we
have already remarked (see for instance \cite{CaDFb}).
\par
Apart from the details of the specific case considered
once a symplectic embedding is given there is a general
formula one can write down for the {\it period matrix}
${\cal N}$ that guarantees symmetry (${\cal N}^T = {\cal N}$)
and the required transformation property~\ref{covarianza}.
This is the result we want to review. It will be useful in
the sequel for comparison with the formulae of special geometry
in the case the considered special manifold is homogeneous
(see appendix C, in particular).
\par
The real symplectic group $Sp(2\bar n ,\IR)$ is defined as the set
of all {\it real} $2\bar n \times 2\bar n$ matrices
$ \Lambda \, = \, \left ( \matrix{ A & B \cr C & D \cr } \right ) $
satisfying equation ~\ref{ortosymp}, namely
\begin{equation}
\Lambda^T \, \IC \, \Lambda \, = \, \IC
\label{condiziona}
\end{equation}
 where
$ \IC  \, \equiv \, \left ( \matrix{ {\bf 0} & \bfone \cr -\bfone &
{\bf 0} \cr } \right ) $
If we relax the condition that the matrix should be real but we
still impose eq.~\ref{condiziona} we obtain the definition
of the complex symplectic group $Sp(2\bar n, \IC)$. It is
a well known fact that the following isomorphism is true:
\begin{equation}
Sp(2\bar n, \IR)  \sim  Usp(\bar n , \bar n)   \equiv
Sp(2\bar n, \IC)   \cap   U(\bar n , \bar n)
\label{usplet}
\end{equation}
By definition an element ${\cal S}\,\in \, Usp(\bar n , \bar n)$
is a complex matrix that satisfies simultaneously eq.~\ref{condiziona}
and a pseudo--unitarity condition, that is:
\begin{equation}
{\cal S}^T \, \IC \, {\cal S} \, = \,  \IC \quad \quad \quad
; \quad \quad \quad
{\cal S}^\dagger \, \IH \, {\cal S} \, =\,  \IH
\label{uspcondo}
\end{equation}
where
$\IH \,  \equiv \, \left ( \matrix{ \bfone & {\bf 0} \cr {\bf 0} & -\bfone
 \cr } \right )$.
The general block form of the matrix ${\cal S}$ is:
\begin{equation}
{\cal S}\, = \, \left ( \matrix{ T & V^\star \cr V & T^\star \cr } \right )
\label{blocusplet}
\end{equation}
and eq.s~\ref{uspcondo} are equivalent to:
\begin{equation}
T^\dagger \, T \, - \, V^\dagger \, V \, =\, \bfone \quad\quad ; \quad\quad
T^\dagger \, V^\star  \, - \,  V^\dagger \, T^\star \, =\, {\bf 0}
\label{relazie}
\end{equation}
The isomorphism of eq.~\ref{usplet} is explicitly realized by
the so called Cayley matrix:
\begin{equation}
{\cal C} \, \equiv \, {\o{1}{\sqrt{2}}} \,
\left ( \matrix{ \bfone & {\rm i}\bfone \cr \bfone & -{\rm i}\bfone
 \cr } \right )
\label{cayley}
\end{equation}
via the relation:
\begin{equation}
{\cal S}\, = \, {\cal C} \, \Lambda \, {\cal C}^{-1}
\label{isomorfo}
\end{equation}
which yields:
\begin{equation}
T \, =\,  {\o{1}{2}}\, \left ( A + D \right ) -
{\o{\rm i}{2}}\, \left ( B-C \right ) \quad \quad ; \quad
\quad
V \, =\, {\o{1}{2}}\, \left ( A - D \right ) -
{\o{\rm i}{2}}\, \left ( B+C \right )
\label{mappetta}
\end{equation}
When we set $V=0$ we obtain the subgroup $U(\bar n) \subset Usp (\bar
n , \bar n)$, that in the real basis is given by the subset of
symplectic matrices of the form
$\left ( \matrix{ A & B \cr -B & A
 \cr } \right )$. The basic idea, to obtain the
general formula for the period matrix, is that the symplectic embedding
of the isometry group ${\cal G}$ will be such that the isotropy
subgroup ${\cal H}\subset {\cal G}$ gets embedded into the maximal
compact subgroup $U(\bar n)$, namely:
\begin{equation}
{\cal G} \,  {\stackrel{\iota_\delta}{\longrightarrow}} \,  Usp (\bar
n , \bar n) \quad\quad ; \quad\quad
{\cal G} \supset {\cal H} \,  {\stackrel{\iota_\delta}{\longrightarrow}} \,
U(\bar n) \subset Usp (\bar n , \bar n)
\label{gruppino}
\end{equation}
If this condition is realized let $L(\phi)$ be a parametrization of
the coset ${\cal G}/{\cal H}$ by means of coset representatives.
Relying on the symplectic embedding of eq.~\ref{gruppino} we obtain
a map:
\begin{equation}
  L(\phi)  \, \longrightarrow  {\cal O}(\phi)\, =  \,
 \left ( \matrix{ U_0(\phi) & U^\star_1(\phi) \cr U_1(\phi)
& U^\star_0(\phi) \cr } \right )\,  \in  \, Usp(\bar n , \bar n)
\label{darstel}
\end{equation}
that associates to $L(\phi)$ a coset representative of $Usp(\bar n ,
\bar n)/U(\bar n)$. By construction if $\phi^\prime \ne \phi$
{\it no} unitary $\bar n \times \bar n$ matrix $W$ {\it can exist}
such that:
\begin{equation}
 {\cal O}(\phi^\prime)  =  {\cal O}(\phi) \,
 \left ( \matrix{ W & {\bf 0} \cr {\bf 0}
& W^\star \cr } \right )
\end{equation}
On the other hand let $\xi \in {\cal G}$ be an element of the
isometry group of ${{\cal G}/{\cal H}}$. Via the symplectic embedding
of eq.~\ref{gruppino} we obtain a $Usp(\bar n, \bar n)$ matrix
\begin{equation}
{\cal S}_ \xi \, = \,
\left ( \matrix{ T_\xi & V^\star_\xi \cr V_\xi & T^\star_\xi \cr } \right )
\label{uspimag}
\end{equation}
such that
\begin{equation}
{\cal S}_ \xi \,{\cal O}(\phi) \, = \, {\cal O}(\xi(\phi)) \,
\left ( \matrix{ W(\xi,\phi) & {\bf 0} \cr {\bf 0}
& W^\star(\xi,\phi) \cr } \right )
\label{cosettone}
\end{equation}
where $\xi(\phi)$ denotes the image of the point
$\phi \in  {{\cal G}/{\cal H}}$ through $\xi$ and $W(\xi,\phi)$ is
a suitable $U(\bar n)$ compensator depending both on $\xi$ and
$\phi$.
Combining eq.s~\ref{cosettone},~\ref{darstel}, with eq.s~\ref{mappetta}
we immediately obtain:
\begin{eqnarray}
 U_0^\dagger \left( \xi(\phi) \right ) +
U^\dagger_1 \left (\xi(\phi) \right)  & = &
  W  \left [ U_0^\dagger \left( \phi \right )   \left (
A^T + {\rm i}B^T \right ) + U_1^\dagger \left( \phi \right )   \left (
A^T - {\rm i}B^T \right ) \right ]   \nonumber\\
 U_0^\dagger \left( \xi(\phi) \right ) -
U^\dagger_1 \left (\xi(\phi) \right)  & = &
 W \, \left [ U_0^\dagger \left( \phi \right )   \left (
D^T - {\rm i}C^T \right ) - U_1^\dagger \left( \phi \right )   \left (
D^T + {\rm i}C^T \right ) \right ]
\label{semitrasform}
\end{eqnarray}
Setting:
\begin{equation}
{\cal N} \, \equiv \, {\rm i} \left [ U_0^\dagger + U_1^\dagger \right
]^{-1} \, \left [ U_0^\dagger - U_1^\dagger \right ]
\label{masterformula}
\end{equation}
and using the result of eq.~\ref{semitrasform} one checks
that the transformation rule~\ref{covarianza} is verified.
It is also an immediate consequence of the analogue of
eq.s~\ref{relazie} satisfied by $U_0$ and $U_1$ that the matrix
in eq.~\ref{masterformula} is symmetric
\begin{equation}
{\cal N}^T \, = \, {\cal N}
\label{massi}
\end{equation}
Eq.~\ref{masterformula} is the master formula derived in 1981 by
Gaillard and Zumino \cite{gaizum}.
It explains the structure of the gauge field
kinetic terms in all $N\ge 3$ extended supergravity theories and
also in those $N=2$ theories where
the  Special K\"ahler manifold  ${\cal SM}$
is a homogeneous manifold ${\cal G}/{\cal H}$.

\subsection{Symplectic embedding of the ${\cal ST}\left [ m,n \right ]$
homogeneous manifolds}
Because of their relevance in superstring compactifications let us
illustrate the general procedure with the following class of
homogeneous manifolds:
\begin{equation}
 {\cal ST}\left [ m,n \right ] \,  \equiv \,
{\o{SU(1,1)}{U(1)}} \, \otimes \, {\o{SO(m,n)}{SO(m)\otimes SO(n)}}
\label{stmanif}
\end{equation}
The isometry group of the ${\cal ST}\left [ m,n \right ]$
manifolds defined in eq.~\ref{stmanif} contains a factor ($SU(1,1)$)
whose transformations act as non--perturbative $S$--dualities and
another factor $(SO(m,n))$ whose transformations act as
$T$--dualities,
holding true at each order in string perturbation theory. The field
$S$ is obtained by combining together the {\it dilaton} $D$ and
the {\it axion} ${\cal A}$:
\begin{eqnarray}
S & = & {\cal A} - {\rm i} \mbox{exp}[D] \nonumber\\
\partial^\mu {\cal A} & \equiv & \varepsilon^{\mu\nu\rho\sigma} \,
\partial_\nu \, B_{\rho\sigma}
\label{scampo}
\end{eqnarray}
while $t^i$ is the name usually given to the moduli--fields of the
compactified target space. Now in string and supergravity
applications $S$ will be identified with the complex coordinate
on the manifold ${\o{SU(1,1)}{U(1)}}$, while  $t^i$  will be
the coordinates of the coset space  ${\o{SO(m,n)}{SO(m)\otimes SO(n)}}$.
The case ${\cal ST}[6,n]$ is the scalar manifold in $N=4$
supergravity, while the case ${\cal ST}[2,n]$ is a very interesting
instance of special K\"ahler manifold appearing in superstring
compactifications.
Although as differentiable and metric manifolds
the spaces ${\cal ST}\left [ m,n \right ]$ are just direct products
of two factors (corresponding to the above mentioned different
physical interpretation of the coordinates $S$ and $t^i$), from the
point of view of the symplectic embedding and duality rotations
they have to be regarded as a single entity. This is even more
evident in the case $m=2,n=\mbox{arbitrary}$,
where the following theorem has been proven by
Ferrara and Van Proeyen \cite{ferratoine}:
${\cal ST}\left [ 2,n \right ]$ are the only special K\"ahler
manifolds with a direct product structure. The definition
of special K\"ahler manifolds is given in the next section,
yet the anticipation of this result
should make clear that the special K\"ahler structure (encoding the
duality rotations in the $N=2$ case)
is not a property of the individual factors
but of the product as a whole. Neither factor
is by itself a special manifold although the product is.
\par
At this point comes  the question
of the correct symplectic embedding. Such a question has two aspects:
\begin{enumerate}
\item{Intrinsically inequivalent embeddings}
\item{Symplectically equivalent embeddings that become inequivalent
after gauging}
\end{enumerate}
 The first issue in the above list is group--theoretical in nature.
When we say that the group ${\cal G}$ is embedded into $Sp(2\bar
n,\IR)$ we must specify how this is done from the point of view
of irreducible representations. Group--theoretically the matter is
settled by specifying how the fundamental representation of
$Sp(2\bar n)$ splits into irreducible representations of ${\cal G}$:
\begin{eqnarray}
& {\bf {2 \bar n}} \, {\stackrel{{\cal G}}{\longrightarrow}}
\oplus_{i=1}^{\ell} \, {\bf D}_i &
\label{splitsplit}
\end{eqnarray}
Once eq.~\ref{splitsplit} is given (in supersymmetric theories
such information is provided by supersymmetry ) the only arbitrariness
which is left is that of conjugation by arbitrary $Sp(2\bar n,\IR)$
matrices. Suppose we have determined an embedding $\iota_\delta$ that
obeys the law in eq.~\ref{splitsplit}, then:
\begin{equation}
\forall \, {\cal S} \, \in \, Sp(2\bar n,\IR) \, : \,
\iota_\delta^\prime \, \equiv \, {\cal S} \circ  \iota_\delta \circ
{\cal S}^{-1}
\label{matrim}
\end{equation}
will obey the same law. That in eq.~\ref{matrim} is a symplectic
transformation that corresponds to an allowed
duality--rotation/field--redefinition in the abelian theory of
type in eq.~\ref{gaiazuma} discussed in the previous subsection. Therefore
all abelian Lagrangians related by such transformations are physically
equivalent.
\par
The matter changes in presence of {\it gauging}. When we switch
on the gauge coupling constant and the electric charges, symplectic
transformations cease to yield physically equivalent theories. This
is the second issue in the above list. The choice of a symplectic
gauge becomes physically significant.
The construction of supergravity theories proceeds in
two steps. In the first step,
one constructs the abelian theory: at that level the only relevant
constraint is that encoded in eq.~\ref{splitsplit} and the choice of
a symplectic gauge is immaterial. Actually one can write the entire
theory in such a way that {\it symplectic covariance} is manifest.
In the second step one {\it gauges} the theory. This {\it breaks
symplectic covariance} and the choice of the correct symplectic gauge
becomes a physical issue. This issue has been recently emphasized
by the results in \cite{FGP1} where it has been shown that
whether N=2 supersymmetry can be spontaneously broken to N=1 or
not depends on the symplectic gauge.
\par
These facts being cleared we proceed to discuss the symplectic
embedding of the ${\cal ST}\left [ m,n \right ]$ manifolds.
\par
Let $\eta$ be the symmetric flat metric with signature
$(m,n)$ that defines the $SO(m,n)$ group, via the relation
\begin{equation}
L \, \in \, SO(m,n) \, \Longleftrightarrow \, L^T \, \eta L \, = \,
\eta
\label{ortogruppo}
\end{equation}
Both in the $N=4$ and in the $N=2$ theory, the number of gauge fields
in the theory is given by:
\begin{equation}
\# \mbox{vector fields} \, = \, m \oplus n
\label{vectornum}
\end{equation}
$m$ being the number of {\it graviphotons} and $n$ the number of
{\it vector multiplets}. Hence we have to embed $SO(m,n)$ into
$Sp(2m+2n,\IR)$ and the explicit form of the decomposition in
eq.~\ref{splitsplit} required by supersymmetry is:
\begin{equation}
{\bf {2m+2n}} \, {\stackrel{SO(m,n)}{\longrightarrow}} \, {\bf { m+n}}
\oplus  {\bf { m+n}}
\label{ortosplitsplit}
\end{equation}
where ${\bf { m+n}}$ denotes the fundamental representation of
$SO(m,n)$. Eq.~\ref{ortosplitsplit} is easily understood in physical
terms. $SO(m,n)$ must be a T--duality group, namely a symmetry
 holding true order by order in perturbation theory. As such it must
 rotate electric  field strengths into electric field strengths and
 magnetic field strengths into magnetic field field strengths. The
 two irreducible representations into which the the fundamental
 representation of the symplectic group decomposes when reduced to
 $SO(m,n)$ correspond precisely to electric and magnetic sectors,
 respectively.
In the {\it simplest  gauge} the symplectic embedding satisfying
eq.~\ref{ortosplitsplit} is block--diagonal and takes the form:
\begin{equation}
\forall \,  L \, \in \, SO(m,n) \quad {\stackrel{\iota_\delta}
{\hookrightarrow}} \quad
\left ( \matrix{ L & {\bf 0}\cr {\bf 0} & (L^T)^{-1} \cr } \right )
\, \in \, Sp(2m+2n,\IR)
\label{ortoletto}
\end{equation}
Consider instead the group $SU(1,1) \sim SL(2,\IR)$. This is the
factor in the isometry group of ${\cal ST}[m,n]$
that is going to act by means of S--duality non perturbative
rotations. Typically it will rotate each electric field strength into
its homologous magnetic one. Correspondingly supersymmetry implies
that its embedding into the symplectic group must satisfy the
following condition:
\begin{equation}
{\bf {2m+2n}} \, {\stackrel{SL(2,\IR)}{\longrightarrow}} \,
\oplus_{i=1}^{m+n} \, {\bf 2}
\label{simposplisplit}
\end{equation}
where  ${\bf 2}$ denotes the fundamental representation of
$SL(2,\IR)$. In addition it must commute with the embedding
of $SO(m,n)$ in eq.~\ref{ortoletto} . Both conditions are fulfilled
by setting:
\begin{equation}
\forall \,   \left ( \matrix{a & b \cr  c &d \cr }\right )
\, \in \, SL(2,\IR) \quad {\stackrel{\iota_\delta}
{\hookrightarrow}} \quad
\left ( \matrix{ a \, \bfone & b \, \eta \cr c \, \eta &
d \, \bfone \cr } \right )
\, \in \, Sp(2m+2n,\IR)
\label{ortolettodue}
\end{equation}
Utilizing eq.s~\ref{isomorfo} the corresponding  embeddings into
the group $Usp(m+n,m+n)$ are immediately derived:
\begin{eqnarray}
\forall \,  L \, \in \, SO(m,n) & {\stackrel{\iota_\delta}
{\hookrightarrow}} & \left ( \matrix{ {\o{1}{2}}  \left ( L+
\eta L \eta \right ) & {\o{1}{2}}  \left ( L-
\eta L \eta \right )\cr {\o{1}{2}}  \left ( L -
\eta L \eta \right ) & {\o{1}{2}}  \left ( L+
\eta L \eta \right ) \cr } \right )  \,
  \, \in \, Usp(m+n,m+n)  \nonumber\\
  \forall \,   \left ( \matrix{t & v^\star \cr  v &t^\star \cr }\right )
\, \in \, SU(1,1) & {\stackrel{\iota_\delta}
{\hookrightarrow}} &  \left ( \matrix{ {\rm Re}t \bfone +{\rm i}{\rm Im}t\eta &
{\rm Re}v \bfone -{\rm i}{\rm Im}v \eta   \cr
{\rm Re}v \bfone +{\rm i}{\rm Im}v\eta &
{\rm Re}t \bfone - {\rm i}{\rm Im}t\eta \cr } \right )
 \, \in \, Usp(m+n,m+n)  \nonumber\\
\label{uspembed}
\end{eqnarray}
where the relation between the entries of the $SU(1,1)$ matrix
and those of the corresponding $SL(2,\IR)$ matrix are provided
by the relation in eq.~\ref{mappetta}.
\par
Equipped with these relations we can proceed to derive the explicit
form of the {\it period matrix} ${\cal N}$.
\par
The homogeneous manifold $SU(1,1)/U(1)$ can be conveniently
parametrized in terms of a single complex coordinate $S$, whose
physical interpretation will be that of {\it axion--dilaton},
according to eq.~\ref{scampo}. The coset parametrization appropriate
for comparison with other constructions (special geometry  or
$N=4$ supergravity) is given
by the family of matrices:
\begin{equation}
  M(S) \, \equiv \, {\o{1}{n(S)} } \, \left (
\matrix{ \bfone & { \o{{\rm i} -S }{ {\rm i} + S } }\cr
{\o{ {\rm i} + {\bar S} }{ {\rm i} -{\bar S} } } & \bfone \cr}
\right )\quad \quad : \quad \quad
n(S) \, \equiv \, \sqrt{ {\o{4 {\rm Im}S } {
 1+\vert S \vert^2 +2 {\rm Im}S } } }
 \label{su11coset}
\end{equation}
To parametrize the coset $SO(m,n)/SO(m)\times SO(n)$ we can instead
take the usual coset representatives
(see for instance~\cite{CaDFb}):
\begin{equation}
L(X) \, \equiv \, \left (\matrix{ \left ( \bfone + XX^T \right )^{1/2}
& X \cr X^T & \left ( \bfone + X^T X \right )^{1/2}\cr } \right )
\label{somncoset}
\end{equation}
where the $m \times n $ real matrix $X$ provides a set of independent
coordinates. Inserting these matrices into the embedding formulae of
eq.s~\ref{uspembed} we obtain a matrix:
\begin{equation}
\iota_\delta \left ( M (S) \right ) \circ  \iota_\delta
\left ( L(X) \right ) \,
 = \, \left ( \matrix{ U_0(S,X) & U^\star_1(S,X) \cr
U_1(S,X)
& U^\star_0(S,X) \cr } \right ) \, \in \, Usp(n+m , n+m)
\label{uspuspusp}
\end{equation}
that inserted into the master formula of eq.~\ref{masterformula}
yields the following result:
\begin{equation}
{\cal N}\, = \, {\rm i} {\rm Im}S \, \eta L(X) L^T(X) \eta
+ {\rm Re}S \, \eta
\label{maestrina}
\end{equation}
Alternatively, remarking that if $L(X)$ is an $SO(m,n)$ matrix
also $L(X)^\prime =\eta L(X) \eta$ is such a matrix and represents
the same equivalence class, we can rewrite ~\ref{maestrina} in the
simpler form:
\begin{equation}
{\cal N}\, = \, {\rm i} {\rm Im}S \,   L(X)^\prime L^{T\prime}
(X)
+ {\rm Re}S \, \eta
\label{maestrino}
\end{equation}
\section{Special K\"ahler Geometry}
\label{LL4}
\setcounter{equation}{0}
The first discovery that the self-interaction of Wess--Zumino multiplets
is governed by K\"ahler geometry is due to Zumino \cite{sugkgeom_1}
(1979). Independently, the parametrization of the coupling of Wess--Zumino
multiplets to supergravity in terms of a real function, later identified with
the K\"ahler potential, was obtained in \cite{sugkgeom_5,sugkgeom_6}
(1978), shortly after that supergravity had been discovered by
Freedman, Ferrara and van Nieuwenhuizen \cite{sugraprimo} (1976)
and recast
in first order formalism by Deser and Zumino \cite{sugrasecondo}  (1976).
\par
The complete form of standard N=1 supergravity, determined by means
of the superconformal calculus, was obtained in \cite{sugkgeom_4}
(1983), while the geometric interpretation of the coupling structure is
due to Bagger and Witten \cite{sugkgeom_2,sugkgeom_3} (1983).
\par
Special K\"ahler geometry in special coordinates was introduced
in 1984--85 by B. de Wit et al. in \cite{specspec1,BDW}
and E. Cremmer et al. in \cite{skgsugra_2}, where the coupling of
N=2 vector multiplets to N=2 supergravity was fully determined. The
more intrinsic definition of special K\"ahler geometry in terms of
symplectic bundles is due to Strominger \cite{skgmat_1} (1990), who
obtained it in connection with the moduli spaces of
Calabi--Yau compactifications. The coordinate-independent description
and derivation of special K\"ahler geometry in the context of N=2
supergravity is due to Castellani, D'Auria, Ferrara \cite{CaDF}
and to D'Auria, Ferrara, Fre' \cite{DFF} (1991).
Recently Ceresole, D'Auria, Ferrara and Van Proeyen \cite{CDFVP} have
shown how one can and in important instances must dispense of the
notion of holomorphic prepotential $F(X)$.
Let us begin by reviewing the notions of K\"ahler and Hodge--K\"ahler
manifolds that are the prerequisites to introduce the notion of
Special K\"ahler manifolds. Once again we do this in order
to fix our notations.

\subsection{Hodge--K\"ahler manifolds}
\def\mom{{M(k, \IC)}}
Consider a {\sl line bundle}
${\cal L} {\stackrel{\pi}{\longrightarrow}} {\cal M}$ over a K\"ahler
manifold. By definition this is a holomorphic
vector bundle of rank $r=1$. For such bundles the only available
Chern class is the first:
\begin{equation}
c_1 ( {\cal L} ) \, =\, \o{i}{2\pi}
\, {\bar \partial} \,
\left ( \, h^{-1} \, \partial \, h \, \right )\, =
\, \o{i}{2\pi} \,
{\bar \partial} \,\partial \, \mbox{log} \,  h
\label{chernclass23}
\end{equation}
where the 1-component real function $h(z,{\bar z})$ is some hermitian
fibre metric on ${\cal L}$. Let $f (z)$ be a holomorphic section of
the line bundle
${\cal L}$: noting that  under the action of the operator ${\bar
\partial} \,\partial \, $ the term $\mbox{log} \left ({\bar \xi}({\bar z})
\, \xi (z) \right )$ yields a vanishing contribution, we conclude that
the formula in eq.~\ref{chernclass23}  for the first Chern class can be
re-expressed as follows:
\begin{equation}
c_1 ( {\cal L} ) ~=~\o{i}{2\pi} \,
{\bar \partial} \,\partial \, \mbox{log} \,\parallel \, \xi(z) \, \parallel^2
\label{chernclass24}
\end{equation}
where $\parallel \, \xi(z) \, \parallel^2 ~=~h(z,{\bar z}) \,
{\bar \xi}({\bar z}) \,
\xi (z) $ denotes
the norm of the holomorphic section $\xi (z) $.
\par
Eq.~\ref{chernclass24} is the starting point for the definition
of Hodge K\"ahler manifolds,
an essential notion in supergravity theory.
\par
A K\"ahler manifold ${\cal M}$ is a Hodge manifold if and
only if there exists
a line
bundle ${\cal L} \, \longrightarrow \, {\cal M}$ such that its
first Chern class equals
the cohomology class of the K\"ahler 2-form K:
\begin{equation}
c_1({\cal L} )~=~\left [ \, K \, \right ]
\label{chernclass25}
\end{equation}
\par
In local terms this means that there is a holomorphic section
$W(z)$ such that we can write
\begin{equation}
K\, =\, \o{i}{2\pi} \, g_{ij^{\star}} \, dz^{i} \, \wedge \,
d{\bar z}^{j^{\star}} \, = \,
\o{i}{2\pi} \, {\bar \partial} \,\partial \, \mbox{log} \,\parallel \,
W(z) \,
\parallel^2
\label{chernclass26}
\end{equation}
Recalling the local expression of the K\"ahler metric
in terms of the K\"ahler potential
$ g_{ij^{\star}} ~=~{\partial}_i \, {\partial}_{j^{\star}}
{\cal K} (z,{\bar z})$,
it follows from eq.~\ref{chernclass26} that if the
manifold ${\cal M}$ is a Hodge manifold,
then the exponential of the K\"ahler potential
can be interpreted as the metric
$h(z,{\bar z})~=~\exp \left ( {\cal K} (z,{\bar z})\right )$
on an appropriate line bundle ${\cal L}$.
\par
This structure is precisely that advocated by the Lagrangian of
$N=1$ matter coupled supergravity:
the holomorphic section $W(z)$ of the line bundle
${\cal L}$ is what, in N=1 supergravity theory, is
named the superpotential and the logarithm of
its norm  $\mbox{log} \,\parallel \, W(z) \, \parallel^2\, = \,
{\cal K} (z,{\bar z})\, + \, \mbox{log} \, | \, W(z) \, |^2  ~=~ G(z,{\bar z})$
is precisely the  invariant function in terms of which one writes the
potential and Yukawa coupling terms of the supergravity action
(see \cite{sugkgeom_4} and for a review \cite{CaDFb}).
\par
\subsection{Special K\"ahler Manifolds: general discussion}
\par
There are in fact two kinds
of special K\"ahler geometry: the local and the rigid one.
The former describes the scalar field sector of vector multiplets
in $N=2$ supergravity while the latter describes the same sector
in rigid $N=2$  Yang--Mills theories. Since $N=2$
includes $N=1$ supersymmetry, local and rigid special
K\"ahler manifolds must be compatible with the geometric structures that are
respectively enforced by local and rigid $N=1$ supersymmetry in the
scalar sector. The distinction between the two cases
deals with the first Chern--class of the line--bundle
${\cal L} {\stackrel{\pi}{\longrightarrow}} {\cal M}$, whose sections
are the possible superpotentials.  In the local theory $c_1({\cal L})
=[K]$ and this restricts ${\cal M}$ to be a Hodge--K\"ahler manifold.
In the rigid theory, instead, we have $c_1({\cal L})=0$. At the level
of the Lagrangian this reflects into a different behaviour of the
fermion fields. These latter are sections of ${\cal L}^{1/2}$ and
couple to the canonical hermitian connection defined on ${\cal L}$:
\begin{equation}
\begin{array}{ccccccc}
{\theta}& \equiv & h^{-1} \, \partial  \, h = {\o{1}{h}}\, \partial_i h \,
dz^{i} &; &
{\bar \theta}& \equiv & h^{-1} \, {\bar \partial}  \, h = {\o{1}{h}} \,
\partial_{i^\star} h  \,
d{\bar z}^{i^\star} \cr
\end{array}
\label{canconline}
\end{equation}
In the local case where
\begin{equation}
\left  [ \, {\bar \partial}\,\theta \,  \right ] \, = \,
c_1({\cal L}) \, = \, [K]
\label{curvc1}
\end{equation}
the fibre metric $h$ can be identified with the exponential of the
K\"ahler potential and we obtain:
\begin{equation}
\begin{array}{ccccccc}
{\theta}& = &  \partial  \,{\cal K} =  \partial_i {\cal K}
dz^{i} & ; &
{\bar \theta}& = &   {\bar \partial}  \, {\cal K} =
\partial_{i^\star} {\cal K}
d{\bar z}^{i^\star}\cr
\end{array}
\label{curvconline}
\end{equation}
In the rigid case,  ${\cal L}$ is instead a flat bundle and its
metric is unrelated to the K\"ahler potential. Actually one can choose
a vanishing connection:
\begin{equation}
\theta \,= \, {\bar \theta} \, = \, 0
\label{rigconline}
\end{equation}
The distinction between rigid and local special manifolds is the
$N=2$ generalization of this difference occurring at the
$N=1$ level. In the $N=2$ case, in addition to the line--bundle
${\cal L}$ we need a flat holomorphic vector bundle ${\cal SV}
\, \longrightarrow \, {\cal M}$ whose sections can be identified
with the superspace {\it fermi--fermi} components of electric and magnetic
field--strengths (see appendix B). In this way, according
to the discussion of
previous sections the diffeomorphisms of the scalar manifolds
will be lifted to produce an action on the gauge--field strengths
as well. In a supersymmetric theory where scalars and gauge fields
belong to the same multiplet this is a mandatory condition.
However  this symplectic bundle structure must be made
compatible with the line--bundle structure already requested
by $N=1$ supersymmetry. This leads to the existence of
two kinds of special geometry. Another essential distinction
between the two kind of geometries arises from the different
number of vector fields in the theory. In the rigid case
this number equals that of the vector multiplets so that
\begin{eqnarray}
\# \, \mbox{vector fields}\, \equiv \, {\bar n} & = & n
\nonumber\\
\# \, \mbox{vector multiplets}\equiv n & = &
\mbox{dim}_{\bf C} \, {\cal M}\nonumber\\
\mbox{rank} \, {\cal SV}   \, \equiv \, 2\bar n & = & 2 n
 \label{rigrank}
\end{eqnarray}
On the other hand,
in the local case, in addition to the vector fields arising
from the vector multiplets we have also the graviphoton coming from
the graviton multiplet. Hence we conclude:
\begin{eqnarray}
\# \, \mbox{vector fields}\, \equiv \, {\bar n} & = & n+1
\nonumber\\
\# \, \mbox{vector multiplets}\equiv n & = &
\mbox{dim}_{\bf C} \, {\cal M}\nonumber\\
\mbox{rank} \, {\cal SV}   \, \equiv \, 2\bar n & = & 2 n+2
 \label{locrank}
\end{eqnarray}
In the sequel we make extensive use of covariant derivatives with
respect to the canonical connection of the line--bundle ${\cal L}$.
Let us review its normalization. As it is well known there exists
a correspondence between line--bundles and
$U(1)$--bundles. If $\mbox{exp}[f_{\alpha\beta}(z)]$ is the transition
function between two local trivializations of the line--bundle
${\cal L} \, \longrightarrow \, {\cal M}$, the transition function
in the corresponding principal $U(1)$--bundle ${\cal U} \,
\longrightarrow {\cal M}$ is just
$\mbox{exp}[{\rm i}{\rm Im}f_{\alpha\beta}(z)]$ and the K\'ahler potentials
in two different charts are related by:
\begin{equation}
{\cal K}_\beta = {\cal K}_\alpha + f_{\alpha\beta}   + {\bar
{f}}_{\alpha\beta}
\label{carte}
\end{equation}.
At the level of connections this correspondence is formulated by
setting:
\begin{equation}
\mbox{ $U(1)$--connection}   \equiv   {\cal Q} \,  = \,   \mbox{Im}
\theta = -{\o{\rm i}{2}}   \left ( \theta - {\bar \theta}
\right)
\label{qcon}
\end{equation}
If we apply the above formula to the case of the $U(1)$--bundle
${\cal U} \, \longrightarrow \, {\cal M}$
associated with the line--bundle ${\cal L}$ whose first Chern class equals
the K\"ahler class, we get:
\begin{equation}
{\cal Q}  =    -{\o{\rm i}{2}} \left ( \partial_i {\cal K}
dz^{i} -
\partial_{i^\star} {\cal K}
d{\bar z}^{i^\star} \right )
\label{u1conect}
\end{equation}
 Let now
 $\Phi (z, \bar z)$ be a section of ${\cal U}^p$.  By definition its
covariant derivative is
\begin{equation}
\nabla \Phi = (d + i p {\cal Q}) \Phi
\end{equation}
or, in components,
\begin{equation}
\begin{array}{ccccccc}
\nabla_i \Phi &=&
 (\partial_i + {1\over 2} p \partial_i {\cal K}) \Phi &; &
\nabla_{i^*}\Phi &=&(\partial_{i^*}-{1\over 2} p \partial_{i^*} {\cal K})
\Phi \cr
\end{array}
\label{scrivo2}
\end{equation}
A covariantly holomorphic section of ${\cal U}$ is defined by the equation:
$ \nabla_{i^*} \Phi = 0  $.
We can easily map each  section $\Phi (z, \bar z)$
of ${\cal U}^p$
into a  section of the line--bundle ${\cal L}$ by setting:
\begin{equation}
\tilde{\Phi} = e^{-p {\cal K}/2} \Phi  \,   .
\label{mappuccia}
\end{equation}
  With this position we obtain:
\begin{equation}
\begin{array}{ccccccc}
\nabla_i    \tilde{\Phi}&    =&
(\partial_i   +   p   \partial_i  {\cal K})
\tilde{\Phi}& ; &
\nabla_{i^*}\tilde{\Phi}&=& \partial_{i^*} \tilde{\Phi}\cr
\end{array}
\end{equation}
Under the map of eq.~\ref{mappuccia} covariantly holomorphic sections
of ${\cal U}$ flow into holomorphic sections of ${\cal L}$
and viceversa.
\subsection{Special K\"ahler manifolds: the local case}
We are now ready to give the definition of local special K\"ahler
manifolds and illustrate their properties.
A first definition that does not  make direct reference to the
symplectic bundle is the following:
\bd
A Hodge K\"ahler manifold is {\bf Special K\"ahler (of the local type)}
if there exists a completely symmetric holomorphic 3-index section $W_{i
j k}$ of $(T^\star{\cal M})^3 \otimes {\cal L}^2$ (and its
antiholomorphic conjugate $W_{i^* j^* k^*}$) such that the following
identity is satisfied by the Riemann tensor of the Levi--Civita
connection:
\begin{eqnarray}
\partial_{m^*}   W_{ijk}& =& 0   \quad   \partial_m  W_{i^*  j^*  k^*}
=0 \nonumber \\
\nabla_{[m}      W_{i]jk}& =&  0
\quad \nabla_{[m}W_{i^*]j^*k^*}= 0 \nonumber \\
{\cal R}_{i^*j\ell^*k}& =&  g_{\ell^*j}g_{ki^*}
+g_{\ell^*k}g_{j i^*} - e^{2 {\cal K}}
W_{i^* \ell^* s^*} W_{t k j} g^{s^*t}
\label{specialone}
\end{eqnarray}
\label{defspecial}
\ed
In the above equations $\nabla$ denotes the covariant derivative with
respect to both the Levi--Civita and the $U(1)$ holomorphic connection
of eq.~\ref{u1conect}.
In the case of $W_{ijk}$, the $U(1)$ weight is $p = 2$.
\par
The holomorphic sections $W_{ijk}$ have two different physical
interpretations in the case that the special manifold is utilized
as scalar manifold in an N=1 or N=2 theory. In the first case
they correspond to the Yukawa couplings of Fermi families
\cite{petropaolo}. In the second case they provide the coefficients
for the anomalous magnetic moments of the gauginos, since they appear
in the Pauli--terms of the $N=2$ effective action.
Out of the $W_{ijk}$ we can construct covariantly holomorphic
sections of weight 2 and - 2 by setting:
\begin{equation}
C_{ijk}\,=\,W_{ijk}\,e^{  {\cal K}}  \quad ; \quad
C_{i^\star j^\star k^\star}\,=\,W_{i^\star j^\star k^\star}\,e^{  {\cal K}}
\label{specialissimo}
\end{equation}
Next we can give the second more intrinsic definition that relies
on the notion of the flat symplectic bundle.
Let ${\cal L}\, \longrightarrow \,{\cal  M}$ denote the complex
line bundle whose first Chern class equals
the K\"ahler form $K$ of an $n$-dimensional Hodge--K\"ahler
manifold ${\cal M}$. Let ${\cal SV} \, \longrightarrow \,{\cal  M}$
denote a holomorphic flat vector bundle of rank $2n+2$ with structural
group $Sp(2n+2,\IR)$. Consider   tensor bundles of the type
${\cal H}\,=\,{\cal SV} \otimes {\cal L}$.
A typical holomorphic section of such a bundle will be
denoted by ${\Omega}$ and will have the following structure:
\begin{equation}
{\Omega} \, = \, {\twovec{{X}^\Lambda}{{F}_ \Sigma} } \quad
\Lambda,\Sigma =0,1,\dots,n
\label{ololo}
\end{equation}
By definition
the transition functions between two local trivializations
$U_i \subset {\cal M}$ and $U_j \subset {\cal M}$
of the bundle ${\cal H}$ have the following form:
\begin{equation}
{\twovec{X}{ F}}_i=e^{f_{ij}} M_{ij}{\twovec{X}{F}}_j
\end{equation}
where   $f_{ij}$ are holomorphic maps $U_i \cap U_j \, \rightarrow
\,\IC $
while $M_{ij}$ is a constant $Sp(2n+2,\IR)$ matrix. For a consistent
definition of the bundle the transition functions are obviously
subject to the cocycle condition on a triple overlap:
\begin{eqnarray}
e^{f_{ij}+f_{jk}+f_{ki}} &=&1 \ \nn\\
M_{ij} M_{jk} M_{ki} &=& 1 \
\end{eqnarray}
Let $i\langle\ \vert\ \rangle$ be the compatible
hermitian metric on $\cal H$
\begin{equation}
i\langle \Omega \, \vert \, \bar \Omega \rangle \, \equiv \,  -
i \Omega^\T \twomat {0} {\bfone} {-\bfone}{0} {\bar \Omega}
\label{compati}
\end{equation}
\bd
We say that a Hodge--K\"ahler manifold ${\cal M}$
is {\bf special K\"ahler of the local type} if there exists
a bundle ${\cal H}$ of the type described above such that
for some section $\Omega \, \in \, \Gamma({\cal H},{\cal M})$
the K\"ahler two form is given by:
\begin{equation}
K= \o{i}{2\pi}
 \partial \bar \partial \, \mbox{\rm log} \, \left ({\rm i}\langle \Omega \,
 \vert \, \bar \Omega
\rangle \right )
\label{compati1} .
\end{equation}
\ed
From the point of view of local properties, eq.~\ref{compati1}
implies that we have an expression for the K\"ahler potential
in terms of the holomorphic section $\Omega$:
\begin{equation}
{\cal K}\,  = \,  -\mbox{log}\left ({\rm i}\langle \Omega \,
 \vert \, \bar \Omega
\rangle \right )\,
=\, -\mbox{log}\left [ {\rm i} \left ({\bar X}^\Lambda F_\Lambda -
{\bar F}_\Sigma X^\Sigma \right ) \right ]
\label{specpot}
\end{equation}
The relation between the two definitions of special manifolds is
obtained by introducing a non--holomorphic section of the bundle
${\cal H}$ according to:
\begin{equation}
V \, = \, \twovec{L^{\Lambda}}{M_\Sigma} \, \equiv \, e^{{\cal K}/2}\Omega
\,= \, e^{{\cal K}/2} \twovec{X^{\Lambda}}{F_\Sigma}
\label{covholsec}
\end{equation}
so that eq.~\ref{specpot} becomes:
\begin{equation}
1 \, = \,  {\rm i}\langle V  \,
 \vert \, \bar V
\rangle  \,
= \,   {\rm i} \left ({\bar L}^\Lambda M_\Lambda -
{\bar M}_\Sigma L^\Sigma \right )
\label{specpotuno}
\end{equation}
Since $V$ is related to a holomorphic section by eq.~\ref{covholsec}
it immediately follows that:
\begin{equation}
\nabla_{i^\star} V \, = \, \left ( \partial_{i^\star} - {\o{1}{2}}
\partial_{i^\star}{\cal K} \right ) \, V \, = \, 0
\label{nonsabeo}
\end{equation}
On the other hand, from eq.~\ref{specpotuno}, defining:
\begin{equation}
U_i   =   \nabla_i V  =   \left ( \partial_{i} + {\o{1}{2}}
\partial_{i}{\cal K} \right ) \, V   \equiv
\twovec{f^{\Lambda}_{i} }{h_{\Sigma\vert i}}
\label{uvector}
\end{equation}
it follows that:
\begin{equation}
\nabla_i U_j  = {\rm i} C_{ijk} \, g^{k\ell^\star} \, {\bar U}_{\ell^\star}
\label{ctensor}
\end{equation}
where $\nabla_i$ denotes the covariant derivative containing both
the Levi--Civita connection on the bundle ${\cal TM}$ and the
canonical connection $\theta$ on the line bundle ${\cal L}$.
In eq.~\ref{ctensor} the symbol $C_{ijk}$ denotes a covariantly
holomorphic (
$\nabla_{\ell^\star}C_{ijk}=0$) section of the bundle
${\cal TM}^3\otimes{\cal L}^2$ that is totally symmetric in its indices.
This tensor can be identified with the tensor of eq.~\ref{specialissimo}
appearing in eq.~\ref{specialone}.
Alternatively, the set of differential equations:
\begin {eqnarray}
&&\nabla _i V  = U_i\\
 && \nabla _i U_j = {\rm i} C_{ijk} g^{k \ell^\star} U_{\ell^\star}\\
 && \nabla _{i^\star} U_j = g_{{i^\star}j} V\\
 &&\nabla _{i^\star} V =0 \label{defaltern}
\end{eqnarray}
with V satisfying eq.s \ref{covholsec}, \ref {specpotuno} give yet
another definition of special geometry. This is actually what one
obtains from the $N=2$ solution of Bianchi identities (see appendix
A).
In particular it is easy to find eq.~\ref{specialone}
as integrability conditions of~\ref{defaltern}
The {\it period matrix} is now introduced via the relations:
\begin{equation}
{\bar M}_\Lambda = {\bar {\cal N}}_{\Lambda\Sigma}{\bar L}^\Sigma \quad ;
\quad
h_{\Sigma\vert i} = {\bar {\cal N}}_{\Lambda\Sigma} f^\Sigma_i
\label{etamedia}
\end{equation}
which can be solved introducing the two $(n+1)\times (n+1)$ vectors
\begin{equation}
f^\Lambda_I = \twovec{f^\Lambda_i}{{\bar L}^\Lambda} \quad ; \quad
h_{\Lambda \vert I} =  \twovec{h_{\Lambda \vert i}}{{\bar M}_\Lambda}
\label{nuovivec}
\end{equation}
and setting:
\begin{equation}
{\bar {\cal N}}_{\Lambda\Sigma}= h_{\Lambda \vert I} \circ \left (
f^{-1} \right )^I_{\phantom{I} \Sigma}
\label{intriscripen}
\end{equation}
As a consequence of its definition the matrix ${\cal N}$ transforms,
under diffeomorphisms of the base K\"ahler manifold exactly as it
is requested by the rule in eq.~\ref{covarianza}.
Indeed this is the very reason
why the structure of special geometry has been introduced. The
existence of the symplectic bundle ${\cal H} \, \longrightarrow \,
{\cal M}$ is required in order to be able to pull--back the action
of the diffeomorphisms on the field
strengths and to construct the kinetic matrix ${\cal N}$.
\par
From the previous formulae it is easy to derive a set of useful
relations among which we quote the following \cite{CDFp}:
\begin{eqnarray}
{\rm Im}{\cal N}_{\Lambda\Sigma}L^\Lambda\bar{L}^\Sigma &=& -{1 \over 2}
\label{normal}\\
\langle V ,U_i \rangle &=& \langle V, U_{i^\star}\rangle = 0
\label{ortogo}\\
 U^{\Lambda\Sigma} \, \equiv \,  f^\Lambda_i \, f^\Sigma_{j^\star} \,
g^{ij^\star} &=&
-{\o{1}{2}} \, \left ( {\rm Im}{\cal N} \right )^{-1 \vert
\Lambda\Sigma} \, -\, {\bar L}^\Lambda L^\Sigma
\label{dsei} \\
g_{ij^\star} &=& -{\rm i} \langle \, U_{i} \, \vert \, {\bar U}_{j^\star}
\, \rangle = -2 f^\La_i \im \cN_{\La\Si} f^\Si_{j^\star}\ ;\\ 
C_{ijk} &=& \langle \, \nabla_i U_{j} \, \vert \, {  U}_{k} \,
\rangle=f^\La_i \del_j \bar \cN_{\La\Si} f^\Si_k =(\cN-\bar\cN)_{\La\Si}
f^\La_i \del_j f^\Si_k
\label{sympinvloc}
\end{eqnarray}
In particular eq.s \ref{sympinvloc} express the K\"ahler metric and
the anomalous magnetic moments in terms of symplectic invariants.
It is clear from our discussion that nowhere we have
assumed the base K\"ahler manifold to be a homogeneous space. So,
in general, special manifolds are not homogeneous spaces. Yet
there is a subclass of homogenous special manifolds. The homogeneous
symmetric ones were classified by Cremmer and Van Proeyen in
\cite{cremvanp} and are displayed in
table~\ref{homospectable}.
It goes without saying that for homogeneous special manifolds the two
constructions of the period matrix, that provided by the master
formula in eq.~\ref{masterformula} and that given by eq.~\ref{intriscripen}
must agree.In Appendix C we shall shortly verify it in the case
of the manifolds
${\cal ST}[2,n]$ that correspond to the second infinite family of
homogeneous special manifolds displayed in table
~\ref{homospectable}.
\par
Anyhow, since special geometry guarantees the existence of a kinetic
period matrix with the correct covariance property it is evident that
to each special manifold we can associate a duality covariant bosonic
Lagrangian of the type considered in eq.~\ref{gaiazuma}. However special
geometry contains more structures than just the period matrix ${\cal
N}$ and the scalar metric $g_{ij^\star}$.
All the other items of the construction do have
a place and play an essential role in the supergravity Lagrangian and
the supersymmetry transformation rules.

\subsection{Special K\"ahler manifolds: the rigid case}
Let ${\cal M}$ be a K\"ahler manifold  with $\mbox{dim}_{\bf C} \,
{\cal M} \, = \, n$ and  let
${\cal L}\, \longrightarrow \,{\cal  M}$ be a {\bf flat}
line bundle $c_1({\cal L})=0$\footnotemark
\footnotetext{the holomorphic sections of ${\cal L}$ would be
the possible superpotentials if ${\cal M}$ were used as scalar
manifold in an $N=1$ globally supersymmetric theory.}.
Let ${\cal SV} \, \longrightarrow \,{\cal  M}$
denote a holomorphic {\bf flat} vector bundle of rank $2n$ with structural
group $ISp(2n,\IR)$. Consider   tensor bundles of the type
${\cal H}\,=\,{\cal SV} \otimes {\cal L}$.
A typical holomorphic section of such a bundle will be
denoted by ${\Omega}$ and will have the following structure:
\begin{equation}
{\Omega} \, = \, {\twovec{{Y}^I}{{F}_ J} } \quad
I,J =1,\dots,n
\label{ololorig}
\end{equation}
By definition
the transition functions between two local trivializations
$U_i \subset {\cal M}$ and $U_j \subset {\cal M}$
of the bundle ${\cal H}$ have the following form:
\begin{equation}
{\twovec{Y}{ F}}_i=e^{{\hat f}_{ij}} {\hat M}_{ij}{\twovec{Y}{F}}_j
\end{equation}
where   ${\hat f}_{ij}\in \IC$ are purely imaginary complex numbers
while $\hat M_{ij}$ denotes the action of an element $(\hat M,c)
\in ISp(2n,\IR)$
on $\Omega$. $\hat M$ is a symplectic matrix $\hat M\in Sp(2n,\IR)$ 
and $c$ is a
$n$-vector:
\begin{equation}
\pmatrix{\hat M & c}\pmatrix{Y\cr F} = {\hat M}\pmatrix{V\cr F}+
\pmatrix{0\cr c }\ .
\end{equation}
For a consistent
definition of the bundle the transition functions are obviously
subject to the cocycle condition on a triple overlap:
\begin{eqnarray}
e^{{\hat f}_{ij}+{\hat f}_{jk}+{\hat f}_{ki}} &=& 1 \ \nonumber\\
{\hat M}_{ij} {\hat M}_{jk} {\hat M}_{ki} &=& \bfone
\end{eqnarray}
Let $i\langle\ \vert\ \rangle$ be the compatible
hermitian metric on $\cal H$
\begin{equation}
{\rm i}\langle \Omega \, \vert \, \bar \Omega \rangle \, \equiv \, -
{\rm i} \Omega^T \twomat {0} { \bfone} {- \bfone}{0} {\bar \Omega}
\label{compatirig}
\end{equation}
\bd
We say that a Hodge--K\"ahler manifold ${\cal M}$
is {\bf special K\"ahler of the rigid type} if there exists
a bundle ${\cal H}$ of the type described above such that
for some section ${\hat \Omega} \, \in \, \Gamma({\cal H},{\cal M})$
the K\"ahler two form is given by:
\begin{equation}
K=  \, - \, \o{i}{2\pi}
 \partial \bar \partial \left ({\rm i}\langle {\hat \Omega} \,
 \vert \, {\hat {\bar \Omega}}
\rangle \right )
\label{compati1rig} .
\end{equation}
\ed
Just as in the local case eq.~\ref{compati1rig}
yields an expression for the K\"ahler potential
in terms of the holomorphic section ${\hat \Omega}$:
\begin{equation}
\begin{array}{ccccc}
{\cal K}& = &  \left ({\rm i}\langle {\hat \Omega} \,
 \vert \, {\hat {\bar \Omega}}
\rangle \right )
&=& \left [ {\rm i} \left ({\bar Y}^I F_I -
{\bar F}_J Y^J \right ) \right ]
\end{array}
\label{specpotrig}
\end{equation}
Similarly defining
\begin{equation}
{\hat U}_i   =   \partial_i {\hat \Omega}   \equiv
\twovec{f^{I}_{i} }{h_{J\vert i}}
\label{uvectorrig}
\end{equation}
one finds:
\begin{equation}
D_i {\hat U}_j  = {\rm i} C_{ijk} \, g^{k\ell^\star} \,
{\hat {\bar U}}_{\ell^\star}
\label{ctensorrig}
\end{equation}
where $D_i$ is the covariant derivative with respect to the
Levi--Civita connection on ${\cal TM}$ and
where $C_{ijk}$ is a totally symmetric holomorphic
section of the bundle ${\cal TM}^3\otimes{\cal L}^2$:
$\partial_{\ell^\star}C_{ijk}=0$.
Just as in the local case we may alternatively define the rigid special
geometry by the following set of differential equations:
\begin{eqnarray}
\partial_{i^\star}\hat{\Omega} = 0\\
{\hat U}_i   =   \partial_i {\hat \Omega}\\
 D_i {\hat U}_j  = {\rm i} C_{ijk} \, g^{k\ell^\star} \,
{\hat {\bar U}}_{\ell^\star}\label{eqfunrig}
\end{eqnarray}.
The integrability condition of eq.~\ref{eqfunrig}  is similar
but  different from eq.~\ref{specialone} due to the replacement of
the
covariant derivative on ${\cal TM}\times{\cal L}$ by that
on ${\cal TM}$, due to the flatness of ${\cal L}$. We get
\begin{eqnarray}
\partial_{m^*}   C_{ijk}& =& 0   \quad   \partial_m  C_{i^*  j^*  k^*}
=0 \nonumber \\
\nabla_{[m}      C_{i]jk}& =&  0
\quad \nabla _{[m}C_{i^*]j^*k^*}= 0 \nonumber \\
{\cal R}_{i^*j\ell^*k}& =& - C_{i^* \ell^* s^*} C_{t k j} g^{s^*t}
\label{specialonerig}
\end{eqnarray}
which are the rigid counterpart of \ref{specialone}.
The definition of the {\it period matrix} is
obtained in full analogy to eq.~\ref{etamedia}:
\begin{equation}
h_{I\vert i} = {\bar {\cal N}}_{IJ} f^J_i
\label{etaalta}
\end{equation}
that yields:
\begin{equation}
{\bar {\cal N}}_{IJ}= h_{I \vert i} \circ \left (
f^{-1} \right )^i_{\phantom{i} J}
\label{intriscripenrig}
\end{equation}
Finally we observe that, exactly as in the local case,
the metric and the magnetic moments can be expressed in terms of the symplectic
sections:
\begin{equation}
\begin{array}{ccccccc}
g_{ij^\star} &=& -{\rm i} \langle \,{\hat U}_{i} \, \vert \,
{\hat {\bar U}}_{j^\star}
\, \rangle & ; &
C_{ijk} &=& \langle \, \partial_i {\hat U}_{j} \, \vert \, {\hat  U}_{k} \,
\rangle
\label{sympinvrig}
\end{array}
\end{equation}

\subsection{Special K\"ahler manifolds: the issue of special
coordinates}
 So far no privileged coordinate system has been chosen on the base
 K\"ahler manifold ${\cal M}$ and no mention has been made
 of the holomorphic prepotential $F(X)$ that is ubiquitous in the $N=2$
 literature. The simultaneous avoidance
 of privileged coordinates and of the prepotential is not
 accidental. Indeed, when the definition of special K\"ahler
 manifolds  is given in intrinsic terms, as we did in the previous
 subsection, the holomorphic prepotential $F(X)$ can be dispensed
 of. Whether a prepotential $F(X)$ exists or not
 depends on the choice of a symplectic gauge which is
 immaterial in the abelian theory but not in the gauged one.
 Actually, in the local case, it appears that some  physically
 interesting cases are precisely instances where $F(X)$ does not
 exist. On the contrary the prepotential $F(X)$ seems to be
 a necessary ingredient in the tensor calculus constructions of
 $N=2$ theories that for this reason are not completely general.
 This happens because tensor calculus uses special coordinates
 from the very start. Let us then see how the notion of  $F(X)$
 emerges if we resort to special coordinate systems.
 \par
 Note that under a K\"ahler transformation ${\cal K} \, \to \,
 {\cal K} + f(z) +{\bar f}({\bar z})$ the holomorphic section transforms,
 in the local case, as $\Omega \, \to \, \Omega \, e^{-f}$, so that
 we have $X^\Lambda \, \to \, X^\Lambda \, e^{-f}$. This means that,
 at least locally, the upper half of $\Omega$ (associated with
 the electric field strengths) can be regarded as a set $X^\Lambda$
 of homogeneous coordinates on ${\cal M}$, provided that the
 jacobian matrix
 \begin{equation}
e^I_{i}(z) = \partial_{i} \left ( {\o{X^I}{X^0}}\right ) \quad ;
\quad a=1,\dots ,n
\label{nonsingcoord}
\end{equation}
is invertible. In this case, for the lower part of the symplectic
section $\Omega$ we obtain $F_\Lambda = F_\Lambda(X)$. Recalling
eq.s~\ref{ortogo}, in particular:

\begin{equation}
0 \, =\,  \langle\,  V \, \vert \, U_i \, \rangle
\, =\, X^\Lambda \, \partial_{i} F_\Lambda - \partial_{i} X^\Lambda \,
F_\Lambda
\label{ortogonalcosa}
\end{equation}
we obtain:
\begin{equation}
X^\Sigma \, \partial_ \Sigma F_ \Lambda (x) \, = \, F_ \Lambda (X)
\label{nipiolbuit}
\end{equation}
so that we can conclude:
\begin{equation}
F_ \Lambda (X) \, = \, {\o{\partial}{\partial X^\Lambda} } F(X)
\label{lattedisoia}
\end{equation}
where $F(X)$ is a homogeneous function of degree 2 of the homogeneous
coordinates $X^\Lambda$. Therefore,when the determinant of the
Jacobian ~\ref{nonsingcoord} is non vanishing,
we can use the {\it special coordinates}:
\begin{equation}
t^I \, \equiv \, {\o{X^I}{X^0}}
\label{speccoord}
\end{equation}
and the whole geometric structure can be derived by a single
holomorphic prepotential:
\begin{equation}
{\cal F}(t) \, \equiv \,  (X^0)^{-2} F(X)
\label{gianduia}
\end{equation}
In particular, eq.~\ref{specpot} for
the K\"ahler potential becomes
\begin{equation}
{\cal K}(t, \bar t) = -\mbox{log} \,\,  {\rm i}\Bigl [
2 \left ( {\cal F} - \bar {\cal F} \right )  -
\left ( \partial_I {\cal F}
+ \partial_{{I^\star}} \bar {\cal F} \right )\left  ( t^I -\bar t^{I^\star}
\right ) \Bigr ]
\label{vecchiaspec}
\end{equation}
while eq.~\ref{sympinvloc} for the magnetic moments simplifies into
\begin{equation}
W_{IJK}= \partial_I \partial_J \partial_K {\cal F}(t)
\label{yuktre}
\end{equation}
Finally we note that in the rigid case the Jacobian from a generic
parametrisation to special coordinates
\begin{equation}
e^I_{i}(z) = \partial_{i} \left ( {\o{X^I}{X^0}}\right )=A+B{\bar{\cal  N}}
\label{nonsingcoord2}
\end{equation}
 cannot have zero eingenvalues, and therefore the function $F$ always
 exist. In this case the matrix $\bar\cN$ coincides with $\frac{\del^2 F}
{\del X^I \del X^J}$.

\section{Hypergeometry}
\label{LL6}
\setcounter{equation}{0}
Next we turn to the hypermultiplet sector of an $N=2$ theory.
Here there are $4$ real scalar fields for each hypermultiplet
and, at least locally, they can be regarded as the four
components of a quaternion. The locality caveat is, in this case,
very substantial because global quaternionic coordinates can
be constructed only occasionally even on those manifolds
that are denominated quaternionic in the mathematical literature
\cite{alekseev}, \cite{gal}.
Anyhow, what is important  is that,
in the hypermultiplet sector, the scalar manifold ${\cal HM}$
has dimension multiple of four:
\begin{equation}
\mbox{dim}_{\bf R} \, {\cal HM} \, = \, 4 \, m \,\equiv \,
4 \, \# \, \mbox{of hypermultiplets}
\label{quatdim}
\end{equation}
and, in some appropriate
sense, it has a quaternionic structure.
\par
As {\it Special K\"ahler} is the collective name given
to the vector multiplet geometry both in the rigid and
in the local case, in the same way  we name {\it Hypergeometry}
that pertaining to the hypermultiplet sector, irrespectively whether
we deal with global or local N=2 theories. Yet in the very same
way as there are two kinds of special geometries, there are also
two kinds of hypergeometries and for a very similar reason.
Supersymmetry requires the existence of a principal
$SU(2)$--bundle
\begin{equation}
{\cal SU} \, \longrightarrow \, {\cal HM}
\label{su2bundle}
\end{equation}
that plays for hypermultiplets the same role played by the
the line--bundle ${\cal L} \, \longrightarrow \, {\cal SM}$
in the case of vector multiplets. As it happens there the
bundle ${\cal SU}$ is {\bf flat} in the {\it rigid case} while its
curvature is proportional to the K\"ahler forms in the
{\it local case}.
\par
The difference with the case of vector multiplets
is that rigid and local hypergeometries were already known in
mathematics prior to their use \cite{dWLVP},
\cite{skgsugra_13}, \cite{DFF}, \cite{sabarwhal},
\cite{vanderseypen}  in the context of $N=2$ supersymmetry
and had the following names:
\begin{eqnarray}
\mbox{rigid hypergeometry} & \equiv & \mbox{HyperK\"ahler geom.}
\nonumber\\
\mbox{local hypergeometry} & \equiv & \mbox{Quaternionic
geom.}
\label{picchio}
\end{eqnarray}
\subsection{Quaternionic, versus HyperK\"ahler manifolds}
Both a quaternionic or a HyperK\"ahler manifold ${\cal HM}$
  is a $4 m$-dimensional real manifold
endowed with a metric $h$:
\begin{equation}
d s^2 = h_{u v} (q) d q^u \otimes d q^v   \quad ; \quad u,v=1,\dots,
4  m
\label{qmetrica}
\end{equation}
and three complex structures
\begin{equation}
(J^x) \,:~~ T({\cal HM}) \, \longrightarrow \, T({\cal HM}) \qquad
\quad
(x=1,2,3)
\end{equation}
that satisfy the quaternionic algebra
\begin{equation}
J^x J^y = - \delta^{xy} \, \bfone \,  +  \, \epsilon^{xyz} J^z
\label{quatalgebra}
\end{equation}
and respect to which the metric is hermitian:
\begin{equation}
\forall   \mbox{\bf X} ,\mbox{\bf Y}  \in   T{\cal HM}   \,: \quad
h \left( J^x \mbox{\bf X}, J^x \mbox{\bf Y} \right )   =
h \left( \mbox{\bf X}, \mbox{\bf Y} \right ) \quad \quad
  (x=1,2,3)
\label{hermit}
\end{equation}
From eq.~\ref{hermit} it follows that one can introduce
a triplet of  2-forms
\begin{equation}
\begin{array}{ccccccc}
K^x& = &K^x_{u v} d q^u \wedge d q^v & ; &
K^x_{uv} &=&   h_{uw} (J^x)^w_v \cr
\end{array}
\label{iperforme}
\end{equation}
that provides the generalization of the concept of K\"ahler form
occurring in  the complex case. The triplet $K^x$ is named
the {\it HyperK\"ahler} form. It is an $SU(2)$ Lie--algebra valued
2--form  in the same way as the K\"ahler form is a $U(1)$ Lie--algebra
valued 2--form. In the complex case the definition of K\"ahler manifold
involves the statement
that the K\"ahler 2--form is closed. At the same time in
Hodge--K\"ahler manifolds (those appropriate to local supersymmetry)
the K\"ahler 2--form can be identified with the curvature of
a line--bundle which in the case of rigid supersymmetry
is flat. Similar steps can be taken also here and lead to two possibilities:
either HyperK\"ahler or Quaternionic manifolds.
\par
Let us  introduce a principal $SU(2)$--bundle
${\cal SU}$ as defined in eq.~\ref{su2bundle}. Let $\omega^x$ denote a
connection on such a bundle.
To obtain either a HyperK\"ahler or a quaternionic manifold
we must impose the condition that the HyperK\"ahler 2--form
is covariantly closed with respect to the connection $\omega^x$:
\begin{equation}
\nabla K^x \equiv d K^x + \epsilon^{x y z} \omega^y \wedge
K^z    \, = \, 0
\label{closkform}
\end{equation}
The only difference between the two kinds of geometries resides
in the structure of the ${\cal SU}$--bundle.
\bd
A HyperK\"ahler manifold is a $4 m$--dimensional manifold with
the structure described above and such that the ${\cal SU}$--bundle
is {\bf flat}
\ed
 Defining the ${\cal SU}$--curvature by:
\begin{equation}
\Omega^x \, \equiv \, d \omega^x +
{1\over 2} \epsilon^{x y z} \omega^y \wedge \omega^z
\label{su2curv}
\end{equation}
in the HyperK\"ahler case we have:
\begin{equation}
\Omega^x \, = \, 0
\label{piattello}
\end{equation}
Viceversa
\bd
A quaternionic manifold is a $4 m$--dimensional manifold with
the structure described above and such that the curvature
of the ${\cal SU}$--bundle
is proportional to the HyperK\"ahler 2--form
\ed
Hence, in the quaternionic case we can write:
\begin{equation}
\Omega^x \, = \, { {\lambda}}\, K^x
\label{piegatello}
\end{equation}
where $\lambda$ is a non vanishing real number.
\par
As a consequence of the above structure the manifold ${\cal HM}$ has a
holonomy group of the following type:
\begin{eqnarray}
{\rm Hol}({\cal HM})&=& SU(2)\otimes {\cal H} \quad (\mbox{quaternionic})
\nonumber \\
{\rm Hol}({\cal HM})&=& \bfone \otimes {\cal H} \quad (\mbox{HyperK\"ahler})
\nonumber \\
{\cal H} & \subset & Sp (2m,\IR)
\label{olonomia}
\end{eqnarray}
In both cases, introducing flat
indices $\{A,B,C= 1,2\} \{\alpha,\beta,\gamma = 1,.., 2m\}$  that run,
respectively, in the fundamental representations of $SU(2)$ and
$Sp(2m,\IR)$, we can find a vielbein 1-form
\begin{equation}
{\cal U}^{A\alpha} = {\cal U}^{A\alpha}_u (q) d q^u
\label{quatvielbein}
\end{equation}
such that
\begin{equation}
h_{uv} = {\cal U}^{A\alpha}_u {\cal U}^{B\beta}_v
\IC_{\alpha\beta}\epsilon_{AB}
\label{quatmet}
\end{equation}
where $\IC_{\alpha \beta} = - \IC_{\beta \alpha}$ and
$\epsilon_{AB} = - \epsilon_{BA}$ are, respectively, the flat $Sp(2m)$
and $Sp(2) \sim SU(2)$ invariant metrics.
The vielbein ${\cal U}^{A\alpha}$ is covariantly closed with respect
to the $SU(2)$-connection $\omega^z$ and to some $Sp(2m,\IR)$-Lie Algebra
valued connection $\Delta^{\alpha\beta} = \Delta^{\beta \alpha}$:
\begin{eqnarray}
\nabla {\cal U}^{A\alpha}& \equiv & d{\cal U}^{A\alpha}
+{i\over 2} \omega^x (\epsilon \sigma_x\epsilon^{-1})^A_{\phantom{A}B}
\wedge{\cal U}^{B\alpha} \nonumber\\
&+& \Delta^{\alpha\beta} \wedge {\cal U}^{A\gamma} \IC_{\beta\gamma}
=0
\label{quattorsion}
\end{eqnarray}
\noindent
where $(\sigma^x)_A^{\phantom{A}B}$ are the standard Pauli matrices.
Furthermore ${ \cal U}^{A\alpha}$ satisfies  the reality condition:
\begin{equation}
{\cal U}_{A\alpha} \equiv ({\cal U}^{A\alpha})^* = \epsilon_{AB}
\IC_{\alpha\beta} {\cal U}^{B\beta}
\label{quatreality}
\end{equation}
Eq.\ref{quatreality}  defines  the  rule to lower the symplectic indices by
means   of  the  flat  symplectic   metrics $\epsilon_{AB}$   and
$\IC_{\alpha \beta}$.
More specifically we can write a stronger
version of eq.~\ref{quatmet}\cite{sugkgeom_3}:
\begin{eqnarray}
({\cal U}^{A\alpha}_u {\cal U}^{B\beta}_v + {\cal U}^{A\alpha}_v {\cal
 U}^{B\beta}_u)\IC_{\alpha\beta}&=& h_{uv} \epsilon^{AB}\nonumber\\
({\cal U}^{A\alpha}_u {\cal U}^{B\beta}_v + {\cal U}^{A\alpha}_v {\cal
U}^{B\beta}_u) \epsilon_{AB} &=& h_{uv} {1\over m} \IC^{\alpha
\beta}
\label{piuforte}
\end{eqnarray}
\noindent
We have also the inverse vielbein ${\cal U}^u_{A\alpha}$ defined by the
equation
\begin{equation}
{\cal U}^u_{A\alpha} {\cal U}^{A\alpha}_v = \delta^u_v
\label{2.64}
\end{equation}
Flattening a pair of indices of the Riemann
tensor ${\cal R}^{uv}_{\phantom{uv}{ts}}$
we obtain
\begin{equation}
{\cal R}^{uv}_{\phantom{uv}{ts}} {\cal U}^{\alpha A}_u {\cal U}^{\beta B}_v =
-\,{{\rm i}\over 2} \Omega^x_{ts} \epsilon^{AC}
 (\sigma_x)_C^{\phantom {C}B} \IC^{\alpha \beta}+
 \IR^{\alpha\beta}_{ts}\epsilon^{AB}
\label{2.65}
\end{equation}
\noindent
where $\IR^{\alpha\beta}_{ts}$ is the field strength of the $Sp(2m)
$ connection:
\begin{equation}
d \Delta^{\alpha\beta} + \Delta^{\alpha \gamma} \wedge \Delta^{\delta \beta}
\IC_{\gamma \delta} \equiv \IR^{\alpha\beta} = \IR^{\alpha \beta}_{ts}
dq^t \wedge dq^s
\label{2.66}
\end{equation}
Eq. ~\ref{2.65} is the explicit statement that the Levi Civita connection
associated with the metric $h$ has a holonomy group contained in
$SU(2) \otimes Sp(2m)$. Consider now eq.s~\ref{quatalgebra},
\ref{iperforme} and~\ref{piegatello}.
We easily deduce the following relation:
\begin{equation}
h^{st} K^x_{us} K^y_{tw} = -   \delta^{xy} h_{uw} +
  \epsilon^{xyz} K^z_{uw}
\label{universala}
\end{equation}
that holds true both in the HyperK\"ahler and in the Quaternionic case.
In the latter case, using eq. \ref{piegatello},
eq.~\ref{universala} can be rewritten as follows:
\begin{equation}
h^{st} \Omega^x_{us} \Omega^y_{tw} = - \lambda^2 \delta^{xy} h_{uw} +
\lambda \epsilon^{xyz} \Omega^z_{uw}
\label{2.67}
\end{equation}
Eq.~\ref{2.67} implies that the intrinsic components of the curvature
 2-form $\Omega^x$ yield a representation of the quaternion algebra.
In the HyperK\"ahler case such a representation is provided only
by the HyperK\"ahler form.
In the quaternionic case we can write:
\begin{equation}
\Omega^x_{A\alpha, B \beta} \equiv \Omega^x_{uv} {\cal U}^u_{A\alpha}
{\cal U}^v_{B\beta} = - i \lambda \IC_{\alpha\beta} (\sigma_x)_A^{\phantom
{A}C}\epsilon _{CB}
\label{2.68}
\end{equation}
\noindent
Alternatively eq.\ref{2.68} can be rewritten in an intrinsic form as
\begin{equation}
\Omega^x =\,-{\rm i}\, \lambda \IC_{\alpha\beta}
(\sigma _x)_A^{\phantom {A}C}\epsilon _{CB} {\cal U}^{\alpha A} \wedge {\cal
U}^{\beta B}
\label{2.69}
\end{equation}
\noindent
whence we also get:
\begin{equation}
{i\over 2} \Omega^x (\sigma_x)_A^{\phantom{A}B} =
\lambda{\cal U}_{A\alpha} \wedge {\cal
U}^{B\alpha}
\label{2.70}
\end{equation}
Homogeneous symmetric quaternionic spaces are displayed in Table
~\ref{quatotable}.

\section{The Gauging}
\label{LL7}
\setcounter{equation}{0}
With the above discussion of HyperK\"ahler and Quaternionic manifolds
we have  completed the review of the geometric structures
involved in the construction of an {\it abelian, ungauged} $N=2$
supergravity or of an abelian $N=2$ rigid gauge theory.
 As we are going to see in the next section,
 the bosonic Lagrangian of $N=2$ supergravity coupled to $n$
abelian vector multiplets and $m$ hypermultiplets is the
following:
\begin{eqnarray}
{\cal L}_{ungauged}^{SUGRA \vert Bose} & =&
\sqrt{-g}\Bigl  [ \,  R[g]
\, + \, g_{i {j^\star}}(z,\bar z )\, \partial^{\mu} z^i \,
\partial _{\mu} \bar z^{j^\star} \,
-  \,\lambda\, h_{uv}(q) \, \partial^{\mu} q^u \,
 \partial _{\mu} q^v
\nonumber\\
& & + \,{\rm i} \,\left(
\bar {\cal N}_{\Lambda \Sigma} {\cal F}^{- \Lambda}_{\mu \nu}
{\cal F}^{- \Sigma \vert {\mu \nu}}
\, - \,
{\cal N}_{\Lambda \Sigma}
{\cal F}^{+ \Lambda}_{\mu \nu} {\cal F}^{+ \Sigma \vert {\mu \nu}} \right )
\, \Bigr ]
\label{ungausugra}
\end{eqnarray}
where the  $n$ complex fields $z^{i}$ span some {\it special K\"ahler
manifold of the local type} ${\cal SM}$ and the $4 m$ real fields
$q^u$ span a quaternionic manifold ${\cal HM}$. By $g_{ij^\star}$
and $h_{uv}$ we have denoted the metrics on these two manifolds.
The proportionality constant between the $SU(2)$ curvature and the
HyperK\"ahler form  appearing in the Lagrangian is fixed
to the value $\lambda=-1$ if we want canonical kinetic terms
for the hypermultiplet scalars. The period matrix
${\cal N}_{\Lambda\Sigma}$ depends only on the special manifold
coordinates $z^{i}, {\bar z}^{j^\star}$ and it is expressed in terms
of the symplectic sections of the flat symplectic bundle by
eq.~\ref{intriscripen}.
On the other hand the bosonic Lagrangian of a rigid $N=2$ abelian gauge
theory containing $n$ vector multiplets and coupled to
$m$ hypermultiplets is the following one:
\begin{eqnarray}
 {\cal L}_{ungauged}^{YM \vert Bose} & =&
 g_{i {j^\star}}(z,\bar z )\, \partial^{\mu} z^i \,
\partial _{\mu} \bar z^{j^\star} \,
+ \,  h_{uv}(q) \, \partial^{\mu} q^u \,
 \partial _{\mu} q^v
\nonumber\\
& & + \,{\rm i} \,\left(
\bar {\cal N}_{IJ} {\cal F}^{- I}_{\mu \nu}
{\cal F}^{- J \vert {\mu \nu}}
\, - \,
{\cal N}_{IJ}
{\cal F}^{+ I}_{\mu \nu} {\cal F}^{+ J \vert {\mu \nu}} \right )
\,     \nonumber \\
\label{ungaugauga}
\end{eqnarray}
where the  $n$ complex fields $z^{i}$ span some {\it special K\"ahler
manifold of the rigid type} ${\cal SM}$ and the $4 m$ real fields
$q^u$ span a HyperK\"ahler manifold ${\cal HM}$. By $g_{ij^\star}$
and $h_{uv}$ we have denoted the metrics on these two manifolds.
   The period matrix
${\cal N}_{IJ}$ depends only on the special manifold
coordinates $z^{i}, z^{j^\star}$ and it is expressed in terms
of the symplectic sections of the flat symplectic bundle by
eq.~\ref{intriscripenrig}.
In both theories there are no electric or magnetic currents and
we have on shell {\it symplectic covariance}. By means of
the first homomorphism in eq.~\ref{spaccoindue} any diffeomorphism
of the scalar manifold can be lifted to a symplectic transformation
on the electric--magnetic field strengths, the {\it period} matrix
transforming, by construction, covariantly as required by
eq.~\ref{covarianza}. Under this lifting any isometry of the scalar manifold
becomes a symmetry of the differential system made by the equations of
motions plus Bianchi identities. There are in fact three type of these
isometries:
\begin{enumerate}
\item{{\it The classical symmetries}, namely those isometries
$\xi \, \in \, {\cal I} \left ( {\cal M}_{scalar} \right )$
whose image in the symplectic group is block--diagonal:
\begin{equation}
\iota_\delta ( \xi ) \, = \, \twomat{A_ \xi}{  0}{  0}{(A^T_ \xi)^{-1}}
\end{equation}
 These transformations are exact ordinary symmetries of the Lagrangian.
They clearly form a subgroup
\begin{equation}
 {\cal C}las \left ( {\cal M}_{scalar} \right ) \, \subset \,
 {\cal I} \left ( {\cal M}_{scalar} \right )
 \label{classym}
\end{equation}
   }
 \item{{\it The perturbative symmetries}, namely those isometries
$\xi \, \in \, {\cal I} \left ( {\cal M}_{scalar} \right )$
whose image in the symplectic group is lower triangular:
\begin{equation}
\iota_\delta ( \xi ) \, = \, \twomat{A_ \xi}{  0}{  C_ \xi}
{(A^T_ \xi)^{-1}}
\label{persym}
\end{equation}
 These transformations map the electric field strengths
 into linear combinations of the electric field strengths and can
 be reduced to linear transformations of the
 gauge potentials. They are almost  invariances of the action.  Indeed
 the only non--invariance comes from the transformation of the period
 matrix
\begin{equation}
{\cal N} \, \longrightarrow  \, (A^T_ \xi)^{-1} \, {\cal N}
(A_ \xi)^{-1}  \, + \, C_ \xi \, (A^T_ \xi)^{-1}
\end{equation}
Denoting collectively all the fields of the theory by  $\Phi$
and utilizing eq.s~\ref{gaiazuma},~\ref{gaiazumadue},~\ref{scripten},
~\ref{lagrapm},~\ref{covarianza}, under a perturbative transformation
the action changes as follows:
\begin{eqnarray}
\int \, {\cal L}(\Phi) \, d^4 x  &  \to &
\int \, {\cal L}(\Phi^\prime) \, d^4 x  +
\,  \Delta\theta_{\Lambda\Sigma} \, \int F^\Lambda \wedge F^\Sigma
\nonumber\\
\Delta\theta _{\Lambda\Sigma} & = &  {1\over 2}\,\left [ C_\xi \, (A^T_\xi)^{-1}
\right ]_{\Lambda\Sigma}
\label{topogatto}
\end{eqnarray}
\par
The added term is a total derivative and does not affect the field
equations. Quantum mechanically, however, it is relevant. It
corresponds to a redefinition of the $theta$--angle. It yields
a symmetry of the path--integral as long as the added term
is an integer multiple of $2 \pi  \hbar $. This consideration
will restrict the possible perturbative transformations to
a discrete subgroup. In any case the group
of perturbative isometries defined by eq.~\ref{persym}  contains
the group of classical isometries as a subgroup:
$ {\cal I} \left ( {\cal M}_{scalar} \right ) \supset
  {\cal P}ert \left ( {\cal M}_{scalar} \right ) \supset
 {\cal C}las \left ( {\cal M}_{scalar} \right )$. }
\item{{\it The non--perturbative symmetries} namely those isometries
$\xi \, \in \, {\cal I} \left ( {\cal M}_{scalar} \right )$
whose image in the symplectic group is of the form:
\begin{equation}
\iota_\delta ( \xi ) \, = \, \twomat{  A_ \xi}{   B_ \xi }{  C_ \xi}
{  D_\xi}
\label{nonpersym}
\end{equation}
with $B_\xi \ne 0$.
These transformations are neither a symmetry of the classical
action nor of the perturbative path integral. Yet they are a
symmetry of the quantum theory. They exchange electric field
strengths with magnetic ones, electric currents with magnetic ones
and hence elementary excitations with soliton states. }
\end{enumerate}
The above discussion of duality symmetries may be
intriguing for the following reason. How can we talk
about non--perturbative symmetries that  exchange electric charges
with magnetic charges if, so far, in the abelian theories described
by eq.s~\ref{ungausugra} and ~\ref{ungaugauga} there are neither
electric nor magnetic couplings? The answer is that the same general
form of abelian theories encoded in these equations can be taken to
represent two quite different things:
\begin{enumerate}
\item {The fundamental theory  prior to the gauging. It is neutral
and abelian since the  non--abelian
interactions and the electric charges are introduced only by the gauging,
but it contains all the fundamental fields.}
\item {The effective theory of the massless modes of the non--abelian
theory. It is abelian and neutral because the only fields which
remain massless are, apart from the graviton, the multiplets in the
Cartan subalgebra ${\cal H} \subset {\cal G}$ of the gauge group and
the neutral hypermultiplets corresponding to flat directions of the
scalar potential.}
\end{enumerate}
What distinguishes the two cases is the type of scalar manifolds
and their isometries.
\par
In case 1)  we have:
\begin{eqnarray}
\mbox{dim}_{\bf C} \, {\cal SM}&= & n \equiv \mbox{dim} \, {\cal G}
\nonumber\\
{\o{1}{4}}\,\mbox{dim}_{\bf R} \, {\cal HM}&= &
{\hat m} \equiv \# \, \mbox{of all hypermul.} 
\label{microscop}
\end{eqnarray}
\par
while in case 2) we have instead:
\begin{eqnarray}
\mbox{dim}_{\bf C} \, {\cal SM}&= & r \equiv \mbox{rank} \, {\cal G}
\nonumber\\
{\o{1}{4}}\,\mbox{dim}_{\bf R} \, {\cal HM}&= &
{  m} \equiv \# \, \mbox{of moduli hypermul.} 
\label{macroscop}
\end{eqnarray}
As far as the gauging of the $N=2$ theory is concerned, the problem
consists in identifying the gauge group $\cal G$ as a subgroup, at
most of dimension $n+1$ of the isometries of the product space
\begin{equation}
{\cal SM}\times {\cal HM}\ .
\end{equation}

Here we shall mainly consider two cases even if more general situations
are possible. The first is when the gauge group $G$ is non abelian, the
second is when it is the abelian group $G=U(1)^{n_V+1}$.
In the first case supersymmetry requires that $G$ be
a subgroup of the isometries of $\cM$, since the scalars (more precisely,
the sections
$L^\La$) must belong to the adjoint representation of $G$. In such case
the hypermultiplet space will generically split into\cite{dere}
\be
n_H=\sum_i n_i R_i+\half\sum_l n^P_l R^P_l
\ee
where $R_i$ and $R^P_l$ are a set of irreducible representations of
$G$ and $R^P_l$ denote pseudoreal representations.

\par In the abelian case, the special manifold is not required to have any
isometry and if the hypermultiplets are charged with respect to the $n_V+1$
$U(1)$'s, then the $\cQ$ manifold should at least have $n_V+1$ abelian
isometries.

As a consequence of gauging the Lagrangians in eq.s~\ref{ungausugra} and
~\ref{ungaugauga} get  modified by the replacement of
ordinary derivatives with covariant derivatives
and by the introduction of new terms that are
of two types:
\begin{enumerate}
\item {fermion--fermion bilinears with scalar field dependent
coefficients}
\item{A scalar potential ${\rm V}$ }
\end{enumerate}
It is particularly nice and rewarding that all the modifications
of the Lagrangian and of the supersymmetry transformation rules
can be described in terms of a very general geometric construction
associated with the action of Lie--Groups on manifolds that
admit a symplectic structure: {\it the momentum map}.
In supersymmetry  indeed,
the geometric notion of {\it momentum map} has an exact correspondence
with the notion of {\it gauge multiplet auxiliary fields}
or {\it $D$--fields}. Next section
is devoted to a review of the momentum map and to its
applications in N=2 theories.
\section{The Momentum Map}
\label{LL8}
\setcounter{equation}{0}
The momentum map is a construction that applies to all manifolds
with a symplectic structure, in particular to K\"ahler, HyperK\"ahler
and Quaternionic manifolds.
\par
Let us begin with the K\"ahler case, namely with the momentum
map of holomorphic isometries. The HyperK\"ahler and quaternionic
case  correspond, instead, to the momentum map of triholomorphic
isometries.
\subsection{Holomorphic momentum map on K\"ahler manifolds}
Let  $g_{i {j^\star}}$ be the K\"ahler metric of a K\"ahler
manifold ${\cal M}$:   it appears in the kinetic
term of the scalar fields: the Wess--Zumino multiplet scalars in N=1
theories, the vector multiplet scalars in N=2 theories.
If the metric $g_{i {j^\star}}$ has a non trivial group of
continuous isometries ${\cal G}$
generated by Killing vectors $k_\Lambda^i$ ($\Lambda=1, \ldots, {\rm dim}
\,{\cal G} )$, then the kinetic
Lagrangian  admits ${\cal G}$ as a group of global space--time
symmetries. Indeed under an infinitesimal variation
\begin{equation}
z^i \to z^i + \epsilon^\Lambda k_\Lambda^i (z)
\end{equation}
${\cal L}_{kin}$ remains invariant. Furthermore if all the
couplings of the scalar fields are performed in a diffeomorphic
invariant way, then any isometry of $g_{i {j^\star}}$ extends
from a symmetry
of ${\cal L}_{kin}$ to a symmetry of the whole Lagrangian.
Diffeomorphic invariance means that the scalar fields can appear only
through the metric, the Christoffel symbol in the covariant derivative
and through the curvature. Alternatively they can appear through
sections of vector bundles constructed over ${\cal M}$. Typical
case is the dependence on the scalar fields introduced by the
{\it period matrix} ${\cal N}$.
\par
Let $k^i_{\Lambda} (z)$ be a basis of holomorphic Killing vectors for
the metric $g_{i{j^\star}}$.  Holomorphicity means the following
differential constraint:
\begin{equation}
\partial_{j^*} k^i_{\Lambda} (z)=0
\leftrightarrow \partial_j k^{i^*}_{\Lambda} (\bar z)=0 \label{holly}
\end{equation}
while the generic Killing equation (suppressing the
gauge index $\Lambda$):
\begin{equation}
\nabla_\mu k_\nu +\nabla_\mu k_\nu=0
\end{equation}
in holomorphic indices reads as follows:
\begin{equation}
\begin{array}{ccccccc}
\nabla_i k_{j} + \nabla_j k_{i} &=&0 & ; &
\nabla_{i^*} k_{j} + \nabla_j k_{i^*} &=& 0
\label{killo}
\end{array}
\end{equation}
where the covariant components are defined as
$k_{j }=g_{j i^*} k^{i^*}$ (and similarly for
$k_{i^*}$).
\par
The vectors $k_{\Lambda}^i$ are generators of infinitesimal
holomorphic coordinate transformations:
\begin{equation}
\delta z^i = \epsilon^\Lambda k^i_{\Lambda} (z)
\end{equation}
which leave the metric invariant.
In the same way as the metric is the derivative of a more fundamental
object, the Killing vectors in a K\"ahler manifold are the
derivatives of suitable prepotentials. Indeed the first of
eq.s~\ref{killo}  is automatically satisfied by holomorphic vectors
and the second equation reduces to the following one:
\begin{equation}
k^i_{\Lambda}=i g^{i j^*} \partial_{j^*} {\cal P}_{\Lambda},
\quad {\cal P}^*_{\Lambda} = {\cal P}_{\Lambda}\label{killo1}
\end{equation}
In other words if we can find a real function ${\cal P}^\Lambda$ such
that the expression $i g^{i j^*} \partial_{j^*}
{\cal P}_{(\Lambda)}$ is holomorphic, then eq.~\ref{killo1} defines a
Killing vector.
\par
The construction of the Killing prepotential can be stated in a more
precise geometrical formulation which involves the notion of
{\it momentum
map}. Let us review this construction which reveals
another  deep connection between supersymmetry and
geometry.
\par
 Consider a K\"ahlerian manifold ${\cal M}$
of real dimension $2n$.
Consider a compact Lie group ${\cal G}$ acting on
 ${\cal M}$  by means of Killing vector
fields ${\bf X}$ which are holomorphic
with respect to the  complex structure
${ J}$ of ${\cal M}$; then these vector
fields preserve also the K\"ahler 2-form
\begin{equation}
\begin{array}{ccc}
\matrix{
{\cal L}_{\scriptscriptstyle{\bf X}}g = 0 & \leftrightarrow &
\nabla_{(\mu}X_{\nu)}=0 \cr
{\cal L}_{\scriptscriptstyle{\bf X}}{  J}= 0 &\null &\null \cr }
  \Biggr \} & \Rightarrow &  0={\cal L}_{\scriptscriptstyle{\bf X}}
K = i_{\scriptscriptstyle{\bf X}}
dK+d(i_{\scriptscriptstyle{\bf X}}
K) = d(i_{\scriptscriptstyle{\bf X}}K) \cr
\end{array}
\label{holkillingvectors}
\end{equation}
Here ${\cal L}_{\scriptscriptstyle{\bf X}}$ and
$i_{\scriptscriptstyle{\bf X}}$
denote respectively the Lie derivative along
the vector field ${\bf X}$ and the contraction
(of forms) with it.
\par
If ${\cal M}$ is simply connected,
$d(i_{{\bf X}}K)=0$ implies the existence
of a function ${\cal P}_{{\bf X}}$ such
that
\begin{equation}
-\o{1}{2\pi}d{\cal P}_{{\bf X}}=
i_{\scriptscriptstyle{\bf X}}K
\label{mmap}
\end{equation}
The function ${\cal P}_{{\bf X}}$ is
defined up to a constant,
which can be arranged so as to make it equivariant:
\begin{equation}
{\bf X} {\cal P}_{\bf Y} =
{\cal P}_{[{\bf X},{\bf Y}]}
\label{equivarianza}
\end{equation}
${\cal P}_{{\bf X}}$ constitutes
then a {\it momentum map}.
This can be regarded as a map
\begin{equation}
{\cal P}: {\cal M} \, \longrightarrow \,
\IR \otimes
{\IG }^*
\end{equation}
where ${\IG}^*$ denotes the dual of the Lie algebra
${\IG }$ of the group ${\cal G}$.
Indeed let $x\in {\IG }$ be the Lie algebra element
corresponding to the Killing
vector ${\bf X}$; then, for a given
$m\in {\cal M}$
\begin{equation}
\mu (m)\,  : \, x \, \longrightarrow \,  {\cal P}_{{\bf X}}(m) \,
\in  \, \IR
\end{equation}
is a linear functional on  ${\IG}$.
 If we expand
${\bf X} = a^\Lambda k_\Lambda$ in a basis of Killing vectors
$k_\Lambda$ such that
\begin{equation}
[k_\Lambda, k_\Gamma]= f_{\Lambda \Gamma}^{\ \ \Delta} k_\Delta
\label{blio}
\end{equation}
we have also
\begin{equation}
{\cal P}_{\bf X}\, = \, a^\Lambda {\cal P}_\Lambda
\end{equation}
In the following we  use the
shorthand notation ${\cal L}_\Lambda, i_\Lambda$ for
the Lie derivative
and the contraction along the chosen basis
of Killing vectors $ k_\Lambda$.
\par
From a geometrical point of view the prepotential,
or momentum map, ${\cal P}_\Lambda$
is the Hamiltonian function providing the Poissonian
realization  of the Lie algebra on the K\"ahler manifold. This
is just another way of stating the already mentioned
{\it  equivariance}.
Indeed  the  very  existence  of the closed 2-form $K$ guarantees that
every K\"ahler space is a symplectic manifold and that we can define  a
Poisson bracket.
\par
Consider Eqs.~\ref{killo1}.
To every generator of the abstract  Lie algebra
${\IG}$ we have associated a function  ${\cal P}_\Lambda$ on
${\cal M}$; the Poisson bracket of
${\cal P}_\Lambda$ with ${\cal P}_\Sigma$ is defined as
follows:
\begin{equation}
\{{\cal P}_\Lambda , {\cal P}_\Sigma\} \equiv 4\pi K
(\Lambda, \Sigma)
\end{equation}
where $K(\Lambda, \Sigma)
\equiv K (\vec k_\Lambda, \vec k_\Sigma)$ is
the value of $K$ along the pair of Killing vectors.
\par
In reference \cite{DFF} we proved the following lemma.
\blem
{\it{The following identity is true}}:
\begin{equation}
\{{\cal P}_\Lambda, {\cal
P}_\Sigma\}=f_{\Lambda\Sigma}^{\ \ \Gamma}{\cal
P}_\Gamma + C_{\Lambda \Sigma} \label{brack}
\end{equation}
{\it{where $C_{\Lambda \Sigma}$ is a constant fulfilling the
cocycle condition}}
\begin{equation}
f^{\ \ \Gamma}_{\Lambda\Pi} C_{\Gamma \Sigma} +
f^{\ \ \Gamma}_{\Pi\Sigma} C_{\Gamma \Lambda}+
f_{\Sigma\Lambda}^{\ \  \Gamma} C_{\Gamma \Pi}=0
\label{cocy}
\end{equation}
\elem
If the Lie algebra ${\IG}$ has a trivial second cohomology group
$H^2({\IG})=0$, then the cocycle $C_{\Lambda \Sigma}$ is a
coboundary; namely we have
\begin{equation}
C_{\Lambda \Sigma} = f^{\ \ \Gamma}_{\Lambda \Sigma} C_\Gamma
\end{equation}
where $C_\Gamma$ are suitable constants. Hence, assuming
$H^2 (\IG)= 0$
we can reabsorb $C_\Gamma$ in  the definition of ${\cal
P}_\Lambda$:
\begin{equation}
{\cal P}_\Lambda \rightarrow {\cal P}_\Lambda+ C_\Lambda
\end{equation}
and we obtain the stronger equation
\begin{equation}
\{{\cal P}_\Lambda, {\cal P}_\Sigma\} =
f_{\Lambda\Sigma}^{\ \  \Gamma} {\cal P}_\Gamma
\label{2.39}
\end{equation}
Note that $H^2({\IG}) = 0$ is true for all semi-simple Lie
algebras.
Using eq.~\ref{brack}, eq.~\ref{2.39}
can be rewritten in components as follows:
\begin{equation}
{i\over 2} g_{ij^*}(k^i_\Lambda k^{j^*}_\Sigma -
k^i_\Sigma k^{j^*}_\Lambda)=
{1\over 2} f_{\Lambda \Sigma}^{\  \  \Gamma} {\cal
P}_\Gamma
\label{2.40}
\end{equation}
Equation~\ref{2.40} is identical with the equivariance condition
in eq.~\ref{equivarianza}.
\par
Comparing the definition of the K\"ahler
potential in eq.~\ref{popov} with the definition of the momentum
function in eq.~\ref{killo1}, we obtain an expression for
the momentum map function in terms of derivatives of the K\"ahler
potential:
\begin{equation}
{\rm i} \, {\cal P}_\Lambda \, =  \,{\o{1}{2}}
\left ( k^{i}_\Lambda \, \partial_i {\cal K} -
k^{i^\star}_\Lambda \, \partial_{i^\star} {\cal K}\right )
\, = \, k^{i}_\Lambda \, \partial_i {\cal K}
\, = \, - k^{i^\star}_\Lambda \, \partial_{i^\star} {\cal K}
\label{kexpresp}
\end{equation}
Eq.~\ref{kexpresp} is true if the K\"ahler potential is exactly
invariant under the transformations of the isometry group ${\cal G}$
and not only up to a K\"ahler transformation as defined
in eq.~\ref{041}. In other words eq.~\ref{kexpresp} is true if
\begin{equation}
0\, = \, {\cal L}^\Lambda \, {\cal K} \, = \,
k^{i}_\Lambda \, \partial_i {\cal K} +
k^{i^\star}_\Lambda \, \partial_{i^\star} {\cal K}
\label{kpotinvariant}
\end{equation}
Not all the isometries of a general K\"ahler manifold have such
a property, but those that in a suitable coordinate frame  display
a linear action on the coordinates certainly do. However,
in Hodge K\"ahler manifolds, eq.~\ref{kpotinvariant} can be replaced
by the following one which is certainly true:
\begin{eqnarray}
0 & = & {\cal L}^\Lambda \, G \, = \,
k^{i}_\Lambda \, \partial_i G +
k^{i^\star}_\Lambda \, \partial_{i^\star} G \nonumber \\
G(z,\bar z) & \equiv &   \mbox{log } \, \parallel  W(z) \parallel^2
\, = \, {\cal K}(z,\bar z ) \, + \, \mbox{Re} \, W(z)
\label{normsect}
\end{eqnarray}
where the {\it superpotential} $W(z)$ is any holomorphic section
of the Hodge line--bundle. Indeed the transformation under
the isometry of the K\"ahler potential is compensated by the
transformation of the superpotential. Consequently, in Hodge--K\"ahler
manifolds eq.~\ref{kexpresp} can be rewritten as
\begin{equation}
{\rm i} \, {\cal P}_\Lambda \, =  \, {\o{1}{2}}
\left ( k^{i}_\Lambda \, \partial_i G -
k^{i^\star}_\Lambda \, \partial_{i^\star} G\right ) \,
= \, k^{i}_\Lambda \, \partial_i G \, = \,
- k^{i^\star}_\Lambda \, \partial_{i^\star} G
\label{gexpresp}
\end{equation}
and holds true for any isometry.
\par
In $N=1$ supersymmetry the K\"ahlerian momentum maps
${\cal P}_\Gamma$ appear as auxiliary fields of the
vector multiplets. For $N=1$ supergravity the scalar manifold
is of the Hodge type and eq.~\ref{gexpresp} can always be employed.
\par
On the other hand,
in $N=2$ supersymmetry the auxiliary fields of the vector multiplets,
that form an $SU(2)$ triplet,
are given by the momentum map of triholomorphic isometries on the
hypermultiplet manifold (HyperK\"ahlerian or quaternionic depending
on the local or rigid nature of supersymmetry). The triholomorphic
momentum map is discussed in the  subsection after the next.
Yet, although not
identified with the auxiliary fields, the holomorphic momentum map
plays a role also in $N=2$ theories in the {\it gauging} of the
$U(1)$ connection ~\ref{u1conect}, as we show shortly from now.
\subsection{Holomorphic momentum map on Special K\"ahler manifolds}
Here the K\"ahler manifold is not only Hodge but it is special.
Correspondingly we can write a formula for  ${\cal P}_\Lambda$
in terms of symplectic invariants. In this context, to distinguish
the holomorphic momentum map from the triholomorphic one
${\cal P}_\Lambda^x$ that carries an $SU(2)$ index $x=1,2,3$,
we adopt the notation ${\cal P}_\Lambda^0$.
The request that the isometry group should be embedded into the
symplectic group is formulated by writing:
\begin{equation}
{\cal L}_\Lambda \, V \,  \equiv  \,  k^i_\Lambda \,\partial_i V \, +   \,
k^{i^\star}_\Lambda \, \partial_{i^\star} V
\, =  \,  T_\Lambda \, V \, + V \, f_\Lambda(z)
\label{pullsectiso}
\end{equation}
where $V$ is the covariantly holomorphic section of the vector bundle
${\cal H} \, \longrightarrow \, {\cal M}$ defined in eq.~\ref{specpotuno},
\begin{equation}
 T_\Lambda \, = \, \left (\matrix
{a_\Lambda & b_\Lambda \cr c_\Lambda & d_\Lambda \cr } \right ) \, \in \,
{\bf Sp}({\bf 2n+2},\IR)
\label{alcyon}
\end{equation}
is some element of the real symplectic Lie algebra and $f_\Lambda(z)$
corresponds to an infinitesimal K\"ahler transformation.
\par
The classical or perturbative isometries ( $ b_\Lambda = 0 $)
that are relevant to the gauging procedure
are normally characterized by
\begin{equation}
 f_\Lambda(z)=0
 \label{parcondicio}
\end{equation}
Under condition~\ref{parcondicio},
recalling eq.s~\ref{specpot} and ~\ref{covholsec},
from eq.~\ref{pullsectiso}
we  obtain:
\begin{equation}
  {\cal L}^\Lambda \, {\cal K} \, = \,
k^{i}_\Lambda \, \partial_i {\cal K} +
k^{i^\star}_\Lambda \, \partial_{i^\star} {\cal K} \, = \, 0
\label{kpotinvariantuno}
\end{equation}
that is identical with eq.~\ref{kpotinvariant}. Hence we can use
eq.~\ref{kexpresp}, that we rewrite as:
\begin{equation}
{\rm i} \, {\cal P}_\Lambda^0 \, = \,
k^{i}_\Lambda \, \partial_i {\cal K} \, =
\, - k^{i^\star}_\Lambda \, \partial_{i^\star} {\cal K}
\label{kexprespprime}
\end{equation}
Utilizing the definition in eq.~\ref{uvector} we easily obtain:
\begin{equation}
k_\Lambda^i \, U^i \, = \, T_\Lambda \, V \, \mbox{exp}[f_\Lambda(z)]
+ {\rm i} \, {\cal P}_\Lambda^0 \, V
\label{passino}
\end{equation}
Taking the symplectic scalar product of eq.~\ref{passino} with ${\bar
V}$ and recalling eq.~\ref{specpotuno} we finally
\footnotemark
\footnotetext{The following and the next two formulae have been
obtained in private discussions of one of us (P.Fr\'e)
with A. Van Proeyen and B. de Wit} get:
\begin{equation}
{\cal P}_\Lambda^0 \, = \,
\langle {\bar V} \, \vert \, T_\Lambda \, V \rangle \, = \,
\langle {  V} \, \vert \, T_\Lambda \, {\bar V } \rangle \, =
\, \mbox{exp}\left [ {\cal K}\right ] \,
\langle { \bar \Omega} \, \vert \, T_\Lambda \, { \Omega } \rangle
\label{passetto}
\end{equation}
In the gauging procedure we are interested in groups the symplectic
image of whose generators is block--diagonal and coincides with
the adjoint representation in each block. Namely
\begin{equation}
T_\Lambda \, = \, \left ( \matrix { f^{\Sigma}_{\phantom{\Sigma}
\Lambda\Delta} & {\bf 0} \cr {\bf 0} &-
f^{\Sigma}_{\phantom{\Sigma}\Lambda\Delta}
\cr} \right )
\label{aggiungirep}
\end{equation}
Then eq.~\ref{passetto} becomes
\begin{equation}
{\cal P}_\Lambda^0\, = \, e^{\cal K} \, \left ( \, F_\Delta \,
f^\Delta_{\phantom{\Delta}\Lambda\Sigma} \, {\bar X}^\Sigma
\, + \, {\bar F}_\Delta \,
f^\Delta_{\phantom{\Delta}\Lambda\Sigma} \, {  X}^\Sigma \right )
\label{pippopluto}
\end{equation}
\subsection{The triholomorphic momentum map on HyperK\"ahler and
Quaternionic manifolds}
Next we turn to a discussion of
isometries of the manifold ${\cal HM}$ associated with hypermultiplets.
As we know, it can be either HyperK\"ahlerian or quaternionic.
For applications to $N=2$ theories we must assume that on ${\cal HM}$
we have an action by triholomorphic isometries of the same
 Lie group ${\cal G}$ that acts on the Special K\"ahler
manifold ${\cal SM}$. This means  that on ${\cal HM}$ we
have Killing vectors
\begin{equation}
\vec k_\Lambda = k^u_\Lambda {\vec \partial\over \partial q^u}
\label{2.71}
\end{equation}
\noindent
satisfying the same Lie algebra  as the corresponding Killing
vectors on ${\cal SM}$. In other words
\begin{equation}
\hat{\vec{k}}_\Lambda =
k^i_\Lambda \vec \partial_i + k^{i^*}_\Lambda
\vec\partial_{i^*} + k_\Lambda^u \vec\partial_u
\label{2.72}
\end{equation}
\noindent
is a Killing vector of the block diagonal metric:
\begin{equation}
\hat g = \left (
\matrix { g_{ij^*} & \quad 0 \quad \cr \quad 0
\quad & h_{uv} \cr } \right )
\label{2.73}
\end{equation}
defined on the product manifold ${\cal SM}\otimes{\cal HM}$.
Triholomorphicity means that the Killing vector fields leave
the HyperK\"ahler structure invariant up to $SU(2)$
rotations in the $SU(2)$--bundle defined by eq.~\ref{su2bundle}.
Namely:
\begin{equation}
\begin{array}{ccccccc}
{\cal L}_\Lambda K^x & = &\epsilon^{xyz}K^y
W^z_\Lambda & ; &
{\cal L}_\Lambda\omega^x&=& \nabla W^x_\Lambda
\end{array}
\label{cambicchio}
\end{equation}
where $W^x_\Lambda$ is an $SU(2)$ compensator associated with the
Killing vector $k^u_\Lambda$. The compensator $W^x_\Lambda$ necessarily
fulfils   the cocycle condition:
\begin{equation}
{\cal L}_\Lambda W^{x}_\Sigma - {\cal L}_\Sigma W^x_\Lambda + \epsilon^{xyz}
W^y_\Lambda W^z_\Sigma = f_{\Lambda \Sigma}^{\cdot \cdot \Gamma}
W^x_\Gamma
\label{2.75}
\end{equation}
In the HyperK\"ahler case the $SU(2)$--bundle is flat and the
compensator can be reabsorbed into the definition of the
HyperK\"ahler forms. In other words we can always find a
map
\begin{equation}
{\cal HM} \, \longrightarrow \, L^x_{\phantom{x}y} (q)
\, \in \, SO(3)
\end{equation}
that trivializes the ${\cal SU}$--bundle globally. Redefining:
\begin{equation}
K^{x\prime} \, = \, L^x_{\phantom{x}y} (q) \, K^y
\label{enfantduparadis}
\end{equation}
the new HyperK\"ahler form  obeys the stronger equation:
\begin{equation}
{\cal L}_\Lambda K^{x\prime} \, = \, 0
\label{noncambio}
\end{equation}
On the other hand,
in the quaternionic case,   the non--triviality of the
${\cal SU}$--bundle forbids to eliminate the $W$--compensator
completely. Due to the identification between HyperK\"ahler
forms and $SU(2)$ curvatures eq.~\ref{cambicchio} is rewritten
as:
\begin{equation}
\begin{array}{ccccccc}
{\cal L}_\Lambda \Omega^x& = &\epsilon^{xyz}\Omega^y
W^z_\Lambda & ; &
{\cal L}_\Lambda\omega^x&=& \nabla W^x_\Lambda
\end{array}
\label{cambiacchio}
\end{equation}
In both cases, anyhow, and in full analogy with the case of
K\"ahler manifolds, to each Killing vector
we can associate a triplet ${\cal
P}^x_\Lambda (q)$ of 0-form prepotentials.
Indeed we can set:
\begin{equation}
{\bf i}_\Lambda  K^x =
- \nabla {\cal P}^x_\Lambda \equiv -(d {\cal
P}^x_\Lambda + \epsilon^{xyz} \omega^y {\cal P}^z_\Lambda)
\label{2.76}
\end{equation}
where $\nabla$ denotes the $SU(2)$ covariant exterior derivative.
\par
As in the K\"ahler case eq.~\ref{2.76} defines a momentum map:
\begin{equation}
{\cal P}: {\cal M} \, \longrightarrow \,
\IR^3 \otimes
{\IG }^*
\end{equation}
where ${\IG}^*$ denotes the dual of the Lie algebra
${\IG }$ of the group ${\cal G}$.
Indeed let $x\in {\IG }$ be the Lie algebra element
corresponding to the Killing
vector ${\bf X}$; then, for a given
$m\in {\cal M}$
\begin{equation}
\mu (m)\,  : \, x \, \longrightarrow \,  {\cal P}_{{\bf X}}(m) \,
\in  \, \IR^3
\end{equation}
is a linear functional on  ${\IG}$.
 If we expand
${\bf X} = a^\Lambda k_\Lambda$ on a basis of Killing vectors
$k_\Lambda$ such that
\begin{equation}
[k_\Lambda, k_\Gamma]= f_{\Lambda \Gamma}^{\ \ \Delta} k_\Delta
\label{blioprime}
\end{equation}
and we also choose a basis ${\bf i}_x \, (x=1,2,3)$ for $\IR^3$
we get:
\begin{equation}
{\cal P}_{\bf X}\, = \, a^\Lambda {\cal P}_\Lambda^x \, {\bf i}_x
\end{equation}
Furthermore we need a generalization of the equivariance defined
by eq.~\ref{equivarianza}
\begin{equation}
{\bf X} \circ {\cal P}_{\bf Y} \,=  \,
{\cal P}_{[{\bf X},{\bf Y}]}
\label{equivarianzina}
\end{equation}
In the HyperK\"ahler case, the left--hand side of eq.~\ref{equivarianzina}
is defined as the usual action of a vector field on a $0$--form:
\begin{equation}
{\bf X} \circ {\cal P}_{\bf Y}\, =  \, {\bf i}_{\bf X} \, d
{\cal P}_{\bf Y}\, = \,
X^u \, {\o{\partial}{\partial q^u}} \, {\cal P}_{\bf Y}\,
\end{equation}
The equivariance condition   implies
that we can introduce a triholomorphic Poisson bracket defined
as follows:
\begin{equation}
\{{\cal P}_\Lambda, {\cal P}_\Sigma\}^x \equiv 2 K^x (\Lambda,
\Sigma)
\label{hykapesce}
\end{equation}
leading to the triholomorphic Poissonian realization of the Lie
algebra:
\begin{equation}
\left \{ {\cal P}_\Lambda, {\cal P}_\Sigma \right \}^x \, = \,
f^{\Delta}_{\phantom{\Delta}\Lambda\Sigma} \, {\cal P}_\Delta^{x}
\label{hykapescespada}
\end{equation}
which in components reads:
\begin{equation}
K^x_{uv} \, k^u_\Lambda \, k^v_\Sigma \, = \, {\o{1}{2}} \,
f^{\Delta}_{\phantom{\Delta}\Lambda\Sigma}\, {\cal P}_\Delta^{x}
\label{hykaide}
\end{equation}
In the quaternionic case, instead, the left--hand side of
eq.~\ref{equivarianzina}
is interpreted as follows:
\begin{equation}
{\bf X} \circ {\cal P}_{\bf Y}\, =  \, {\bf i}_{\bf X}\,  \nabla
{\cal P}_{\bf Y}\, = \,
X^u \, {\nabla_u} \, {\cal P}_{\bf Y}\,
\end{equation}
where $\nabla$ is the $SU(2)$--covariant differential.
Correspondingly, the triholomorphic Poisson bracket is defined
as follows:
\begin{equation}
\{{\cal P}_\Lambda, {\cal P}_\Sigma\}^x \equiv 2 K^x (\Lambda,
\Sigma)  - { {\lambda}} \, \varepsilon^{xyz} \,
{\cal P}_\Lambda^y  \, {\cal P}_\Sigma^z
\label{quatpesce}
\end{equation}
and leads to the Poissonian realization of the Lie algebra
\begin{equation}
\left \{ {\cal P}_\Lambda, {\cal P}_\Sigma \right \}^x \, = \,
f^{\Delta}_{\phantom{\Delta}\Lambda\Sigma} \, {\cal P}_\Delta^{x}
\label{quatpescespada}
\end{equation}
which in components reads:
\begin{equation}
K^x_{uv} \, k^u_\Lambda \, k^v_\Sigma \, - \,
{ \o{\lambda}{2}} \, \varepsilon^{xyz} \,
{\cal P}_\Lambda^y  \, {\cal P}_\Sigma^z\,= \,  {\o{1}{2}} \,
f^{\Delta}_{\phantom{\Delta}\Lambda\Sigma}\, {\cal P}_\Delta^{x}
\label{quatide}
\end{equation}
Eq.~\ref{quatide}, which is the most convenient way of
expressing equivariance in a coordinate basis, plays a fundamental
role in the construction of the supersymmetric action, supersymmetry
transformation rules and of the superpotential for $N=2$ supergravity
on a general quaternionic manifold. It is also very convenient
to retrieve the rigid supersymmetry limit. Indeed, using physical
units, we may set $\lambda = {{\hat \lambda} \over {\mu^2}}$ where
$\mu$ is the Planck mass (see section 9); letting  $\mu \to \infty$
 eq.~\ref{quatide} reduces to eq.~\ref{hykaide}.
Eq.~\ref{quatide} was introduced in the physical literature in
\cite{DFF} where the general form of $N=2$ supergravity
beyond the limitations of tensor calculus was given.
\subsection{Gauging of the composite connections}
Using the concepts and the geometric structures introduced
in the previous sections the form of the Lagrangian and of the
transformation rules for $N=2$ supergravity
can now be given.  The essential thing is that the fermions
of the theory, behave as sections of the bundles we have
introduced so far.
In particular he gravitino field $\psi_\mu^A$ apart from being
a spinor--valued 1--form on space--time, behaves as a section
of the bundle  ${\cal L}\otimes{\cal SU}$. The gaugino field
$\lambda^{i\vert A}$ apart from being a section of the spinor
bundle, behaves as a section of
${\cal L} \otimes {\cal TSM}\otimes {\cal SU}$.
Finally the hyperino field $\zeta^\alpha$ is a section
of the rank $2m$ vector bundle with structural group $Sp(2m,\IR)$
that one obtains by deleting the $SU(2)$ part of the holonomy
group on ${\cal HM}$. In other words it is a section of the bundle
  ${\cal THM}\otimes {\cal SU}^{-1}$. Correspondingly the covariant
derivatives
of the fermions appearing in the action and in the transformation
rules involves the composite connections ${\cal Q}$ ,
$\Gamma^{i}_{\phantom{i}j}$, $\omega^x$
and $\Delta^{\alpha\beta}$ defined on these bundles.
Gauging just modifies these composite connections by means of
Killing vectors and momentum map functions. Explicitly we
have:
\begin{equation}
\begin{array}{cccccc}
{\cal TSM} & : & \mbox{tangent bundle} &
 \Gamma^{i}_{\phantom{i}j}& \to &{\hat \Gamma}^{i}_{\phantom{i}j} =
 \Gamma^{i}_{\phantom{i}j} +
 g\, A^\Lambda\, \partial_j k^i_\Lambda \cr
{\cal L} & : & \mbox{line bundle} &
{\cal Q} &\to &{\hat {\cal Q}}= {\cal Q} + g\, A^\Lambda\, {\cal
P}^0_\Lambda \cr
{\cal SU} & : & \mbox{$SU(2)$ bundle} &
\omega^x &\to &{\hat \omega}^x = \omega^x + g\, A^\Lambda\, {\cal
P}^x_\Lambda \cr
{\cal SU}^{-1}\otimes{\cal THM} & : & \mbox{$Sp(2m)$ bundle} &
\Delta^{\alpha\beta} &\to &{\hat  \Delta}^{\alpha\beta}=
\Delta^{\alpha\beta}  + g\, A^\Lambda\,
 \partial_u k_\Lambda^v \, {\cal U}^{u \vert  \alpha A}
 \, {\cal U}^\beta_{v \vert A} \cr
 \end{array}
\label{compogauging}
\end{equation}
Correspondingly the gauged curvatures are:
\begin{eqnarray}
 {\hat R}^{i}_{\phantom{i}j} & = & R^i_{\phantom{i}j\ell^\star k}
 \, \nabla {\bar z}^{\ell^\star} \wedge \nabla z^k \, + \,
 g\, F^\Lambda\, \partial_j k^i_\Lambda \nonumber\\
 {\hat {K}} &= & {K}_{ij^\star} \, \nabla {\bar z}^{i} \wedge
 \nabla z^{j^\star} \,  + \, g\, F^\Lambda\, {\cal
P}^0_\Lambda \nonumber\\
{\hat \Omega}^x &=& \Omega^x_{uv} \, \nabla q^u
\wedge \nabla q^v \, +\,  g\, F^\Lambda\, {\cal
P}^x_\Lambda \nonumber\\
{\hat  {\IR}}^{\alpha\beta}&=&
\IR^{\alpha\beta}_{uv} \,\nabla q^u
\wedge \nabla q^v  \,  +\,  g\, A^\Lambda\,
 \partial_u k_\Lambda^v \, {\cal U}^{u \vert  \alpha A}
\, {\cal U}^\beta_{v \vert A}
\label{compogaugcurv}
\end{eqnarray}

\section{The Complete N=2 Supergravity Theory}
\label{omniapanta}
\setcounter{equation}{0}
In this section we write the supersymmetric
invariant action and supersymmetry transformation
rules for a completely general $N=2$ supergravity.
\par
Such a theory includes
 \begin{enumerate}
\item{{\it the gravitational multiplet}, described by the
vielbein 1--form $V^a$, $(a=0,1,2,3)$,
the spin--connection 1--form $\omega^{ab}$,
the $SU(2)$ doublet of gravitino 1-forms $\Psi^A , \Psi_A$
($A=1,2$ and the upper or lower position of the index denotes
left, respectively right chirality), the graviphoton 1-form
$A^0$}
\item{ $n$ {\it vector multiplets}. Each vector multiplet
contains a gauge boson 1--form $A^I$ ($I=1,\dots,n$), a
doublet of gauginos (0--form spinors) $\lambda^{iA}$,
$\lambda^{{i}^\star}_A$,  and a complex scalar field  (0--form)
$z^i$ ($i=,1,\dots,n$). The scalar fields $z^i$ can be regarded
as coordinates on a special manifold ${\cal SM}$ which can be
chosen {\it arbitrarily}.
\begin{equation}
 {\rm dim}_{\bf C} \, {\cal SM}  \, = \, n
 \label{specdim}
\end{equation}
 }
\item{ $m$ {\it hypermultiplets}. Each hypermultiplet contains
a doublet of 0--form spinors, that is the hyperinos $\zeta^\alpha$
($\alpha=1,\dots,2 m$ and here the  lower or upper position of the index 
denotes
left, respectively right chirality) and four real scalar fields $q^u$ ($u=1,
\dots,4 m$), that can be regarded as coordinates of a quaternionic
manifold $ {\cal HM} $ which can be chosen {arbitrarily}.
\begin{equation}
 {\rm dim}_{\IQ} \, {\cal HM}_m \, = \, m  \qquad \,
 {\rm dim}_{\IR} \, {\cal HM}_m  = \, 4 m
 \label{quatdimbis}
\end{equation}
 As explained in the previous sections any quaternionic manifold
 has a holonomy group:
 \begin{equation}
{\cal H}ol \left (  {\cal HM}_m \right ) \, \subset  \, SU(2) \, \otimes
\, Sp(2 m ,\IR )
\label{ololobis}
\end{equation}
and the index $\alpha$ of the hyperinos transforms in the
{\it fundamental} representation of $Sp(2 m, \IR)$ }
 \end{enumerate}
Using the information collected in the previous sections we
can immediately write down the definition of the
curvatures and covariant derivatives for all the fields.
The definition of curvatures in the gravitational sector is given by:
\begin{eqnarray}
T^a & \equiv & {\cal D}V^a- {\rm i} \, \bar\psi_A\wedge
\gamma^a\psi^A\label{torsdef}\\
\rho_A &  \equiv & d\psi_A-{1\over 4} \gamma_{ab} \,
\omega^{ab}\wedge\psi_A+
{{\rm i} \over 2} {\hat {\cal Q}}\wedge \psi_A +
{\hat \omega}_A^{~B}\wedge \psi_B
\equiv \nabla \psi_A \label{gravdefdown} \\
\rho^A & \equiv & d\psi^A-{1\over 4} \gamma_{ab} \, \omega^{ab}\wedge\psi^A
-{{\rm i} \over 2} {\hat {\cal Q}}\wedge\psi^A
+{\hat \omega}^{A}_{\phantom{A}B} \wedge \psi^B \equiv \nabla \psi^A
\label{gravdefup} \\
R^{ab} & \equiv & d\omega^{ab}-\omega^a_{\phantom{a}c}\wedge \omega^{cb}
\label{riecurv}
\end{eqnarray}
where $\omega_A^{\ B}=\frac{\rm i}{2}\omega^x (\sigma_x)_A^B$ and
$\omega^A_{\ B}=\epsilon^{AC}\epsilon_{DB}\omega_C^{\ D}$, and where
 the gauged connections for the ${\cal SU}$ and ${\cal L}$
bundles were introduced in eq.s~\ref{compogauging}. In all the
above formulae the pull--back on space--time through the maps
\begin{equation}
\begin{array}{ccccccc}
z^i & : & M_4 \, \longrightarrow \, {\cal SM} &; &
q^u & : & M_4 \, \longrightarrow \, {\cal HM}
\end{array}
\label{pollobecco}
\end{equation}
is obviously understood. In this way the composite connections
become 1--forms on space--time.
\par
In the vector multiplet sector the curvatures and covariant
derivatives are:
\begin{eqnarray}
\nabla z^i &=& dz^i \, + \, g \, A^\Lambda \, k_\Lambda^i
(z)\label{zcurv}\\
\nabla {\bar z}^{{i}^\star} &=& d{\bar z}^{{i}^\star} \,
+ \, g \, A^\Lambda \, k_\Lambda^{{i}^\star}
(\bar z)\label{zcurvb}\\
\nabla\lambda^{iA} &\equiv & d\lambda^{iA}-{1\over 4} \gamma_{ab} \,
\omega^{ab} \lambda^{iA} -{{\rm i} \over 2} {\hat {\cal Q}}\lambda^{iA}+
{\hat \Gamma}^i_{\phantom{i}j}\lambda^{jA}+{\hat \omega}^{A}_{~B} \wedge
\lambda^{iB}
\label{lamcurv}
\nonumber\\
\nabla\lambda^{{i}^\star}_A &\equiv &d\lambda^{
{i^\star}}_A-{1\over 4} \gamma_{ab} \,
\omega^{ab}\lambda^{{i}^\star}_A+{{\rm i} \over 2}
{\hat{\cal Q}}\lambda^{{i}^\star}_A+
{\hat \Gamma}^{{i}^\star}_{\phantom{\bar
{\imath}}{{j}^\star}}\lambda^{{j}^\star}_A
+{\hat \omega}_{A}^{~B} \wedge
\lambda^{{{i}^\star}}_B
\label{lamcurvb}
\nonumber\\
F^\Lambda &\equiv & dA^\Lambda \, +\,{1\over 2} \, g\,
f^\Lambda_{\phantom{\Lambda}\Sigma\Gamma}\, A^\Sigma\,
\wedge\, A^\Gamma\, +\,
\bar L^\Lambda \bar\psi_A\wedge\psi_B
\epsilon^{AB}+L^\Lambda\bar\psi^A\wedge \psi^B\epsilon_{AB}
\label{Fcurv}
\end{eqnarray}
where   the gauged Levi--Civita connection
${\hat \Gamma}^i_{\phantom{i}j}$
on ${\cal SM}$ is also given by eq.~\ref{compogauging}
and where $L^\Lambda=e^{\cK \over 2} X^\Lambda$ is the upper half (electric)
of the symplectic section of ${\cal H}$ introduced in equation
~\ref{covholsec}. The lower part $M_\Lambda$ of such a symplectic section
would appear in the magnetic field strengths if we did introduce
them.
\par
Finally in the hypermultiplet sector  the covariant derivatives
are:
\begin{eqnarray}
{\cal U}^{A \alpha} & \equiv & {\cal U}^{A \alpha}_v \nabla q^v
\,\equiv\, {\cal U}^{A \alpha}_v \left (d\,q^v\, +\,
g\,A^{\Lambda}\,k^v_{\Lambda}(q)\right)
\label{ucurv}\\
\nabla \zeta _{\alpha} & \equiv & d \zeta _{\alpha} \,-\,{1 \over 4}
\omega ^{ab}\,\gamma _{ab}\,\zeta _{\alpha}
-{{\rm i} \over 2} {\hat {\cal Q}}\,\zeta _{\alpha}\,+
{\hat {\Delta}}_{\alpha}^{\phantom{\alpha}{\beta}}\zeta _{\beta}
\label{iperincurv}\\
\nabla \zeta^{\alpha} & \equiv & d \zeta^{\alpha} \,-\,{1 \over 4}
\omega ^{ab}\,\gamma _{ab}\,\zeta^{\alpha}
+{{\rm i} \over 2} {\hat {\cal Q}}\,\zeta^{\alpha}\,+
{\hat {\Delta}}^{\alpha}_{\phantom{\alpha}{\beta}}\zeta^{\beta}
\label{iperincurvb}
\end{eqnarray}
where ${\hat {\Delta}}_{\alpha}^{\phantom{\alpha}{\beta}}$
is the gauged Levi--Civita connection on
${\cal HM}$ defined in eq.~\ref{compogauging}, satisfying
 the condition to be $Sp(2m,\IR)$ Lie--algebra valued and
\begin{equation}
{\hat {\Delta}}_{\alpha}^{\phantom{\alpha}{\beta}}\,\equiv\,
{\hat {\Delta}}^{\gamma \beta}\,\IC_{\gamma \alpha}\,;
\, {\hat {\Delta}}^{\alpha}_{\phantom{\alpha}{\beta}}
\,\equiv\,\IC_{\beta \gamma}\,{\hat {\Delta}}^{\alpha \gamma}
\end{equation}
\par
Let us note that the definition of the generalized curvatures as given
in eq.s~\ref{torsdef}-~\ref{riecurv} and~\ref{Fcurv} has been chosen
in such a way that when all the
p-forms are extended to superforms in superspace they give the correct
supercurvatures of the $N=2$ superalgebra; that means that if we set
all supercurvatures to zero the corresponding equations represent
the $N=2$ superalgebra in dual form.
Given these definitions our next task is to write down the space-time
Lagrangian and the supersymmetry transformation laws of the fields.
The method employed for this construction is based on the geometrical
  approach: for a review see \cite{CaDFb}.
The rheonomic derivation of the N=2 theory is explained in Appendix A.
Actually one
solves the Bianchi identities in $N=2$ superspace and then constructs
the rheonomic superspace Lagrangian in such a way that
the superspace "curvatures" given by the  solution of the Bianchi identities
are reproduced by the variational equations of motion derived from the
  Lagrangian.
After this procedure is completed the space-time Lagrangian is
immediately retrieved by restricting the superspace p-forms to
space-time.
\par
Using the results of Appendix B one finds the space-time
$N=2$ supergravity action
that can be split in the  following way:
\ba
S &=& \int\sqrt{-g}\,d^4 \, x \left[ \cL_{k}+\cL_{4f}+\cL'_g\right]\ ,
\nonumber\\
\cL_k  &=&\cL_{kin}^{inv}+\cL_{Pauli} \ ,\nonumber\\
\cL_{4f} &=&\cL_{4f}^{inv}+\cL_{4f}^{non\, inv}\ ,\nonumber\\
\cL'_g &=&\cL_{mass}-V(z,\bar z , q)\ ,
\label{lagr}
\ea
where $\cL_{kin}^{inv}$ consists of the true kinetic terms as well as 
Pauli-like terms containing the derivatives of the scalar fields.
The modifications due to the gauging are contained not only in
$\cL'_g$ but also in the gauged covariant derivatives in the rest of
the lagrangian. We collect the various terms of (\ref{lagr}) in the table
below.
\begin{center}
\begin{tabular}{c}
\null\\
\hline
\null \\
{\it N=2 Supergravity lagrangian}\\
\null \\
\hline
\end{tabular}
\end{center}
\vskip 0.1cm
\ba
\cL_{kin}^{inv} &=& -\half R + 
g_{ij^\star}\nabla^\mu z^i \nabla_\mu \bar z^{j^\star}+
h_{uv}\nabla_\mu q^u \nabla^\mu q^v + 
{{\epsilon^{\mu\nu\lambda\sigma}}\over{\sqrt{-g}}}
\left( \bar\Psi^A_\mu\gamma_\sigma \rho_{A\nu\lambda} 
-  \bar\Psi_{A\mu} \gamma_\sigma \rho^A_{\nu\lambda} \right )
\nonumber\\
&-& {{\rm i}\over2}g_{ij^\star} \left(\bar\lambda^{iA}\gamma^\mu
\nabla_\mu\lambda^{j^\star}_A
+\bar\lambda^{j^\star}_A \gamma^\mu \nabla_\mu\lambda^{iA}\right ) 
-{\rm i}\left (\bar\zeta^\alpha\gamma^\mu\nabla_\mu\zeta_\alpha
+\bar\zeta_\alpha\gamma^\mu \nabla_\mu \zeta^\alpha \right) \nonumber \\
&+& {\rm i}\left(
\bar {\cal N}_{\Lambda\Sigma}{\cal F}^{-\Lambda}_{\mu\nu}{\cal F}^{-\Sigma
\mu\nu} - 
{\cal N}_{\Lambda\Sigma} {\cal F}^{+\Lambda}_{\mu\nu}{\cal F}^{+ \Sigma
{\mu\nu}}\right )+ \Big\{ -g_{ij^\star}
 \nabla_\mu \bar z^{j^\star}\bar\Psi^\mu_A\lambda^{i A}\nonumber\\
&-& 2 {\cal U}^{A\alpha}_u \nabla_\mu q^u 
\bar\Psi_A^\mu \zeta _\alpha
+ g_{ij^\star}  \nabla _\mu \bar z^{j^\star}
\bar\lambda^{iA}\gamma^{\mu\nu}\Psi_{A\nu}
+ 2{\cal U}^{\alpha A}_u\nabla_\mu q^u
\bar\zeta_\alpha \gamma^{\mu\nu}
\Psi_{A\nu}+{\rm h.c.}\Big\}\nonumber\\
\ \\
\cL_{Pauli} &=& \Big\{ {\cal F}^{-\Lambda}_{\mu\nu} \left(
\im{\cal N}\right )_{\Lambda\Sigma}
{\lbrack} 4 L^\Sigma  \bar\Psi^{A\mu}
\Psi^{B\nu}\epsilon_{AB}-4{\rm i} 
{\bar f}^\Sigma_{i^\star}\bar\lambda^{i^\star}_A\gamma^\nu
\Psi_B^\mu\epsilon^{AB}+\nonumber\\
&+& \half
\nabla_i f^\Sigma_j
\bar\lambda^{iA} \gamma^{\mu\nu} \lambda^{jB}\epsilon_{AB}- 
L^\Sigma \bar\zeta_\alpha\gamma^{\mu\nu} \zeta_\beta 
\IC^{\alpha\beta}
{\rbrack}+{\rm h.c.}\Big\}\\
\nonumber\\
{\cal L}_{4f}^{inv} & = & {{\rm i}\over 2} \left(
g_{ij^\star}\bar\lambda^{iA}\gamma_\sigma
  \lambda^{j^\star}_B -
2 \delta^A_B \bar\zeta^\alpha\gamma_\sigma \zeta_\alpha\right )
\bar\Psi_{A \mu} \gamma_\lambda \Psi^B_\nu
{\epsilon^{\mu \nu \lambda \sigma} \over \sqrt{-g}}\nonumber \\
&-& \,{1 \over 6}\,
\left ( C_{ijk} \bar {\lambda}^{iA} \gamma^{\mu} \Psi ^B_{\mu} \,
\bar {\lambda}^{jC} \lambda ^{kD}\,
\epsilon _{AC} \epsilon _{BD} +h.c.\right)\nonumber \\
&-& 2 \bar {\Psi}^A_{\mu} \Psi ^B_{\nu} \,\bar {\Psi}_A^{\mu} \Psi _B^{\nu}
+2g_{i{{j}^\star}}\,\bar {\lambda}^{iA} \gamma _{\mu} \Psi ^B_{\nu} \,
\bar {\lambda}^{{i}^\star}_A \gamma^{\mu} \Psi _B^{\nu}   \nonumber\\
 &+&  {1 \over 4}
\left (R_{i{{j}^\star}l{{k}^\star}}\,  + \,
g_{i{{k}^\star}} \, g_{l{{j}^\star}}\,
 -\, {3 \over 2}\,g_{i{{j}^\star}} \, g_{l{{k}^\star}}\right )
\bar {\lambda}^{iA}\lambda^{lB} \bar {\lambda}^{{j}^\star}_A\lambda^
{{k}^\star}_B \,
\nonumber \\
& +& {1 \over 4} \,g_{i{{j}^\star}} \,
\bar {\zeta}^{\alpha} \gamma _{\mu} \zeta _{\alpha}\,
\bar {\lambda}^{iA} \gamma ^{\mu} \lambda^{{j}^\star}_A\,
+ \, {1 \over 2} \, {\cal R}^{\alpha}_{\beta ts}
\, {\cal U}^t_{A \gamma}\,{\cal U}^s_{B \delta}
 \epsilon ^{AB} \, C ^{\delta \eta}
\bar {\zeta}_{\alpha}\,\zeta _{\eta}\,\bar {\zeta}^{\beta}\,
\zeta ^{\gamma} \nonumber \\
& -& \left[{{\rm i} \over 12} \nabla _m \, C_{jkl}
\bar {\lambda}^{jA}\lambda^{mB} \bar {\lambda}^{kC}\lambda^{lD}
\epsilon _{AC} \epsilon _{BD} +\,h.\,c.\right] \nonumber \\
& +& g_{i{{j}^\star}}\,
\bar {\Psi}^A_{\mu} \lambda ^{{j}^\star}_A  \,
\bar {\Psi}_B^{\mu} \lambda ^{i B}\,+\,
2  \bar {\Psi}^A_{\mu} \zeta ^{\alpha} \bar {\Psi}_A^{\mu}
 \zeta _{\alpha}\,+ \left ( \epsilon _{AB}\, \IC_{\alpha  \beta} \,
 \bar {\Psi}^A_{\mu} \zeta ^{\alpha} \,
  \bar {\Psi}^{B \vert \mu} \zeta ^{\beta} \,+ \,h. c.\right)
\label{4ferminv} 
\\
{\cal L}_{4f}^{non \,inv} & = &   \Big\{
\left (\im {\cal N} \right )_{\Lambda \Sigma}\, \Bigl[
2 L^{\Lambda}\, L^{\Sigma}\left( \bar {\Psi}^A_{\mu}
 \Psi ^B_{\nu}\right)^-
\left( \bar {\Psi}^C_{\mu} \Psi ^D_{\nu}\right)^- \epsilon _{AB} \,
 \epsilon _{CD} \nonumber\\
&-& 8 {\rm i}\,L^{\Lambda}{\bar f}^{\Sigma}_{{i}^\star}\left( 
\bar {\Psi}^A_{\mu} \Psi ^B_{\nu}\right)^-
\left(\bar {\lambda}^{{i}^\star}_A \gamma^{\nu}
 \Psi _B^{\mu}\right)^-\nonumber\\
&-& 2 {\bar f}^{\Lambda}_{{i}^\star}
 {\bar f}^{\Sigma}_{{j}^\star}
\left(\bar {\lambda}^{{i}^\star}_A \gamma^{\nu} \Psi _B^{\mu}\right)^-
\left(\bar {\lambda}^{{j}^\star}_C \gamma _{\nu} \Psi _{D \vert \mu}\right)^-
\epsilon^{AB}\,\epsilon^{CD} \nonumber\\
&+&{{\rm i} \over 2}
L^{\Lambda}{\bar f}^{\Sigma}_{{\ell}^\star} \,g^{k {\ell}^\star}\,C_{ijk}
\left( \bar {\Psi}^A_{\mu} \Psi ^B_{\nu}\right)^-
\bar {\lambda}^{iC} \gamma^{\mu \nu} \lambda ^{jD}\,
\epsilon _{AB} \,\epsilon _{CD}
\nonumber \\
& +& {\bar f}^{\Lambda}_{ m^\star}{\bar f}^{\Sigma}_{{\ell}^\star}
 \,g^{k {\ell}^\star}\,C_{ijk}
\left(\bar {\lambda}^{m^\star}_A \gamma _{\nu} \Psi _{B \mu}\right)^-
\bar {\lambda}^{iA} \gamma^{\mu \nu} \lambda ^{jB}\nonumber\\
&-& L^{\Lambda} L^{\Sigma}\left( \bar {\Psi}^A_{\mu} \Psi ^B_{\nu}\right)^-
\bar {\zeta}_{\alpha} \gamma^{\mu \nu} \zeta _{\beta} \,
\epsilon _{AB}\, \IC^{\alpha  \beta} \nonumber\\
 & + & {\rm i} L^{\Lambda}{\bar f}^{\Sigma}_{{i}^\star}
\left(\bar {\lambda}^{{i}^\star}_A \gamma^{\nu} \Psi _B^{\mu}\right)^-
\bar {\zeta}_{\alpha} \gamma_{\mu \nu} \zeta _{\beta} \,
\epsilon^{AB} \IC^{\alpha  \beta}
\nonumber \\
& -& {1 \over 32}\,
C_{ijk}\,C_{lmn} g^{k{r^\star}} \, g^{n{s^\star}} \,
{\bar f}^{\Lambda}_{r^\star} \, {\bar f}^{ \Sigma}_{s^\star} \,
\bar {\lambda}^{iA} \, \gamma _{\mu \nu} \,\lambda^{jB} \,
\bar {\lambda}^{kC} \, \gamma^{\mu \nu}\,\lambda^{lD}
\, \epsilon _{AB} \epsilon _{CD}
\nonumber \\
& - & {1 \over 8}\,
L^{\Lambda} \nabla _i f^{\Sigma}_j
\bar {\zeta}_{\alpha} \gamma _{\mu \nu} \zeta _{\beta}\,
\bar {\lambda}^{iA} \gamma ^{\mu \nu} \lambda^{jB}\,
\,\epsilon _{AB} \,\IC^{\alpha  \beta}
\nonumber \\
&+& {1\over 8}\,
L^{\Lambda} \,L^{\Sigma}
\bar {\zeta}_{\alpha} \gamma _{\mu \nu} \zeta _{\beta}\,
\bar {\zeta}_{\gamma} \gamma^{\mu \nu} \zeta _{\delta}\,
\IC^{\alpha  \beta}\,\IC^{\gamma \delta} \Bigr]   +{\rm h.c.}\Big\}
\label{4fermnoninv}\\
\nonumber\\
{\cal L}_{mass}
&=&  g\Big[ 2 S_{AB} \bar\Psi^A_\mu \gamma^{\mu\nu}\Psi^B_\nu +
{\rm i} g_{ij^\star} W^{iAB} \bar\lambda^{j^\star}_A\gamma_\mu \Psi_B^\mu+
 2{\rm i} N^A_\alpha\bar\zeta^\alpha\gamma_\mu \Psi_A^\mu\nonumber\\
&+&
{\cal M}^{\alpha\beta}{\bar \zeta}_\alpha
\zeta_\beta +{\cal M}^{\alpha}_{\phantom{\alpha}iB}
{\bar\zeta}_\alpha \lambda^{iB} + {\cal M}_{iA\ell B}
{\bar \lambda}^{iA} \lambda^{\ell
B}  \Big] + \mbox{h.c.} \\
{\rm V}\bigl ( z, {\bar z}, q \bigr )
&=& g^2 \Bigl[\left(g_{ij^\star} k^i_\Lambda k^{j^\star}_\Sigma+4 h_{uv}
k^u_\Lambda k^v_\Sigma\right) \bar L^\Lambda L^\Sigma
+ g^{ij^\star} f^\Lambda_i f^\Sigma_{j^\star}
{\cal P}^x_\Lambda{\cal P}^x_\Sigma
-3\bar L^\Lambda L^\Sigma{\cal P}^x_\Lambda
{\cal P}^x_\Sigma\Bigr]\ .\\
\nonumber
\end{eqnarray}
where  $\cF^{\pm\La}_{\mu\nu}=\half (\cF^\La_{\mu\nu}\pm
{{\rm i}\over2}\epsilon^{\mu\nu\rho\sigma}\cF_{\rho\sigma}^\La )$ and
$(...)^-$ denotes the self dual part of
the fermion bilinears. The mass--matrices are given by:
\begin{eqnarray}
S_{AB}&=&{{\rm i}\over2} (\sigma_x)_A^{\phantom{A}C} \epsilon_{BC}
{\cal P}^x_{\Lambda}L^\Lambda \nonumber\\
W^{iAB}&=&\epsilon^{AB}\,k_{\Lambda}^i \bar L^\Lambda\,+\,
{\rm i}(\sigma_x)_{C}^{\phantom{C}B} \epsilon^{CA} {\cal P}^x_{\Lambda}
g^{ij^\star} {\bar f}_{j^\star}^{\Lambda}\nonumber\\
N^A_{\alpha}&=& 2 \,{\cal U}_{\alpha u}^A \,k^u_{\Lambda}\,
\bar L^{\Lambda}\nonumber\\
{\cal M}^{\alpha\beta}  &=&-
\, {\cal U}^{\alpha A}_u \, {\cal U}^{\beta B}_v \, \varepsilon_{AB}
\, \nabla^{[u}   k^{v]}_{\Lambda}  \, L^{\Lambda} \nonumber\\
{\cal M}^{\alpha }_{\phantom{\alpha} iB} &=&-
4 \, {\cal U}^{\alpha}_{B  u} \, k^u_{\Lambda} \,
 f^{\Lambda}_i \nonumber\\
{\cal M}_{iA\vert \ell B} &=& \,{1 \over 3}  \,
\Bigl ( \varepsilon_{AB}\,  g_{ij^\star}   k^{j^\star}_ \Lambda  
 f_\ell^\Lambda +
{\rm i}\bigl ( \sigma_x \epsilon^{-1} \bigr )_{AB} \, {\cal P}^x_ \Lambda
\, \nabla_\ell f^\Lambda _i \Bigr )
\label{pesamatrice}
\end{eqnarray}
The coupling constant $g$ in $\cL'_{g}$ is just a symbolic notation to 
remind that these terms are entirely
due to the gauging and vanish in the ungauged theory, where also all
 gauged covariant
derivatives reduce to ordinary ones. Note that in general
there is not a single coupling constant, but rather there are
as many independent coupling constants as the number of factors in
the gauge group. The normalization of the kinetic term for the quaternions
depends on the scale $\lambda$ of the quaternionic manifold, appearing in
eq. (\ref{su2curv}), for which we have chosen the value $\lambda=-1$.

\par Furthermore, using the geometric approach, the form of the supersymmetry
transformation laws is also easily deduced from the solution of the
Bianchi identities in superspace (see Appendix A). One gets
\begin{center}
\begin{tabular}{c}
\null\\
\hline
\null \\
{\it Supergravity transformation rules of the Fermi  fields}\\
\null \\
\hline
\end{tabular}
\end{center}
\vskip 0.2cm
\begin{eqnarray}
\delta\,\Psi _{A \mu} &=& {\cal D}_{\mu}\,\epsilon _A\,
 -{1 \over 4}
 \left(\partial _i\,{K} \bar {\lambda}^{iB}\epsilon _B\,-\,
 \partial _{{i}^\star}\,{ K}
 \bar {\lambda}^{{i}^\star}_B \epsilon^B \right)\Psi _{A  \mu}\nonumber\\
&&-{\omega}_{A  v}^{\phantom{A}B}\,
{\cal U}_{C \alpha}^v\,
\left(\epsilon^{CD}\,\IC^{\alpha  \beta}\,\bar
{\zeta}_{\beta}\,\epsilon _D\,+\,
\,\bar {\zeta}^{\alpha}\,\epsilon^C \right)\Psi _{B  \mu}\nonumber\\
&& +\,\left ( A_{A}^{\phantom{A} {\nu}B}
\eta_{\mu \nu}+A_{A}^{\prime \phantom{A}
  {\nu} B}\gamma_{\mu \nu} \right ) \epsilon _B \,
\nonumber\\
&& + \left [ {\rm i} \, g \,S_{AB}\eta _{\mu \nu}+
\epsilon_{AB}( T^-_{\mu \nu}\, + \,U^+_{\mu \nu} )
\right ] \gamma^{\nu}\epsilon^B
 \label{trasfgrav}\\
\delta\,\lambda^{iA}&=&
{1 \over 4} \left(\partial _j\,{K} \bar {\lambda}^{jB}\epsilon _B\,-\,
\partial _{{j}^\star}\, {K}
 \bar {\lambda}^{{j}^\star}_B \epsilon^B \right)\lambda^{iA}\nonumber\\
&&-{\omega}^{A}_{\phantom{A}B  v}\,
{\cal U}_{C \alpha}^v\,
\left(\epsilon^{CD}\,\IC^{\alpha  \beta}\,\bar
{\zeta}_{\beta}\,\epsilon _D\,+\,
\,\bar {\zeta}^{\alpha}\,\epsilon^C \right)\lambda^{iB}\nonumber\\
&&-\,\Gamma^i_{\phantom{i}{jk}}
\bar {\lambda}^{kB}\epsilon _B\,\lambda^{jA}
+ {\rm i}\, \left (\nabla _ {\mu}\, z^i -\bar {\lambda}^{iA}\psi
_{A \mu}\right)
\gamma^{\mu} \epsilon^A \nonumber\\
&&+G^{-i}_{\mu \nu} \gamma^{\mu \nu} \epsilon _B \epsilon^{AB}\,+\,
D^{iAB}\epsilon _B
\label{gaugintrasfm}\\
\delta\,\zeta _{\alpha}&=&-\Delta _{\alpha  v}^{\phantom{\alpha}\beta}\,
{\cal U}_{\gamma A}^v\,
\left(\epsilon^{AB}\,\IC^{\gamma \delta}\,\bar
{\zeta}_{\delta}\,\epsilon _B\,+\,
\,\bar {\zeta}^{\gamma}\,\epsilon^A \right)\zeta _{\beta}\nonumber\\
&&+ {1 \over 4} \left(\partial _i\,{ K} \bar {\lambda}^{iB}\epsilon _B\,-\,
 \partial _{{i}^\star}\,
{ K} \bar {\lambda}^{{i}^\star}_B
 \epsilon^B \right)\zeta _{\alpha}\nonumber\\
 &&+\, {\rm i}\,\left(
{\cal U}^{B \beta}_{u}\, \nabla _{\mu}\,q^u\,
-\epsilon^{BC}\,\IC^{\beta\gamma}\,\bar
{\zeta}_{\gamma}\,\psi _C\,-\,
\,\bar {\zeta}^{\beta}\,\psi^B
\right)\,\gamma^{\mu} \epsilon^A
\epsilon _{AB}\,\IC_{\alpha  \beta}
\,+\,g\,N_{\alpha}^A\,\epsilon _A \label{iperintrasf}
\end{eqnarray}
\begin{center}
\begin{tabular}{c}
\null\\
\hline
\null \\
{\it Supergravity transformation rules of the Bose  fields}\\
\null \\
\hline
\end{tabular}
\end{center}
\vskip 0.2cm
\ba
\delta\,V^a_{\mu}&=& -{\rm i}\,\bar {\Psi}_{A 
\mu}\,\gamma^a\,\epsilon^A -{\rm i}\,\bar {\Psi}^A _
\mu\,\gamma^a\,\epsilon_A\\
\delta\,A^\Lambda _{\mu}&=& 
2 \bar L^\Lambda \bar \psi _{A\mu} \epsilon _B
\epsilon^{AB}\,+\,2L^\Lambda\bar\psi^A_{\mu}\epsilon^B \epsilon
_{AB}\nonumber\\
&+&\left({\rm i} \,f^{\Lambda}_i \,\bar {\lambda}^{iA}
\gamma _{\mu} \epsilon^B \,\epsilon _{AB} +{\rm i} \,
{\bar f}^{\Lambda}_{{i}^\star} \,\bar\lambda^{{i}^\star}_A
\gamma _{\mu} \epsilon_B \,\epsilon^{AB}\right)\label{gaugtrasf}\\
\delta\,z^i &=& \bar{\lambda}^{iA}\epsilon _A \label{ztrasf}\\
\delta\,z^{{i}^\star}&=& \bar{\lambda}^{{i}^\star}_A \epsilon^A
\label{ztrasfb}\\
  \delta\,q^u &=& {\cal U}^u_{\alpha A} \left(\bar {\zeta}^{\alpha}
  \epsilon^A + \IC^{\alpha  \beta}\epsilon^{AB}\bar {\zeta}_{\beta}
  \epsilon _B \right)
 \ea
where we have:
\begin{center}
\begin{tabular}{c}
\null\\
\hline
\null \\
{\it Supergravity values of the auxiliary  fields}\\
\null \\
\hline
\end{tabular}
\end{center}
\vskip 0.2cm
\begin{eqnarray}
A_{A}^{\phantom{A}  \mu B}
&=&-{{\rm i} \over 4}\, g_{{{k}^\star}\ell}\,
\left(\bar {\lambda}^{{k}^\star}_A \gamma^{\mu} \lambda^{\ell B}\,
-\,\delta^B_A\,
\bar\lambda^{{k}^\star}_C \gamma^{\mu} \lambda^{\ell C}\right)\label{Adef}\\
A_{A}^{\prime \phantom{A} \mu B}
&=&{{\rm i} \over 4}\, g_{{{k}^\star}\ell}\,\left(\bar{\lambda}^{{k}^\star}_A
 \gamma^{\mu}
\lambda^{\ell B}-{1\over 2}\, \delta^B_A\, \bar\lambda^{{k}^\star}_C
\gamma^{\mu} \lambda^{C\ell}\right) \, -\,  {{\rm i} \over 4}\,
 \delta _A^B \,\bar \zeta _{\alpha}
\gamma^{\mu} \zeta^{\alpha}\label{A'def}
\end{eqnarray}
\begin{eqnarray}
T^-_{\mu\nu} &=& 2{\rm i}
\left (\im {\cal N}\right)_{\Lambda\Sigma} L^{\Sigma}
\left ({\tilde{F}}_{\mu\nu}^{\Lambda -} +{1\over 8} \nabla_i 
\,f^{\Lambda} _j \,
\bar \lambda^{i A} \gamma_{\mu\nu} \, \lambda^{jB} \,\epsilon_{AB}
-{1\over 4} \, \IC^{\alpha  \beta}\,{\bar\zeta}_{\alpha}\gamma _{\mu\nu} \,
\zeta _{\beta}\, L^{\Lambda}
\right )\label{T-def}\\
T^+_{\mu\nu} &=& 2 {\rm i}
\left (\im {\cal N}\right)_{\Lambda\Sigma} {\bar L}^{\Sigma}
\left({\tilde{F}}_{\mu\nu}^{\Lambda +} +{1\over 8} \nabla_{{i}^\star} \,\bar
f^\Lambda_{{j}^\star}\,
\bar \lambda^{{i}^\star}_A \gamma _{\mu\nu} \, \lambda^{{j}^\star}_B
 \epsilon^{AB}
-{1 \over 4}\, \IC_{\alpha  \beta}\,{\bar\zeta}^{\alpha}\gamma _{\mu\nu} \,
\zeta ^{\beta}\, {\bar L}^{\Lambda}
\right)
\label{T+def}
\end{eqnarray}
\begin{eqnarray}
U^-_{\mu\nu} &=& -{{\rm i} \over 4} \, \IC^{\alpha  \beta}\,
{\bar\zeta}_{\alpha}\gamma _{\mu\nu} \,
\zeta _{\beta}\label{U-def}\\
U^+_{\mu\nu} &=& -{{\rm i} \over 4} \,
 \IC_{\alpha  \beta}\,{\bar\zeta}^{\alpha}\gamma _{\mu\nu} \,
\zeta^{\beta}\label{U+def}
\end{eqnarray}
\begin{eqnarray}
G^{i-}_{\mu\nu} &=& - g^{i{{j}^\star}}
 \bar f^\Gamma_{{j}^\star}
\left (\im  {\cal N}\right)_{\Gamma\Lambda}
\Bigl ( {\tilde {F}}^{\Lambda -}_{\mu\nu} + {1\over 8}
\nabla_{k}  f^{\Lambda}_{\ell} \bar \lambda^{kA}
\gamma_{\mu\nu} \, \lambda^{\ell B} \epsilon_{AB}
  \nonumber\\
&\null& \qquad \qquad  -
{1\over 4} \, \IC^{\alpha  \beta}\,{\bar\zeta}_{\alpha}\gamma _{\mu\nu}
\, \zeta _{\beta}\, L^{\Lambda}
\Bigr )\label{G-def}\\
G^{{{i}^\star}+}_{\mu\nu} &=& - g^{{{i}^\star}j} f^{\Gamma}_j
\left (\im {\cal N}\right)_{\Gamma\Lambda}
\Bigl ( {\tilde{F}}^{\Lambda +}_{\mu\nu} +{1\over 8}
\nabla _{{k}^\star} \bar f^{\Lambda}_{{\ell}^\star} \bar \lambda^{{k}^\star}_A
\gamma _{\mu\nu} \, \lambda^{{\ell}^\star}_B \epsilon^{AB}\nonumber\\
&\null& \qquad \qquad - {1 \over 4}\,
\IC_{\alpha  \beta}\,{\bar\zeta}^{\alpha}\gamma _{\mu\nu} \,
\zeta ^{\beta}\, {\bar L}^{\Lambda}
\Bigr )\label{G+def}
\end{eqnarray}
\begin{eqnarray}
D^{iAB} &=& {{\rm i} \over 2}g^{i{{j}^\star}}
 C_{{{j}^\star}{{k}^\star}{{\ell}^\star}} \bar\lambda^{{k}^\star}_C
\lambda^{{\ell}^\star}_D
\epsilon^{AC} \epsilon^{BD}\,+\,W^{iAB}\label{3.21a}
\end{eqnarray}
In eqs. (\ref {T-def}), (\ref{T+def}), (\ref{G-def}), (\ref{G+def})
 we have denoted by ${\tilde{F}}_{\mu\nu}$
the supercovariant field strength defined by:
\begin{equation}
\tilde{F}^\Lambda_{\mu\nu} = \cF^\Lambda_{\mu\nu}\,
+\,L^{\Lambda}\bar\psi^A_\mu\psi^B_{\nu}
\,\epsilon_{AB} \,+\bar L^{\Lambda}\bar{\psi}_{A \mu} \psi_{B \nu}
\epsilon^{AB}\,
-{\rm i} \,f^{\Lambda}_i \,\bar {\lambda}^{iA} \gamma_{[\nu}
 \psi^B_{\mu]}\,\epsilon _{AB}\,
-{\rm i} \,{\bar f}^{\Lambda}_{{i}^\star} \,\bar\lambda^{i^\star}_A
\gamma_{[\nu} \psi _{B \mu]} \,\epsilon^{AB}\ .
\end{equation}

Let us make some observation about the structure of the Lagrangian
and of the transformation laws.
\par
i) We note that all the terms of the Lagrangian are given in terms of
purely geometric objects pertaining to the Special and quaternionic
geometries. Furthermore the Lagrangian does not rely on the existence
of a prepotential function $F=F\left(X\right)$ and it is valid for
any choice of the quaternionic manifold.
\par
ii) The Lagrangian is not invariant under symplectic duality
transformations. However, in absence of gauging ($g=0$), if we
restrict the Lagrangian to configurations where the vectors are
on shell, it becomes symplectic invariant (ref). This allows us
to fix the terms appearing in ${\cal L}^{non\,inv}_{4\,ferm}$ in
a way independent from supersymmetry arguments.
\par
Here we  report only the results of the application of the method of
\cite{CDFVP},\cite{ferscherzum}in our case.
For a complete treatment see \cite{CDFVP},\cite{ferscherzum}.
The non-invariant part of the Lagrangian is:
\begin{equation}
{\cal L}\,=\, {\cal L}^{vectors}_{Kin}\,+
\,{\cal L}^{non\,inv}_{Pauli}
\,+\,{\cal L}^{non\,inv}_{4f}
\end{equation}
where: ${\cal L}^{vectors}_{Kin}\,=\,{\rm i}
\left(\bar{\cal N}_{\Lambda\Sigma}
{\cal F}^{-\Lambda}{\cal F}^{-\Sigma}\,-\,h.c.\right)$. The part
${\cal L}^{non\,inv}_{4f}$ of the 4--fermi Lagrangian
is fixed by the requirement of
on-shell vector invariance.Indeed, imposing the equation of m
motion for the gauge fields,with straightforward calculations
one finds that${\cal L}^{non\,inv}$ can be written as follows:
\begin{equation}
{\cal L}^{non\,inv}_{on\,shell}\,=\,{1 \over 2}\left(
{\cal F}^{-\Lambda}{\cal
H}^-_{\Lambda}\,+\,h.c.\right)\,+\,{\cal L}^{non\,inv}_{4\,ferm}
\end{equation}
where:
\begin{equation}
{\cal H}^-_{\Lambda \vert \mu\nu}\,=\, \left (
{\cal N}-\bar{\cal N}\right )_{\Lambda\Sigma}
{\tau}^{-\Sigma}_{\mu\nu}
\end{equation}
with:
\begin{eqnarray}
{\tau}^{-\Sigma}_{\mu\nu} &=&
 \Bigl[
-2{\rm i}\,L^{\Sigma } \left(\bar {\Psi}^{A \vert \mu}
\Psi ^{B \vert \nu}\right)^- \epsilon _{AB}
\,-\,2
{\bar f}^{\Sigma}_{{i}^\star}\,\left(\bar {\lambda}^{{i}^\star}_A
\gamma^{\nu}
\Psi _B^{\mu}\right)^-\,
\epsilon ^{AB}
\nonumber \\
 &&- \, {{\rm i} \over 4}
\nabla _i f^{\Sigma}_j\,
\bar {\lambda}^{iA} \gamma^{\mu \nu} \lambda ^{jB} \,\epsilon _{AB} \,
+ \,{{\rm i} \over 2}
L^{\Sigma} \bar {\zeta}_{\alpha} \gamma^{\mu \nu} \zeta _{\beta}
\,C ^{\alpha \beta}
\Bigr]
\end{eqnarray}
From duality arguments it then follows (\cite{CDFVP},\cite{ferscherzum})
that the non invariant 4 fermion terms can be written as the
following perfect square:
  \begin{equation}
  {\cal L}^{noninv}_{4ferm}=+{{\rm i}\over 4}{\cal H}^-_{\Lambda \vert
  \mu\nu}
  \,{\tau}^{-\Lambda \vert \mu\nu}\,+\,h.c.= +{{\rm i}\over 4}
  \left ( {\cal N}-\bar{\cal N}\right )_{\Lambda\Sigma}
  {\tau}^{-\Lambda}_{\mu\nu}{\tau}^{-\Sigma \vert \mu\nu}\,+\,h.c.
\end{equation}
This result was in fact employed as a useful consistency check in the
calculations to construct the Lagrangian.
\par
iii) We note that the field strengths ${\cal F}^{\Lambda\,-}_{\mu\nu}$
originally
 introduced in the Lagrangian are the free gauge field
strengths.The interacting field strengths which are supersymmetry
eigenstates are defined as the objects appearing in the
transformation laws of the gravitinos and gauginos
fields,respectively,namely the bosonic part of $T^-_{\mu\nu}$ and
$G^{-\,i}_{\mu\nu}$ defined in eq.s ~\ref{T-def},~\ref{G-def}
\section{Comments on the scalar potential}
A general Ward identity\cite{ward} of $N$-extended supergravity establishes the
following formulae for the scalar potential $V(\phi)$ of the theory
(in  appropriate normalizations for the generic fermionic shifts $\delta
\chi^a$)
\be
Z_{ab} \delta_A \chi^a \delta^B\bar \chi^b-3 \cM_{AC}\bar\cM^{CB}=
\delta^A_{\ B}  V(\phi)\ \ \ A,B=1,\ldots,N
\ee
where $\delta_A\chi^a$ is the extra contribution, due to the gauging,
 to the spin $\half$
supersymmetry variations of the scalar vev's, $Z_{ab}$ is the (scalar
 dependent) kinetic term normalization and $\cM_{AC}$ is the (scalar
dependent) gravitino mass matrix. Since in the case at hand ($N=2$) all
terms in question are expressed in terms of Killing vectors and
prepotentials, contracted with the symplectic sections, we will be
able to derive a completely geometrical formula for $V(z, \bar z,q)$.
The relevant terms in the fermionic transformation rules are
\begin{eqnarray}
\delta \psi_{A\mu}&=&ig S_{AB} \gamma_\mu \epsilon^B\ ,\nonumber\\
\delta\lambda^{iA}&=& g W^{iAB}\epsilon_B\ ,\nonumber\\
\delta\zeta _ \alpha &=& g N^A_\alpha \epsilon_A\ .
\end{eqnarray}
In our normalization the previous Ward identity gives
\begin{equation}
V=(g_{ij^\star}k^i_\La k^{j^\star}_\Si +4 h_{uv} k^u_\La k^v_\Si) 
\bar L^\La L^\Si+(U^{\La\Si}-3 \bar L^\La L^\Si )
\cP^x_\La \cP^x_\Si\ .
\label{ssette}
\end{equation}
with $U^{\La\Si}$ is defined in (\ref{dsei}).
Above, the first two terms  
are related to the gauging of isometries of ${\cal SK}\otimes
\cQ$. For an abelian group, the first term  is
absent. The negative term is  the gravitino mass
contribution, while the one in $U^{\La\Si}$ is the gaugino shift
contribution due to the quaternionic prepotential.

Eq. (\ref{ssette}) can be rewritten in a suggestive form as
\be
V=(k_\La ,k_\Si )\bar L^\La L^\Si+(U^{\La\Si}-3 \bar L^\La L^\Si )
(\cP_\La^x \cP_\Si^x-\cP_\La\cP_\Si)\ ,
\label{sotto}
\ee
where
 \begin{equation}
(k_\La , k_\Si )=\pmatrix{k^i_\La ,&k^{i^\star}_\La ,&k^u_\La}
\pmatrix{0&g_{ij^\star}&0\cr g_{i^\star j}&0&0\cr0&0&2h_{uv}}
\pmatrix{k^j_\Si\cr k^{j^\star}_\Si\cr k^v_\Si}
\label{snove}
\end{equation}
is the scalar product of the Killing vector and we have used eq.
(\ref{killo1}) and the relation
\begin{equation}
k^i_\La L^\La=k^{i^\star}_\La \bar L^\La=\cP_\La L^\La=\cP_\La \bar L^\La=0\ .
\end{equation}
  $\cP^x_\La$ are
the quaternionic (triplet) prepotentials and $U^{\La\Si}, L^\La $
are special geometry data.

In a theory with only abelian vectors, the potential may still be 
non-zero due to Fayet-Iliopoulos terms:
\be
\cP^x_\La=\xi_\La^x\ ({\rm constant});\ \ \
\epsilon^{xyz}\xi^y_\La \xi^z_\Si=0 \ .
\ee
In this case
\be
V(z,\bar z)= (U^{\La\Si}-3 \bar L^\La L^\Si)\xi^x_\La \xi^x_\Si\ .
\ee
Examples with $V(z,\bar z)=0$ but non-vanishing gravitino mass (with
$N=2$ supersymmetry broken to $N=0$) were given in \cite{flatpotn2},
 then generalizing to $N=2$ the no scale models of $N=1$ supergravity
 \cite{flatpotn1}.
These models were obtained by taking a $\xi^x_\Lambda=(\xi_0,0,0)$ .
In this case the expression

\be
V= U^{00}-3 \bar L^0 L^0
\ee
reduces to
\be
V=(\partial _i K g^{i j^\star} \partial _{j^\star} K -3)e^K
\ee
which is the $N=1$ supergravity potential, with solution ( $V=0$) the cubic
 holomorphic prepotential

\be
F(X)=d_{ABC}\frac{X^A X^B X^C}{X^0}\ \ \ \ A=1,\ldots,n \ .
\ee

Another solution is obtained by taking the $\frac{SU(1,1)}{U(1)}
\otimes
\frac{SO(2,n)}{SO(n)}$ coset in the $SO(2,n)$ symmetric parametrization of
the symplectic sections ($X^\Lambda,F_\Lambda=\eta_{\La\Si} S X^\Si\ ;
\ X^\La X^\Si \eta_{\La\Si}=0\ ,\eta_{\La\Si}=(1,1,-1,\ldots,-1)$)
where a prepotential $F$ does not exist. In this case
\be
U^{\La\Si}-3 \bar L^\La L^\Si=-\frac{1}{i(S-\bar S)}\eta_{\La\Si}
\label{genov}
\ee
where we have used the fact that
\be
\cN_{\La\Si}=(S-\bar S) (\Phi_\La \bar\Phi_\Si+\bar\Phi_\La \Phi_\Si)
+\bar S\eta_{\La\Si}\ \ \ ,\ \Phi^\La=\frac{X^\La}{(X^\La \bar X_\La)^{1/2}}
\ .
\ee
The identity (\ref{genov}) allows one to prove that the tree level potential
of an arbitrary heterotic string compactification (including orbifolds with
twisted hypermultiplets) is semi-positive definite provided we don't gauge the
graviphoton and the gravidilaton vectors ($\rm i.e.$ $\cP^x_\La=0$ for
$\La=0,1$, $\cP^x_\La\neq 0$ for $\La=2,\ldots,n_V$). On the other hand, it
also proves that tree level supergravity breaking may only occurr if
$\cP^x_\La\neq 0$ for $\La=0,1$. This instance is related to models with
Scherk-Schwarz mechanism studied in the literature \cite{scsc,ant}. 

A vanishing potential can be obtained if $\xi^x_\La=(\xi_\La,0,0)$ with
\be
\xi_\La \xi_\Si \eta^{\La\Si}=0\ .
\ee
In this case we may also consider the gauge group to be $U(1)^{p+2}\otimes
 G(n_V-p)$ and introduce $\xi _ \Lambda \,=\, \left(\xi_0,\ldots,\xi_{p+1},0,
\ldots,0 \right)$
 such that
$\xi_\La \xi_\Si \eta^{\La\Si}=0$ where $\eta^{\La\Si}$ is the $SO(2,p)$
Lorentzian metric. The potential is now:
\be
V=k^i_\La g_{i j^\star} k^{j^\star}_\Si \bar L^\La L^\Si\ \ ;\ \
(U^{\La\Si}-3\bar L^\La L^\Si) \cP^{x}_\La \cP^{x}_\Si=0
\ee
where $k^i_\La L^{\Lambda}=0$ for $\La \leq p+1$. The gravitino have
equal mass
\be \mid m_{3/2} \mid \simeq e^{K/2} \mid \xi_\La X^\La \mid
\ee
with $\xi_\La \xi_\Si \eta^{\La\Si}=0$ , $\La=0, \ldots, p+1$.

It is amusing to note that the gravitino mass, as a function of the
$O(2,p)/O(2)\otimes O(p)$ moduli and of the F-I terms, just coincides with
 the central charge formula for the level $N_L=1$ in heterotic string
(H-monopoles), if the F-I terms are identified with the $O(2,p)$ lattice
electric charges.

Note that, because of the special form of the gauged $\hat Q$,
 $\hat \omega^x$, we see that whenever $\cP_\La\neq 0$ the gravitino is
charged with respect to the $U(1)$ factor and whenever $\cP^x_\La \neq 0$
 the gravitino is charged with respect to the $SU(2)$ factor of the
 $U(1)\otimes SU(2)$ automorphism group 
of the supersymmetry algebra. In the case of
$U(1)^p$ gauge fields with non-vanishing F-I terms $\xi^x_p=(0,0,\xi_p)$ the
 gauge field $A^\La_\mu \xi_\La=A_\mu$ gauge a $U(1)$ subgroup 
of $SU(2)_L$ susy algebra.

Models with breaking of $N=2$ to $N=1$ \cite{FGP1}
necessarily require $k^u_\La$ not to be zero. The
minimal model where this happens with $V=0$ is the one based on
\be
{\cal SK}\otimes \cQ=\frac{SU(1,1)}{U(1)}\otimes \frac{SO(4,1)}{SO(4)}\ ,
\label{sdodi}
\ee
where a $U(1)\otimes U(1)$ isometry of $\cQ$ is gauged. In this case
the vanishing of $V$ requires a compensation of the 
$\delta\lambda ,\delta\zeta$
variations with the gravitino contribution
\be
4 k^u_\La k^v_\Si h_{uv}+ U^{\La\Si} \cP^x_\La
\cP^x_\Si=3\bar L^\La L^\Si \cP^x_\La \cP^x_\Si\ .
\label{stredi}
\ee
The moduli space of vacua satisfying (\ref{stredi}) is a four dimensional
subspace of (\ref{sdodi}).

\par
One may wonder where are the explicit mass terms for hypermultiplets. In
$N=2$ supergravity, since the hypermultiplet mass is a central charge, 
which is gauged,
such term corresponds to the gauging of a $U(1)$ charge. This is best seen
if we consider the case where no vector multiplets (and then gauginos)
are present. In this case $L^\La=L^0=1$ and the potential becomes
\be
V=4 h_{uv} k^u k^v-3 \cP^x \cP^x
\ee
where $k^u$ is the Killing vector of a $U(1)$ symmetry of $\cQ$,
 gauged by the graviphoton and  $\cP^x$ is the associated prepotential. For
$\frac{SO(4,1)}{SO(4)}$ this reproduces the Zachos model \cite{zachos}.
 The gauged $U(1)$ in this model is contained
in $SU_R(2)$ which commutes with the symmetry $SU_L(2)$ in the decomposition of
$SO(4)=SU_L(2)\otimes SU_R(2)$. This model has a local minimum at vanishing
hypermultiplet vev at which $U(1)$ is unbroken, and the extrema (at $u=1$)
(maxima) which break $U(1)$. The extremal model is when both $n_H=n_V=0$.
Still we may have a pure F-I term
\be
V=-3 \xi^2\ \ \ \xi=(\xi,0,0)
\ee
This corresponds to the gauging of a $U(1)\subset SU(2)_L$ and gravitinos
have charged coupling. This model corresponds to anti-De Sitter $N=2$
supergravity \cite{fredas}.

\section{The rigid limit: N=2 matter coupled Yang--Mills theory}
\setcounter{equation}{0}
In this section we consider the rigid limit of matter coupled
N=2 supergravity. The aim is that of obtaining the most general
form of matter coupled N=2 super Yang--Mills theory. By this we
mean the rigid supersymmetric N=2 theory of $n$ vector multiplets
coupled to $m$ hypermultiplets interacting through a generic
{\it rigid special manifold} and a generic {\it hyperK\"ahler
manifold}. Such a theory, in general, is not  renormalizable:
renormalizability obtains only in the case of a {\it flat} special
manifold and a {\it flat} hyperK\"ahler manifold. Yet it is
very interesting as an effective low energy lagrangian. Seiberg
Witten lagrangian \cite{SW12}, is just an instance in this
general class. One
could derive this type of theory by direct methods solving
Bianchi identities in flat superspace and then constructing
the corresponding rheonomic action. It is however much simpler
to derive it through a suitable {\it scaling limit} from the N=2
supergravity theory. The contraction parameter is obviously the
Planck mass $\mu$ and the limit must be performed in such a way
that local special geometry flows to rigid special geometry and
quaternionic geometry flows to hyperK\"ahler geometry. We already
know how this can happen: the curvature of the line and $SU(2)$
bundles must flow to zero in the limit. In the next subsection
we describe the appropriate rescalings. Then in a further
subsection we report the final result written in space--time
component formalism for the benefit of the reader who does not
want to be involved with the rheonomy formalism.
\par
\subsection{Planck mass rescalings}
We begin with the special geometry sector.
Here we consider the covariantly holomorphic symplectic section
~\ref{covholsec} and we write:
\begin{equation}
V \, \equiv \, \left ( \matrix { L^\Lambda \cr M_\Sigma \cr }
\right ) \,\equiv \, \exp \left [ {\cal K}/2 \right ] \,
\left ( \matrix{ X^\Lambda \cr F_\Sigma \cr } \right ) \,
= \,  \exp \left [ {\hat {\cal K}}/(2\mu^2) \right ] \,
\left ( \Omega_0 + {1\over \mu} {\hat \Omega}  +{1\over {\mu^3}} \Omega_3
\right )
\label{scalauno}
\end{equation}
where:
\begin{equation}
\Omega_0 \, = \, \left ( \matrix{ {1\over{\sqrt{2}}} \cr {\bf 0} \cr
- {{\rm i}\over{\sqrt{2}}}\cr {\bf 0} \cr} \right ) \qquad
{\hat \Omega} \, = \, \left ( \matrix{ 0 \cr Y^I \cr
0   \cr F_J \cr } \right ) \qquad
\Omega _ 3 \, = \, \left ( \matrix{ {\hat Y}_0 \cr {\bf 0} \cr
{\hat F}_0 \cr {\bf 0} \cr} \right )
\label{scaladue}
\end{equation}
The hatted objects are those that survive in the infinite
Planck mass limit $\mu \, \to \, \infty $. Recalling eq.~\ref{specpot}
 we obtain:
\begin{eqnarray}
{\hat {\cal K}} & = & - \, \lim _{\mu \to \infty} \, \mu^2
\, \log \, \left [ {\rm i} \, \langle   \Omega \vert {\bar \Omega}
\rangle \, \right ] \nonumber\\
& =& - \, \lim _{\mu \to \infty} \, \mu^2
\, \log \, \Biggl [ 1 + {{\rm i}\over{\mu^2}} \, \left ( {\bar Y}^I F_I
- {\bar F}_J   Y^J \right ) + {{\sqrt {2}}\over{\mu^3}}
\left ( \mbox{Re} Y^0 - \mbox{Im} F_0 \right )   \nonumber\\
&\null & + \,
{{\rm i}\over{\mu^6}} \left ( {\bar Y}^0 F_0- {\bar F}_0 Y^0 \right )
\Biggr  ] \nonumber \\
& =& -\, {\rm i} \, \left ( {\bar Y}^I F_I
- {\bar F}_J  Y^J \right )\nonumber\\
& = & -\, {\rm i} \, \langle  {\hat \Omega} \vert {\hat {\bar
\Omega}} \rangle \, \equiv \, {\rm i}  {\hat \Omega}^T \,
\left ( \matrix{ {\bf 0} & \bfone \cr -\bfone & {\bf 0} \cr } \right ) \,
{\hat  {\bar \Omega }}
\label{ritrigpot}
\end{eqnarray}
which reproduces eq.~\ref{specpotrig} for the K\"ahler potential
of rigid special geometry. An observation here is in order. The
last line in eq.~\ref{ritrigpot} still differs from eq.~\ref{specpotrig}
in one respect: the symplectic metric and the symplectic sections
in ~\ref{ritrigpot} are $(2n+2)$--dimensional while those in
eq.~\ref{compati1rig} are $2n$--dimensional. Yet the entries of the
symplectic sections in the two additional dimensions are always zero
so that we can safely reduce the bundle and its structural group
from $Sp(2n+2,\IR)$ to $Sp(2n,\IR)$.
\par
Let us next consider the symplectic vector $U_i$ defined in eq.
~\ref{uvector}. Using the above rescalings we obtain:
\begin{equation}
U_i \, = \, {1\over\mu} \, {\hat U}_i \, + \, {1\over{\mu^2}}
\left ( \matrix { {1 \over {2 \sqrt{2}}} \, \partial_i {\hat {\cal
K}} \cr {\bf 0} \cr {{-\rm i} \over {2 \sqrt{2}}} \,
\partial_i {\hat {\cal K}} \cr {\bf 0} \cr } \right ) +
{1\over{\mu^3}}  \,
\left ( \matrix{ \dots \cr \dots \cr \dots \cr \dots \cr }
\right )
\label{rituvector}
\end{equation}
where
\begin{equation}
{\hat U}_i \, = \, \left ( \matrix{ 0 \cr \partial_i Y^I \cr 0 \cr
\partial_i F_J \cr } \right )\, = \, \partial_i \, {\hat \Omega}
\label{rituvecrig}
\end{equation}
So we have retrieved eq.~\ref{uvectorrig}, apart from the identically zero
extra entries. Hence we can set:
\begin{equation}
g_{ij^\star} \, = \, {1\over{\mu^2}} \, {\hat g}_{ij^\star}
\label{ritmetric}
\end{equation}
which is consistent with
\begin{equation}
{\hat g}_{ij^\star} \, = \, - \, {\rm i} \, \langle {\hat U}_i \vert
{\hat U}_{j^\star} \rangle
\label{ritinvrig}
\end{equation}
that reproduces the first of eq.s~\ref{sympinvrig}: the second of
such equations is retrieved by setting:
\begin{equation}
C_{ijk} \, = \, {1\over{\mu^2}} \, {\hat C}_{ijk} \, + \,
{\cal O} \left ( {1\over{\mu^3}}\right ) \quad \Longrightarrow \quad
{\hat C}_{ijk} \, = \, \langle {\partial_i \hat U}_j \vert
{\hat U}_{k} \rangle
\label{ursula}
\end{equation}
Finally we observe that the Levi--Civita connection $\Gamma^i_{jk}$
is not rescaled by any power of the Planck mass since it contains
a metric and an inverse metric (see eq.~\ref{curvelievi}). This
implies the following rescaling for the Riemann tensor of the
special manifold:
\begin{equation}
R_{ij^\star k \ell^\star } \, = \, g_{ip^\star} \,
R^{p^\star}_{\phantom{p^\star}j^\star k \ell^\star} \, =
\, {1\over{\mu^2}} \,
{\hat R}_{ij^\star k \ell^\star }
\label{riscarie}
\end{equation}
and the fundamental identity of local special geometry
~\ref{specialone} becomes
\begin{equation}
{\hat R}_{ij^\star k \ell^\star } \, = \, {1\over{\mu^2}} \,
( {\hat g}_{i j^\star} {\hat g}_{k \ell^\star}  +
{\hat g}_{k j^\star} {\hat g}_{i \ell^\star}) \, + \,
{\hat C}_{i k s} \, {\hat C}_{t^\star  j^\star \ell^\star}
\, {\hat g}^{s t^\star}
\label{ritspecialone}
\end{equation}
that in the limit $\mu \, \to \, \infty$ reproduces the
fundamental identity of rigid special geometry (eq.
~\ref{specialonerig}).
\par
Summarizing we have:
\vskip 0.2cm
\centerline{{\it Rescalings in the Special geometry sector}}
\vskip 0.1cm
\begin{center}
\begin{tabular}{lcr}\hline
\null & \null \\
$L^0 \, \to \, {1\over 2} \, + \, {\cal O}\left ( {1\over{\mu^2}} \right )$
& $L^I \, \to \, {1\over{\mu}} \, Y^I + \, {\cal O}\left ( {1\over{\mu^2}}
\right )$    \\
$g_{ij^\star} \, \to \, {1\over{\mu^2}} \,
{\hat g}_{ij^\star} \, + \, {\cal O}\left ( {1\over{\mu^3}}
\right )$  &
$C_{ijk} \, \to \, {1\over{\mu^2}} \,
{\hat C}_{ijk} \, + \, {\cal O}\left ( {1\over{\mu^3}}
\right )$\\
\null & \null \\
${ R}_{ij^\star k \ell^\star } \, \to \,  \, {1\over{\mu^2}} \,
{\hat R}_{ij^\star k \ell^\star } \, + \, {\cal O}\left ( {1\over{\mu^3}}
\right )$
& $ z^i \, \to \, {\hat z}^i $\\
\null & \null \\
$ f^0_i \, \to \, {1\over{\mu^2}} \, {\hat f}^0_i + \,
{\cal O}\left ( {1\over{\mu^3}}
\right )$ & $
f^I_i \, \to \, {1\over{\mu }} \, {\hat f}^I_i \,
+ \, {\cal O}\left ( {1\over{\mu^3}} \right )$ \\
\null & \null \\
$\Gamma^{i}_{jk} \, \to \, {\hat \Gamma}^{i}_{jk}$ & $
{\cal Q} \, \to \,  {1\over{\mu^2}} \, {\hat {\cal Q}}$ \\
\null & \null \\
\hline
\end{tabular}
\end{center}
\begin{equation}
\null \label{tabrescspec}
\end{equation}
Next we consider the rescalings in the quaternionic manifold
sector. Here we set
\vskip 0.2cm
\centerline{{\it Rescalings in the quaternionic manifold
sector}}
\vskip 0.1cm
\begin{center}
\begin{tabular}{lcr}\hline
\null & \null \\
 ${\cal U}^{\alpha A} \, \to \, {1\over{\mu}} \, {\hat {\cal U}}^{\alpha A}
 $ & $ h_{uv} \, \to \, {1\over{\mu^2}} \, {\hat h}_{uv} $ &
 $ q^u \, \to \, \hat q^u $    \\
 \null & \null & \null \\
 $ K^x \, \to \,  {1\over{\mu^2}} \, {\hat K}^x$ &
${\Omega}^x \, \to \, {\hat \Omega}^x $ & ${\cal P}^x_\Lambda \,
\to \, {1\over{\mu^2}} \, {\hat {\cal P}}^x_\Lambda $\\
\null & \null & \null \\
\hline
\end{tabular}
\end{center}
\begin{equation}
\null \label{tabrescquat}
\end{equation}
Using these rescalings the quaternionic algebra ~\ref{universala}
is satisfied by the rescaled hyperK\"ahler structures ${\hat
K}^x_{uv}$ as much as by the unrescaled ones ${ K}^x_{uv}$:
however the relation ~\ref{piegatello} between the $SU(2)$ curvatures
and the hyperK\"ahler structures ${\hat K}^x_{uv}$ becomes:
\begin{equation}
{\hat \Omega}^x \, = \, {\lambda\over{\mu^2}} \, {\hat K}^x
\label{riscaomegone}
\end{equation}
and in the limit $\mu \, \to \, \infty$ we obtain ${\hat \Omega}^x =
0$, as indeed we expect in the case of a hyperK\"ahler manifold.
Indeed we can rephrase this result by saying that, upon restoration of
physical units, the $SU(2)$--curvature scale is
\begin{equation}
\lambda \, = \, {{\hat \lambda}\over{\mu^2}}
\label{curvone}
\end{equation}
and in the infinite Planck mass limit  goes to zero. Indeed
when we fixed   $\lambda = -1$ to obtain canonical
kinetic terms this value had to be interpreted in squared
Planck mass units (namely ${\hat \lambda}= -1$).
Eq.s~\ref{tabrescquat} are consistent with the definition
\begin{equation}
 {\bf i}_\Lambda {\hat K}^x {\cal P}^x_\Lambda \, = \,
 \nabla {\hat {\cal P}}^x_\Lambda \, = \, d {\hat {\cal P}}^x_\Lambda
 \label{limitone}
\end{equation}
of the triholomorphic momentum map on hyperK\"ahler manifolds.
The last equality in eq.~\ref{limitone} is justified by the vanishing
of the $SU(2)$ curvature that is obtained in the limit $\mu \, \to \,
\infty$. Finally the rescaled form of the quaternionic
equivariance eq.~\ref{quatpesce} is
\begin{equation}
\{{\cal P}_\Lambda, {\cal P}_\Sigma\}^x \equiv 2 K^x (\Lambda,
\Sigma)  - {\o{\lambda}{\mu^2}} \, \varepsilon^{xyz} \,
{\cal P}_\Lambda^y  \, {\cal P}_\Sigma^z
\label{quatpescione}
\end{equation}
and in the infinite Planck mass limit it flows into the
equivariance condition of momentum maps for hyperK\"ahler manifolds,
that is eq.~\ref{hykapesce}.
\par
To complete our rigid limit programme we have to prescribe the
appropriate Planck mass rescalings for the space--time fields
and the fermions. These are as follows:
\vskip 0.2cm
\centerline{{\it Rescalings of space--time fields and fermions}}
\vskip 0.1cm
\begin{center}
\begin{tabular}{lcr}\hline
\null & \null \\
 $V^a \, \to \, {1\over{\mu}} \, {\hat V}^{a}
 $ & $ g_{\mu\nu} \, \to \, {1\over{\mu^2}} \, \hat {g}_{\mu\nu} $ &
 $ x^\mu \, \to \, \hat x^\mu $    \\
 \null & \null & \null \\
$\omega^{ab} \, \to \, \hat \omega^{ab}$  &
$A^0 \, \to \, {1\over{\mu}} {\hat A}^0 $ &
$A^I \, \to \, {1\over{\mu^2}} {\hat A}^I $\\
\null & \null & \null \\
$ \psi_A \, \to \,  {1\over{\sqrt{\mu}}} \, {\hat \psi}_A$ & $
\lambda^{iA} \, \to \, \sqrt{\mu} \, {\hat \lambda}^{iA}$ &
$ \zeta^\alpha \, \to \, {1\over{\sqrt{\mu}}} \,{\hat \zeta}^\alpha$
\\
\null & \null & \null \\
\hline
\end{tabular}
\end{center}
\begin{equation}
\null \label{tabrescmink}
\end{equation}
Utilizing the rescalings of eq.s~\ref{tabrescspec},~\ref{tabrescquat}
and~\ref{tabrescmink} in the curvature
definitions~\ref{torsdef},~\ref{gravdefdown},\break
~\ref{gravdefup},
~\ref{riecurv},~\ref{zcurv},~\ref{zcurvb},~\ref{lamcurv},~\ref{lamcurvb},
~\ref{Fcurv},~\ref{ucurv},~\ref{iperincurv},~\ref{iperincurvb} and
in the curvature rheonomic parametrization given in Appendix B, by
performing the limit $\mu \, \to \, \infty$ we obtain the rheonomic
parametrization and curvature definition of the rigid theory. Indeed
the first four equations~\ref{paramtors},~\ref{paramgrav},~\ref{paramgravb}
become:
\begin{eqnarray}
T^a & \equiv & dV^a \, - \, \omega^{ab} \, \wedge \, V^c \, \eta_{bc}
\, = \, 0 \nonumber\\
\rho_A & \equiv &  d\psi_A \, - \, {1\over 4} \, \omega^{ab} \, \wedge \,
\gamma_{ab} \, \psi_A  \, = \, 0 \nonumber\\
\rho^A  & \equiv &  d\psi^A \, - \, {1\over 4} \, \omega^{ab} \, \wedge \,
\gamma_{ab} \, \psi^A  \, = \, 0 \nonumber\\
 R^{ab} & \equiv & d\omega^{ab} \, - \, \omega^{ac} \, \wedge \,
 \omega^{cd} \,  \eta_{cd} \, = \, 0
\label{rifsupspace}
\end{eqnarray}
that are the structural equations of N=2 rigid superspace if
they are completed with
\begin{equation}
F^0 \,  \equiv \,  dA^0 \, + \, {1\over{\sqrt{2}}} \, \left [
{\bar \psi}_A \, \wedge \, \psi_B \, \varepsilon^{AB} \, + \,
{\bar \psi}^A \, \wedge \, \psi^B \, \varepsilon_{AB} \right ] \, =
\, 0
\label{centcharge}
\end{equation}
Eq.~\ref{centcharge} is precisely what we obtain
in the $\mu \, \to \, \infty $ limit from the case
$\Lambda=0$ of eq.s~\ref{gaugparam} and ~\ref{Fcurv}. Algebraically
eq.~\ref{centcharge} tells us that the graviphoton one--form is
the dual of the central charge generator. The case $\Lambda = I$
of the same equations provides the definition and rheonomic parametrization
of the  Yang--Mills curvatures in rigid superspace:
\begin{eqnarray}
F^I &\equiv & dA^I \, +\,{1\over 2} \, g\,
f^I_{\phantom{I}JK}\, A^J\,
\wedge\, A^K, +\,
\bar Y^I \bar\psi_A\wedge\psi_B
\epsilon^{AB}+Y^I\bar\psi^A\wedge \psi^B\epsilon_{AB}\nonumber\\
\null & = & F^I_{ab} V^a\wedge V^b + \left( {\rm i} \,f^{I}_i \,
\bar {\lambda}^{iA}
\gamma_a \psi^B \,\epsilon _{AB} +{\rm i} \,{\bar f}^{I}_{i^\star}
\,\bar\lambda^{i^\star}_A
\gamma_a \psi _B \,\epsilon^{AB} \right)\wedge V^a
\label{Fcurvrig}
\end{eqnarray}
From the $\mu \, \to \, \infty $ limit of eq.~\ref{gauginparam}
and~\ref{gauginparamb} we obtain the gaugino curvature
parametrizations:
\begin{eqnarray}
\nabla\lambda^{iA} &=&\nabla _a\lambda ^{iA} V^a+
{\rm i}\, Z^i_a \gamma^a \psi^A+
G^{-i}_{ab} \gamma^{ab} \psi _B \epsilon^{AB}+
D^{i\vert AB} \psi _B
\nonumber\\
\nabla\lambda^{i^\star}_A &=&
\nabla_a \lambda^{i^\star}_A V^a+ {\rm i} \,\bar Z^{i^\star}_a
\gamma^a\psi _A+G^{+{i^\star}}_{ab} \gamma^{ab} \psi^B \epsilon _{AB}
+ D^{i^\star}_{\phantom{i^\star}AB}\,  \psi^B
\label{gauginparambrig}
\end{eqnarray}
where $Z^i_a$ and $Z^{i^\star}_a$ are defined by eq.~\ref{scalparam}
and its complex conjugate that survive unmodified in the limit while
$G^{\pm{i^\star}}_{ab}$  and the auxiliary fields $D^{i\vert AB}$,
$D^{i^\star}_{\phantom{i^\star}AB}$ are given in eq.s~\ref{auxilcampi}.
As usual the rheonomic parametrizations correspond to the
supersymmetry transformation rules that we have collected in the
next subsection  together with the space--time action for the benefit
of those readers who doe  not want to get involved with the rheonomy
formalism. Also the rheonomic parametrizations~\ref{iperparam},
~\ref{iperinparam},~\ref{iperinparamb} mantain the same form in
the rigid limit, but the hyperino shifts $N^\alpha_A ,N^A_\alpha$
are now given by eq.s~\ref{auxilcampi}. Using the same scaling limit
one obtains the rigid rheonomic action (which we do not report) from
which one retrieves the space--time action reported in the next
subsection.
\subsection{Summary of the rigid N=2 Yang--Mills theory}
Let us then summarize our results by writing the final most general form of 
N=2 matter coupled Yang--Mills theory.  Such a theory arises
from a generic choice of the rigid special manifold
${\cal SM}_{rig}$, a generic choice of the Hyperk\"ahler manifold
${\cal HM}_{rig}$ and a generic choice of the {\it gauging}.
\par
Let:
\begin{equation}
{\cal F}^I \, \equiv \, d A^I \, + \,  {\o{1}{2}}\, f^{I}_{\phantom{I}JK}
A^J \, \wedge A^K \, { {=}} \, {\cal F}^I_{\mu\nu} \,
dx^\mu \, \wedge \, dx^\nu
\label{ymcurv}
\end{equation}
be the field--strengths of the gauge group ${\cal G}$. Let $z^i$
be the coordinates of the rigid special manifold
${\cal SM}_{rig}$, whose complex dimension
$n$ equals the real dimension  of the gauge group and let $q^u$ be the
$4 \, m$ coordinates of the Hyperk\"ahler manifold
${\cal HM}_{rig}$. In addition let $\lambda^{iA}, \lambda^{i^\star}_{A}$
be the two chiral projections of the gaugino field
and $\zeta^\alpha , \zeta_\alpha$ the two chiral projections of the
hyperino field. Let us moreover define   :
\begin{eqnarray}
&& \mbox{\it the covariant derivatives of the Bose fields}\nonumber\\
\nabla_\mu z^i &=& \partial_\mu z^i \, + \, g \,A^I_\mu \, k^i_I \nonumber\\
\nabla_\mu z^i &=& \partial_\mu {\bar z}^i \, + \, g \, \,A^I_\mu \,
k^{i^\star}_I \nonumber\\
\nabla_\mu q^u & = & \partial_\mu q^u \, + \, g \, \,A^I_\mu \,
k_I^u \nonumber\\
&& \mbox{$\qquad\qquad\qquad\qquad${and}$\qquad\qquad$} \nonumber\\
&& \mbox{\it the covariant derivatives of the Fermi fields }\nonumber\\
\nabla_\mu \lambda^{iA} & = & \partial_\mu \lambda^{iA} \, +
\, \left ( {\Gamma}^{i}_{jk} \, \nabla_\mu z^j \, + \, g \,
A^I_\mu \, \partial_j k^i_I \right ) \, \lambda^{jA} \nonumber\\
\nabla_\mu \lambda^{i^\star}_A & = & \partial_\mu \lambda^{i^\star}_A \, +
\, \left ( {\bar \Gamma}^{i^\star}_{j^\star k^\star} \,
\nabla_\mu {\bar z}^{j^\star} \, + \, g \,
A^I_\mu \, \partial_{j^\star} k^{i^\star}_I \right ) \,
\lambda^{j^\star}_A \nonumber\\
\nabla_\mu \zeta^{\alpha} & = & \partial_\mu \zeta^\alpha \,
+ \,   \left ( {\Delta}^{\alpha\beta}_u \, \nabla_\mu q^u \, +
\, g \,
A^I_\mu \, \partial_u k^v_I \, {\cal U}^{u\vert \alpha A}
\, {\cal U}^{\beta B}_v \,\varepsilon_{AB} \right ) \, \IC_{\beta\gamma}
\, \zeta^\gamma\nonumber\\
\nabla_\mu \zeta_{\gamma} & = & \partial_\mu \zeta_\gamma \,
+ \,    \IC_{\gamma\alpha}
\left ( {\Delta}^{\alpha\beta}_u \, \nabla_\mu q^u \, +
\, g \,
A^I_\mu \, \partial_u k^v_I \, {\cal U}^{u\vert \alpha A}
\, {\cal U}^{\beta B}_v \,\varepsilon_{AB} \right )
\, \zeta_\beta
\end{eqnarray}
In terms of these field strengths and derivatives and of all
the geometric structures pertaining to rigid special manifolds
and to hyperK\"ahler manifolds discussed in previous sections
the most general N=2 supersymmetric invariant lagrangian has the
following form:
\begin{center}
\begin{tabular}{c}
\null\\
\hline
\null \\
{\it  Matter coupled N=2 Yang Mills action}\\
\null \\
\hline
\end{tabular}
\end{center}
\vskip 0.2cm
\begin{equation}
  {\cal L}  \,  = \,    {\cal L}_{kin} \, + \,
 {\cal L}_{Pauli}
  \, + \,
 {\cal L}_{massmatrix} \, + \, {\cal L}_{potential}  \, + \, {\cal L}_{4fermi}
\label{riglagratut}
\end{equation}
where
\begin{eqnarray}
\lefteqn{ {\cal L}_{kin}\, = \, {\rm i} \, \Bigl (
{\bar {\cal N}}_{IJ} \, {\cal F}^{I-}_{\mu\nu} \, {\cal F}^{J-\vert \mu\nu}
\, - \, {  {\cal N}}_{IJ} \,
{\cal F}^{I+}_{\mu\nu} \, {\cal F}^{J+\vert \mu\nu}  \Bigr )
}\nonumber\\
&& \mbox{}\, + \,  g_{ij^\star} \, \nabla^\mu z^i \, \nabla_\mu {\bar
z}^{j^\star} \, + \,  h_{uv} \, \nabla^\mu q^u \, \nabla_\mu  q^v \,
\nonumber\\
&& \mbox{} \,- \,{\o{\rm i}{2}} \, g_{ij^\star} \,
\Bigl ( {\bar \lambda}^{iA} \, \gamma^\mu \, \nabla_\mu
\lambda^{j^\star}_A \, + \,
{\bar \lambda}^{j^\star}_A \, \gamma^\mu \, \nabla_\mu
\lambda^{iA} \Bigr ) \nonumber\\
&& \mbox{} \, - \, \mbox{}  {\rm i}
\Bigl ( {\bar \zeta}^{\alpha} \, \gamma^\mu \, \nabla_\mu
\zeta_{\alpha} \, + \,
{\bar \zeta}_{\alpha} \, \gamma^\mu \, \nabla_\mu
\zeta^{\alpha}\Bigr )
\label{kinerige}\\
\lefteqn{ {\cal L}_{Pauli}\, = \, {\rm i}
{\o{1}{2}} \, C_{ijk} \, \Bigl ( g^{k\ell^\star}{\bar f}^J_{\ell^\star}\,
\mbox{Im}{\cal N}_{IJ} \, {\cal F}_{\mu\nu}^{-I} \Bigr )
\, {\bar \lambda}^{iA} \, \gamma^{\mu\nu} \,
\lambda^{jB} \, \varepsilon_{AB} }\nonumber\\
&& \mbox{}\, - \,{\rm i} {\o{1}{2}} \,
{\bar C}_{i^\star j^\star k^\star} \, \Bigl (
g^{k^\star\ell}{  f}^J_{\ell }\,
\mbox{Im}{\cal N}_{IJ} \,
{\cal F}_{\mu\nu}^{+I} \, \Bigr ) \,
{\bar \lambda}^{i^\star}_{A} \, \gamma^{\mu\nu} \,
\lambda^{j^\star}_B \, \varepsilon^{AB}
\label{paulirige}\\
\lefteqn{ {\cal L}_{massmatrix} \, = \, {\cal M}^{\alpha \vert \beta}\,
{\bar \zeta}_\alpha
\, \zeta_\beta    \, + \, {\cal M}_{\alpha \vert \beta} \, {\bar \zeta}^\alpha
\, \zeta^\beta }\nonumber\\
&&\mbox{} \, + \, {\cal M}^{\alpha\vert }_{\phantom{\alpha}\vert iB}\,
{\bar \zeta}_\alpha \, \lambda^{iB} \, + \,
{\cal M}_{\alpha \vert  i^\star }^{\phantom{\alpha} \vert
\phantom{i^\star}B}
\, {\bar \zeta}^\alpha \, \lambda^{i^\star}_B
\nonumber\\
&& \mbox{} \, + \, {\cal M}_{iA\vert \ell B}
\, {\bar \lambda}^{iA} \lambda^{\ell
B}  \, + \, {\cal
M}^{\phantom{i^\star}A\vert\phantom{\ell^\star}B}_{i^\star\phantom{A}
\vert \ell^\star \phantom{B}}\,
{\bar \lambda}^{i^\star}_A \lambda^{\ell^\star}_B
\label{massmatrixrig}\\
\lefteqn { {\cal L}_{potential} \, = \,- \, {\rm V}\bigl ( z, {\bar
z}, q \bigr ) }
\label{potrig}\\
\lefteqn{ {\cal L}_{4fermi} \, = \, {\o{1}{4}} \, R_{ij^\star \ell
k^\star} \, {\bar \lambda}^{iA}\, \lambda^{\ell B} \,
{\bar \lambda}^{j^\star}_{A} \, \lambda^{k^\star}_{B} }\nonumber\\
&&\mbox{} \, + \, {\o{1}{2}} \, \IR^\alpha_{\phantom{\alpha}\beta\vert
ts}\, {\cal U}^t_{A\gamma} \,{\cal U}^s_{B\delta} \,
\varepsilon^{AB} \, \IC^{\delta\eta} \, {\bar \zeta}_\alpha
\, \zeta_\eta \,   {\bar \zeta}^\beta
\, \zeta^\gamma \nonumber\\
&& \mbox{} \,- \, {\o{1}{32}}\,\mbox{Im}{\cal N}_{IJ} \,
   C_{ijk} \, C_{\ell m n} \,
\, g^{kr^\star}   g^{ns^\star}\, {\bar f}^I_{r^\star}
 {\bar f}^J_{s^\star}   {\bar \lambda}^{iA}
\gamma_{\mu\nu}  \lambda^{jB} \, {\bar \lambda}^{\ell C}
\gamma^{\mu\nu}  \lambda^{m D} \, \varepsilon_{AB}
\varepsilon_{CD}\nonumber\\
&& \mbox{} \,-  \,{\o{1}{32}} \,\mbox{Im}{\cal N}_{IJ} \,
 {\bar C}_{i^\star j^\star k^\star }
{\bar C}_{\ell^\star m^\star n^\star} \,
\, g^{k^\star r}   g^{n^\star s}
{  f}^I_{  r}
{  f}^J_{  s} \, {\bar \lambda}^{i^\star}_{A}
\gamma_{\mu\nu}  \lambda^{j^\star}_{B} \, {\bar \lambda}^{k^\star}_{C}
\gamma^{\mu\nu}  \lambda^{\ell^\star}_{D} \, \varepsilon^{AB}
\varepsilon^{CD}
\label{4fermirig}\\
\end{eqnarray}
where the mass--matrices and the scalar potential are given by:
\begin{center}
\begin{tabular}{c}
\null\\
\hline
\null \\
{\it  N=2 Yang Mills mass matrices and scalar potential }\\
\null \\
\hline
\end{tabular}
\end{center}
\vskip 0.2cm
\begin{eqnarray}
 {\cal M}^{\alpha \vert \beta}  &=&-   \, g \,
{\cal U}^{\alpha A}_u \, {\cal U}^{\beta B}_v \, \varepsilon_{AB}
\, \nabla^{[u} \, k^{v]}_I \, Y^I \nonumber\\
{\cal M}_{\alpha \vert \beta} &=&-\, g \,
{\cal U}_{\alpha A\vert u} \,
{\cal U}_{\beta B \vert v} \,
\varepsilon^{AB}
\, \nabla^{[u} \, k^{v]}_I \, {\bar Y}^I \nonumber\\
 {\cal M}^{\alpha\vert }_{\phantom{\alpha}\vert iB} &=&
  \, 4 \, g \, {\cal U}^{\alpha A}_u \, k^u_I \, f^I_i \,
\varepsilon_{AB} \nonumber\\
{\cal M}_{\alpha \vert  i^\star }^{\phantom{\alpha} \vert
\phantom{i^\star}B} &=&-
  \, 4 \, g \, {\cal U}_{\alpha A\vert u}
\, k^u_I \, {\bar f}^I_{i^\star} \,   \varepsilon^{AB}\nonumber\\
{\cal M}_{iA\vert \ell B} &=& {\o{1}{3}} \, g \,
\Bigl ( \varepsilon_{AB}\,  g_{ij^\star}   k^{j^\star}_I   f_\ell^I +
{\rm i} (\sigma_x)_A^{\phantom{A}C} \epsilon_{BC} \, {\cal P}^x_I
\, \nabla_\ell f^I_i \Bigr )\nonumber\\
{\cal
M}^{\phantom{i^\star}A\vert\phantom{\ell^\star}B}_{i^\star\phantom{A}
\vert \ell^\star \phantom{B}}
&=&     {\o{1}{3}} \, g \,
\Bigl ( \varepsilon^{AB}\,  g_{i^\star j}   k^{j}_I
{\bar f}_{\ell^\star}^I -
{\rm i}\epsilon^{AC}(\sigma_x)_C^{\phantom{C}B}  \, {\cal P}^x_I
\, \nabla_{\ell^\star} {\bar f}^I_{i^\star} \Bigr )
\label{pesamatricerig}\\
{\rm V}\bigl ( z, {\bar
z}, q \bigr )&=& g^2 \, \left ( \, g_{ij^\star} k^i_I \, k^{j^\star}_J
\, + \, 4 \, h_{uv} \, k^u_I k^v_J \right) \, {\bar Y}^I \, Y^J
\nonumber\\
&~& + \, g^{ij^\star} \, f^I_{i} \, {\bar f}^J_{j^\star} \,
\sum_{x=1}^{3} \, {\cal P}^x_I \, {\cal P}^x_J
\label{potentissrig}
\end{eqnarray}
The coupling constant in front of the mass--matrices and of the
potential is just a symbolic notation to remind the reader
that these terms are entirely
due to the gauging and vanish in the ungauged theory. In general
there is not a single coupling constant rather there are
as many independent coupling constants as mutually commuting
subgroups in the gauge group. For instance if ${\cal G}$ is a
product or $r$ $U(1)$--factors, there are $r$ independent
coupling constants that can be reabsorbed into the definition
of the killing vectors $k_I^i, k_I^u$.
\par
The supersymmetry transformation rules with respect to which
the lagrangian ~\ref{riglagratut} is invariant are the following
ones:
\vfill
\eject
\begin{center}
\begin{tabular}{c}
\null\\
\hline
\null \\
{\it N=2 rigid transformation rules of Bose fields}\\
\null \\
\hline
\end{tabular}
\end{center}
\vskip 0.2cm
\begin{eqnarray}
\delta \, A_\mu^I &=& + \, {\rm i} \, \left ( \, f^I_i \,
{\bar \lambda}^{iA} \, \gamma_\mu \, \epsilon^B \, \varepsilon_{AB}
\, + \, {\bar f}^{I}_{i^\star}
{\bar \lambda}{i^\star}_{\phantom{i^\star}A} \, \gamma_\mu \,
\epsilon_B \, \varepsilon^{AB} \, \right ) \nonumber\\
\delta z^i & = & + \,{\bar \lambda}^{iA} \, \epsilon_A \nonumber\\
\delta {\bar z}^{i^\star} & = & + \, {\bar \lambda}^{i^\star}_A \,
\epsilon^A \nonumber\\
{\cal U}^{\alpha A}_u (q) \, \delta q^u & = &
\varepsilon^{AB} \, \IC^{\alpha\beta} \, {\bar \epsilon_B} \zeta_\beta
+ {\bar \epsilon}^A \zeta^\alpha \label{bosetransformazie}
\end{eqnarray}
\begin{center}
\begin{tabular}{c}
\null\\
\hline
\null \\
{\it  N=2 rigid transformation rules of Fermi fields}\\
\null \\
\hline
\end{tabular}
\end{center}
\vskip 0.2cm
\begin{eqnarray}
\delta {  \lambda}^{iA}& = &{\rm i} \, \nabla_{\mu} z^i \, \gamma^\mu
\, \epsilon^A \, + \, G^{-i}_{\mu\nu} \, \gamma^{\mu\nu} \,
\epsilon_B \, \varepsilon^{AB}  \, + \, D^{i\vert AB} \, \epsilon_{B}
\nonumber\\
\delta {  \lambda}^{i^\star}_A& = &{\rm i} \, \nabla_{\mu}
{\bar z}^{i^\star} \,
\gamma^\mu
\, \epsilon_A \, + \, G^{+i^\star}_{\mu\nu} \, \gamma^{\mu\nu} \,
\epsilon^B \, \varepsilon_{AB}  \, + \, D^{i^\star}_{
\phantom{i^\star}\vert AB} \, \epsilon^{B}
\nonumber\\
\delta {\zeta}_{\alpha}&=& {\rm i}\, {\cal U}^{\beta B}_u \,
\nabla_\mu q^u \, \gamma^\mu  \epsilon^A \, \varepsilon_{AB} \,
\IC_{\alpha\beta} \, + \, N^A_\alpha \, \epsilon_A \nonumber\\
\delta {\zeta}^{\alpha}&=& {\rm i}\, {\cal U}_{\beta B\vert u} \,
\nabla_\mu q^u \, \gamma^\mu  \epsilon_A \, \varepsilon^{AB} \,
\IC^{\alpha\beta} \, + \, N_A^\alpha \, \epsilon^A
\label{fermitransformazie}
\end{eqnarray}
where:
\begin{center}
\begin{tabular}{c}
\null\\
\hline
\null \\
{\it  N=2 rigid values of the auxiliary fields}\\
\null \\
\hline
\end{tabular}
\end{center}
\vskip 0.2cm
\begin{eqnarray}
G^{-i}_{\mu\nu} & = & {\rm i} \,  g^{ij^\star}
{\bar f}^I_{j^\star} \, \mbox{Im}{\cal N}_{IJ} \,
\left ( {\cal F}^{-J}_{\mu\nu} + {\o{1}{8}} \nabla_k f^J_{\ell}
\, {\bar \lambda}^{kA} \, \gamma_{\mu\nu} \, \lambda^{\ell B}
\, \varepsilon_{AB} \, \right )\nonumber\\
G^{+i^\star}_{\mu\nu} & = &  \, {\rm i} \,  g^{i^\star j}
{  f}^I_{j } \, \mbox{Im}{\cal N}_{IJ} \,
\left ( {\cal F}^{+J}_{\mu\nu} + {\o{1}{8}} \nabla_{k^\star}
{\bar f}^J_{\ell^\star}
\, {\bar \lambda}^{k^\star}_A \, \gamma_{\mu\nu} \,
\lambda^{\ell^\star}_B
\, \varepsilon^{AB} \, \right )\nonumber\\
D^{i\vert AB} & = & Y^{i\vert AB} + W^{i\vert [AB]} + W^{i\vert (AB)}
\nonumber\\
D^{i^\star}_{\phantom{i^\star}\vert AB} & = &
Y^{i^\star}_{\phantom{i^\star}\vert AB} +
W^{i^\star}_{\phantom{i^\star}\vert [AB]}
+ W^{i^\star}_{\phantom{i^\star}\vert (AB)}\nonumber\\
Y^{i\vert AB} & = & {\rm i} {\o{1}{2}} \, g^{ij^\star}\,
{\bar C}_{  j^\star k^\star \ell^\star } \, {\bar
\lambda}^{k^\star}_C \, \lambda^{\ell^\star}_D  \varepsilon^{AC}
\, \varepsilon^{BD} \nonumber\\
Y^{i^\star}_{\phantom{i^\star}\vert AB} & = &
- {\rm i} {\o{1}{2}} \, g^{i^\star j}\,
{  C}_{ j  k \ell  } \, {\bar
\lambda}^{k  C} \, \lambda^{\ell D}  \varepsilon_{AC}
\, \varepsilon_{BD}\nonumber\\
 W^{i\vert [AB]} & = & \varepsilon^{AB} \, k^i_I {\bar
 Y}^I \nonumber\\
 W^{i^\star}_{\phantom{i^\star}\vert [AB]} & = &
 \varepsilon_{AB} \, k^{i^\star}_I {
 Y}^I \nonumber\\
 W^{i\vert (AB)} & = & - {\rm i} \, \epsilon^{AC}(\sigma_x)_C^{\phantom{C}B}\,
 {\cal P}^x_I g^{ij^\star} {\bar f}_{j^\star}^I \nonumber\\
W^{i^\star}_{\phantom{i^\star}\vert (AB)} & = &
{\rm i} \, (\sigma_x)_A^{\phantom{A}C} \epsilon_{BC}\, {\cal P}^x_I
g^{i^\star j} f_{j }^I \nonumber\\
 N^A_\alpha & = & 2 \, {\cal U}^A_{\alpha\vert u} \, k^u_{I} \,
 {\bar Y}^I \nonumber\\
 N_A^\alpha & = & -2 \, {\cal U}^\alpha_{A\vert u} \, k^u_{I} \,
 {  Y}^I
\label{auxilcampi}
\end{eqnarray}

\subsection{The renormalizable microscopic theory}
As an exemplification of the general formalism and for the sake of its
intrinsic interest, in this subsection we consider the case of the
renormalizable microscopic  N=2 (matter coupled) Yang--Mills theory.
The theory is specified by the choice of the following
geometrical data:
\begin{enumerate}
\item { A flat rigid special manifold ${\cal SM}_{flat}$ describing
the vector multiplet couplings}
\item{A flat Hyperk\"ahler manifold ${\cal HM}_{flat}$ describing the
hypermultiplet couplings}
\end{enumerate}
Let us briefly discuss these geometries and the corresponding
form of the Lagrangian.
\vskip 0.2cm
\leftline{\underline {\it Flat rigid special geometry}}
\vskip 0.2cm
In the vector multiplet sector the appropriate geometry is
described as follows. Let $\theta$ be the theta--angle, $1/g^2$ the
inverse of the squared gauge coupling constant, and
${\bf g}_{IJ}$ the
constant Killing metric on the gauge Lie algebra. Define the complex
parameter:
\begin{equation}
\tau \, = \, \theta \, + \, \mbox{i} \, {\o{1}{g^2}}
\label{tauparam}
\end{equation}
and choose as holomorphic section of the flat symplectic bundle
the following one:
\begin{equation}
{\hat \Omega} \, = \, \left ( \matrix{ Y^I \cr \tau \, {\bf g}_{IJ}
\, Y^J } \right ) \qquad \qquad I,J=1,\dots , n=\mbox{dim}{\cal G}
\label{olorenorma}
\end{equation}
In this case the upper half of the holomorphic section ~\ref{olorenorma}
can be taken as coordinates on the manifold (the special
coordinates):
\begin{equation}
z^i \, \equiv \, Y^I \, .
\label{specialrenorma}
\end{equation}
The action of the gauge group on these coordinates is obviously the
adjoint action:
\begin{equation}
\delta_I \, Y^J \, = \, f^J_{\phantom{Y}IK} \, Y^K
\label{aggiunto}
\end{equation}
where $ f^J_{\phantom{Y}IK}$ are the structure constants of the gauge
Lie algebra:
\begin{equation}
\left [ t_I \, , \, t_J \right ] \, = \, f^K_{\phantom{K}IJ} \, t_K
\label{algebret}
\end{equation}
$t_I$ being a basis of  generators.
Hence using eq.s~\ref{sympinvrig} and ~\ref{intriscripenrig} we obtain
\begin{equation}
\matrix{
 {\cal N}_{IJ} & = &  {\bar \tau} \, {\bf g}_{IJ}  &  \null  &
 g_{ij^\star}  & = &  2 \, \mbox{Im} \tau \, {\bf g}_{IJ}    \cr
 \mbox{Im}{\cal N}_{IJ}  & = &  - \,  \mbox{Im}  {  \tau}\,
{\bf g}_{IJ}  &
 \null  &
 f^I_i & = &  \delta^I_i    \cr
 C_{ijk}  & = &  0   &
 \null  &
 k_I^j  &  =  &  f^J_{\phantom{Y}IK} \, Y^K   \cr}
\end{equation}
\vskip 0.2cm
\leftline{\underline {\it Flat HyperK\"ahler geometry}}
\vskip 0.2cm
In the hypermultiplet sector we arrange the $4 m$ coordinates $q^u$
of  ${\cal HM}_{flat} \, = \, \IR^{4m}$ into a $4 m$ column vector:
\begin{equation}
 {\bf q} \, \equiv \, q^{a \vert t} \qquad \qquad \cases { a=0,1,2,3 \cr
 t=1,2,\dots \, m \cr }
 \label{qfieldo}
\end{equation}
that is regarded as an element of the tensor product
 $\IR^4 \otimes \IR^m \, \sim \, \IR^{4m}$. Let
 \begin{equation}
 \matrix {
J^{+ \vert 1}& = & \left ( \matrix{ 0 & 1 & 0 & 0 \cr -1 & 0 & 0 & 0
\cr 0 & 0 & 0 & 1 \cr 0 & 0
& -1 & 0 \cr  } \right ) & \quad &
J^{- \vert 1}& = & \left ( \matrix{ 0 & 1 & 0 & 0 \cr
-1 & 0 & 0 & 0 \cr
0 & 0 & 0 & -1 \cr
0 & 0 & 1 & 0 \cr  } \right ) \cr
\null & \null & \null & \null & \null & \null & \null \cr
J^{+ \vert 2}& = & \left ( \matrix{ 0 & 0 & -1 & 0
                  \cr 0 & 0 & 0 & 1 \cr 1 & 0 & 0 & 0 \cr
                  0 & -1 & 0 & 0 \cr  } \right )  & \quad &
J^{- \vert 2} & = & \left (
\matrix{ 0 & 0 & -1 & 0 \cr 0 & 0 & 0 & -1
\cr 1 & 0 & 0 & 0
\cr 0 & 1 & 0 & 0 \cr  } \right ) \cr
\null & \null & \null & \null & \null & \null & \null \cr
J^{+ \vert 3}& = & \left ( \matrix{ 0 & 0 & 0 & 1 \cr
                   0 & 0 & 1 & 0 \cr
                   0 & -1 & 0 & 0 \cr
                   -1 & 0 & 0 & 0 \cr  } \right ) & \quad &
J^{- \vert 3} & = & \left (
 \matrix{ 0 & 0 & 0 & 1
\cr 0 & 0 & -1 & 0
\cr 0 & 1 & 0 & 0
\cr -1 & 0 & 0 & 0 \cr  } \right )
\cr}
\end{equation}
be the two triplets of self--dual and antiself dual 't Hooft matrices
satisfying the quaternionic algebra:
\begin{eqnarray}
J^{\pm\vert x} \, J^{\pm\vert y} & = & - \, \delta^{xy} \,
\bfone_{4\times 4} \,
+ \, \varepsilon^{xyz} \, J^{\pm\vert z}\nonumber\\
J^{\pm\vert x}_{ab} &=& \pm \, {\o{1}{2}} \, \varepsilon_{abcd} \,
J^{\pm\vert x}_{cd}\nonumber\\
0 &=& \left [ J^{+\vert x} \, , \, J^{-\vert y} \right ]  \qquad
\forall x, y
\end{eqnarray}
Let, furthermore
\begin{equation}
  e_a \, =\, \cases{e_0 \, = \,
  \left ( \matrix{ 1 & 0\cr 0 & 1 \cr} \right ) \cr
  e_x \, = \, \cases{ e_1 \, = \,
\left ( \matrix{ 0 & -{\rm i}\cr -{\rm i} & 0 \cr} \right ) \cr
e_2 \, = \,
\left ( \matrix{ 0 & -1\cr 1 & 0 \cr} \right ) \cr
e_3 \, = \,
\left ( \matrix{ -{\rm i} & 0\cr 1 & {\rm i} \cr} \right ) \cr}\cr}
\label{quatunit}
\end{equation}
be a complete basis of two matrices for the expansion of a generic
quaternion:
\begin{equation}
Q \, \equiv \, q^a \, e_a
\end{equation}
$e_x$, being the three imaginary units.
The flat HyperK\"ahler metric and the corresponding triplet of
HyperK\"ahler 2--forms are given by:
\begin{eqnarray}
ds^2 & \equiv & h_{uv} \, dq^u \,  dq^v \, = \, d{\bf q}^T \, 
\left (\bfone_{4\times 4} \otimes \bfone_{m\times m} \right ) \, d{\bf q}
\nonumber\\
K^x & = &  d{\bf q}^T \, \wedge \, \left ( J^{+ \vert x}
\otimes \bfone_{m \times m} \right ) \, d{\bf q}
\label{spiatto}
\end{eqnarray}
Alternatively in the above formula one can use the triplet of
antiself dual t'Hooft matrices to define the HyperK\"ahler structure.
Using the identities:
\begin{equation}
\label{ide3}
\left\{\begin{array}{l}
J^{+\vert x}_{ab} = {\o{1}{2}} \hskip 3pt \mbox{tr} (e_a {\bar e}_b e_x^T)\\
J^{-\vert x}_{ab} = - \,{\o{1}{2}} \hskip 3pt \mbox{tr} (e_a e_x^T{\bar e}_b )
\end{array}\right.
\end{equation}
and rearranging the $4m$ coordinates $q^{a\vert t}$ into an
$m$-vector of quaternions:
\begin{equation}
{\bf Q} \, = \, \left ( \matrix {Q^1 \, = \, q^{a \vert 1} \, e_a \cr
Q^2 \, = \, q^{a \vert 2} \, e_a \cr
\dots \cr
Q^t \, = \, q^{a \vert t} \, e_a \cr
\dots \cr } \right )
\label{quatvector}
\end{equation}
eq.s~\ref{spiatto} can be rewritten as follows:
\begin{eqnarray}
ds^2 & = &  {\o{1}{2}} \hskip 3pt \mbox{tr}  \, \left (
d {\bf Q}^\dagger  \, \bfone_{m\times m} d {\bf Q} \right
)\nonumber\\
K & = &   \,{\o{1}{2}} \, d{\bf Q}^T \, \wedge \,
\bfone_{m\times m} d{\bar {\bf Q}}  \, = \,
{\o{1}{2}} \, K^x \, e_x^T
\label{riscrivo}
\end{eqnarray}
The action of the gauge group ${\cal G}$ on the hypermultiplets is
assumed to be linear and be generated by a set of $4m \times 4m $
matrices $T_I$. Namely we set:
\begin{equation}
\delta_I {\bf q} \, = \, T_I \, {\bf q} \, \quad\longrightarrow \quad
k^u_I \, = \, \left ( T_I \right )^{u}_{\phantom{u}v} \, q^v
\label{linearazione}
\end{equation}
In order for this action to be an isometry of the Euclidean diagonal
metric ~\ref{spiatto} it is necessary and sufficient that the
matrices $T_I$ belong to the orthogonal Lie algebra $SO(4m)$, namely:
\begin{equation}
T_I^T = - T_I
\end{equation}
The action of ${\cal G}$ however is not only required to be isometrical
but also to be triholomorphic. This means:
\begin{equation}
\ell_{I} \, K^x \, \equiv \, {\bf i}_I \, d K^x \, + \, d\, {\bf i}_I \,
K^x \, = \, d\, {\bf i}_I \,
K^x  \, = \, 0
\end{equation}
A straightforward calculation yields:
\begin{equation}
d\, {\bf i}_I \, K^x  \, =  - \, d{\bf q}^T \, \wedge \, \left [
T_I \, , \, J^{+\vert x}\otimes \bfone_{m \times m} \right ] \, d{\bf
q}
\label{commutazia}
\end{equation}
so that the triholomorphicity condition is that the generators $T_I$
should commute with the tensor product  of the 't Hooft matrices with
the unit matrix in $m$--dimensions. When this last condition is
verified we can write the momentum maps as:
\begin{equation}
{\cal P}^x_I \, = \, {\bf q}^T \,
J^{+\vert x}\otimes \bfone_{m \times m} \, T_I \,{\bf q}
\label{unmomento}
\end{equation}
Alternatively using the quaternionic notation we have:
\begin{equation}
 {\bf P}_I \, = \, {\o{1}{2}} \, {\cal P}^x_I \, e_x^T \, = \,
  \, {\o{1}{2}} \,{\bf Q}^T \bfone_{m \times m} \, T_I \, {\bar {\bf Q}}
\end{equation}
\vskip 0.2cm
\leftline{\underline {\it The lagrangian}}
\vskip 0.2cm
Using these ingredients the lagrangian of the microscopic renormalizable
theory is immediately retrieved from the general formulae of the
previous subsection. It is convenient to set:
\begin{equation}
\matrix{ {\bf Y} & \equiv & Y^I \, t_I & \null &
 {\bar {\bf Y}} & \equiv &{\bar Y}^I \, t_I \cr
 {\bf F}_{\mu\nu} & \equiv & {  F}^I_{\mu\nu}  \, t_I & \null &
 \mbox{tr} \left ( t_I t_J \right ) & \equiv & {\bf g}_{IJ} \cr }
\end{equation}
$t_I$ denoting a basis of generators of the gauge group
and in this condensed notation we obtain:
\begin{equation}
{\cal L}^{microscopic}_{N=2 YM} \, = \, {\cal
L}^{microscopic}_{bosonic} \, + \, {\cal
L}^{microscopic}_{fermionic}
\end{equation}
where the bosonic lagrangian is:
 \begin{eqnarray}
{\cal L}^{microscopic}_{bosonic} & = &
- \, \mbox{Im}\tau \, \mbox{tr} \,
\left ( {\bf F}_{\mu\nu} \, {\bf F}_{\mu\nu} \right )
\, + \, {\o{1}{2}} \,
\mbox{Re}\tau \, \mbox{tr} \, \left ( {\bf F}_{\mu\nu} \,
{\bf F}_{\rho\sigma} \right )
\, \varepsilon^{\mu\nu\rho\sigma} \nonumber\\
&& + 2\, \mbox{Im}\tau \, \mbox{tr} \, \left ( \nabla_\mu {\bf Y}  \,
\nabla_{\mu} {\bar {\bf Y}} \right ) \, + \, \nabla_\mu {\bf q}^T \,
\, \nabla_\mu {\bf q}  \,
- \, V({\bf Y},{\bf q}) \\
&& \nonumber\\
V({\bf Y},{\bf q})& = & 2 \, \mbox{Im}\tau \, \mbox{tr}\,
\left ( \left [{\bf  Y}  \, , \,  {\bar {\bf Y}} \right ] \right )^2
  \, - \,
2 \, {\bf q}^T \,\left \{  {\bf Y} \, , \, {\bar {\bf Y}} \, \right \}
{\bf q}
\nonumber\\
&& + \, {\o{1}{2 \mbox{Im}\tau}} \, \sum_{x=1}^{3} \,
{\cal P}^x_I \, {\cal P}^x_J \, {\bf g}^{IJ}
\end{eqnarray}
The formula for the scalar potential exhibits in a clear fashion the
flat directions associated with the moduli fields ${\bf Y}$ in the
Cartan subalgebra ${\cal H}$ of the gauge algebra. Actually the
potential is just homogeneous of degree four  in all the scalar fields
as expected from renormalizability.
\par
The fermionic part of the lagrangian also simplifies very much since
it just contains  the kinetic part and the mass terms induced by the
gauging.  The Pauli terms and the 4--fermi terms are all zero,
since the tensor $C_{ijk}$ vanishes and the
Riemann tensors of the special and HyperK\"ahler manifolds also vanish.
The evaluation of the mass matrices is straightforward by inserting
the explicit form of the Killing vectors and of the momentum maps
into eq.s~\ref{pesamatricerig}. The only item that is still missing
in such a calculation is the explicit form of the quaternionic vielbein.
This is very easily given. We set:
\begin{equation}
{\cal U}^{A \alpha} \, \equiv \, {\cal U}^{A \alpha}_{b \vert s} \,
dq^{b \vert s} \, = \, d{\bf Q} \, = \, dq^{a \vert t} \,
( e_a )^{A}_{\phantom{A}B}
\label{scordo1}
\end{equation}
and we identify the symplectic index $\alpha$ running on $2m$ values
with the pair of indices $B,t$ ( $B=1,2$; $t=1,\dots,m$). In this way
we obtain:
\begin{equation}
  {\cal U}^{A \phantom{B}\vert t}_{ \phantom{A} B   \vert b \vert s} \, = \,
  \delta^{t}_{s} \, (e_b)^{A}_{\phantom{A}B}
\label{scordo2}
\end{equation}
\appendix
\section*{Appendix A: The solution of the Bianchi identities
and the supersymmetry transformation laws}
\label{rheo}
\setcounter{equation}{0}
\setcounter{section}{1}
\par
In this Appendix we describe the geometric approach for the
derivation of the $N=2$
supersymmetry transformation laws of the physical fields.
As it will appear in the following this requires
the preliminary solution of Bianchi identities in superspace.
\par
The first step to perform is to extend the physical fields to
superfields in $N=2$
superspace: that means that the space--time 1-forms $\omega^{a\,b}$,
$V^a$,$\psi^A$, $\psi_A$, $A^{\Lambda}$ and the space--time 0--forms
$\lambda^{iA}$, $\lambda_{A}^{i^\star}$, $z^i$, $z^{i^\star}$,
$\zeta^{\alpha}$, $\zeta_{\alpha}$, $q^u$ defined in section 8 are promoted
to 1--superforms
and 0--superforms in $N=2$ superspace, respectively.
\par
The definition of the superspace curvatures actually coincides with
that given in eq.s~\ref{torsdef}--\ref{iperincurvb} provided all the
$p$--forms ($p=0,1,2$) are thought as $p$--superforms (here and in the
following by "curvatures" we mean not only 2--forms, but also the
1--forms defined as covariant differentials of the 0--form
superfields).\\
We note that the  definition of superspace curvatures in
the gravitational sector, namely:
\begin{eqnarray}
T^a & \equiv & {\cal D}V^a- {\rm i} \, \bar\psi_A\wedge
\gamma^a\psi^A\label{torsdefap}\\
\rho_A &  \equiv & d\psi_A-{1\over 4} \gamma_{ab} \,
\omega^{ab}\wedge\psi_A+
{{\rm i} \over 2} {\hat {\cal Q}}\wedge \psi_A +
{\hat \omega}_A^{~B}\wedge \psi_B
\equiv \nabla \psi_A \label{gravdefdownap} \\
\rho^A & \equiv & d\psi^A-{1\over 4} \gamma_{ab} \, \omega^{ab}\wedge\psi^A
-{{\rm i} \over 2} {\hat {\cal Q}}\wedge\psi^A
+{\hat \omega}^{A}_{\phantom{A}B} \wedge \psi^B \equiv \nabla \psi^A
\label{gravdefupap} \\
R^{ab} & \equiv & d\omega^{ab}-\omega^a_{\phantom{a}c}\wedge \omega^{cb}
\label{riecurvap}\\
F^0 & \equiv & dA^0 \, +\,
\bar L^0 \bar\psi_A\wedge\psi_B
\epsilon^{AB}+L^0\bar\psi^A\wedge \psi^B\epsilon_{AB}
\label{Fcurvap}
\end{eqnarray}
where $F^0$ denotes the graviphoton,
has been chosen in such a way that by setting $R^{ab}=T^a=\rho^A=\rho
_A=F^0=0$ , deleting the composite connections $\hat {\cal Q}\,,\,\hat
{\omega}^A_B$ and normalising $L^0 \left(0,0 \right)=1$ we obtain the
Maurer--Cartan equations of the $N=2$ Poincar\'e superalgebra where the one
forms $\omega^{ab}, V^a, \psi ^A,\psi _A, A^0$ are dual to the corresponding
generators of the group.
\par
The next step is to write down the Bianchi identities for all the
curvatures and to solve them in superspace. Applying the $d$ operator to eq.s
 ~\ref{torsdefap}--~\ref{riecurvap} and
~\ref{zcurv}--~\ref{iperincurvb} one finds:
\begin{eqnarray}
&{\cal D}T^a& +\quad R^{ab}\wedge V^b-{\rm i}\bar\psi^A\wedge \gamma^a \rho_A +
{\rm i}\bar\rho^A\wedge \gamma^a \psi_A  =0\label{torsbi}\\
&  \nabla\rho_A &+\quad{1\over 4} \gamma_{ab} \,
R^{ab}\wedge\psi_A- {{\rm i} \over 2} {\hat K}\wedge \psi_A -
{{\rm i} \over 2}{\hat R}_A^{~B}\wedge \psi_B = 0
\label{gravbidown} \\
&  \nabla\rho^A& +\quad{1\over 4} \gamma_{ab} \, R^{ab}\wedge\psi^A +
{{\rm i} \over 2} {\hat K}\wedge\psi^A -
{\hat R}^{A}_{\phantom{A}B} \wedge \psi^B = 0
\label{gravbiup} \\
& {\cal D}R^{ab}& =\quad 0 \label{riebi}\\
& \nabla^2 z^i& -\quad g \left( F^{\Lambda} -\bar L^\Lambda
\,\bar\psi_A\wedge\psi_B
\epsilon^{AB}-\bar L^\Lambda \,\bar\psi^A\wedge\psi^B
\epsilon_{AB}\right)\,k_\Lambda^i(z)= 0 \label{zbi}\\
&  \nabla^2 z^{i^\star}&-\quad g \left( F^{\Lambda} -\bar L^\Lambda
\,\bar\psi_A\wedge\psi_B
\epsilon^{AB}-\bar L^\Lambda \,\bar\psi^A\wedge\psi^B
\epsilon_{AB}\right)\,k_\Lambda^{i^\star}= 0 \label{zstarbi}\\
&  \nabla^2\lambda^{iA}&+\quad {1\over 4} \gamma_{ab} \,
R^{ab} \lambda^{iA} +{{\rm i} \over 2} {\hat K}\lambda^{iA}+
{\hat R}^i_{\phantom{i}j}\lambda^{jA}-{{\rm i}\over 2}{\hat R}^{A}_{~B} \wedge
\lambda^{iB}= 0 \label{lambi}\\
&  \nabla^2\lambda^{i^\star}_A&+\quad{1\over 4} \gamma_{ab} \,
R^{ab} \lambda^{i^\star}_A -{{\rm i} \over 2} {\hat K}\lambda^{i^\star}_A+
{\hat R}^{i^\star}_{\phantom{i^\star}j^\star}\lambda^{j^\star}_A-
{{\rm i}\over 2}{\hat R}^{B}_{~A} \wedge \lambda^{i^\star}_B = 0
\label{lamstarbi}\\
&  \nabla F^\Lambda& -\quad
\nabla\bar L^\Lambda \wedge\bar\psi_A\wedge\psi_B
\epsilon^{AB}-\nabla L^\Lambda \wedge\bar\psi^A \wedge\psi^B\epsilon_{AB}
\nonumber\\
&\null&+ \quad 2 \bar L^\Lambda \bar\psi_A\wedge\rho_B \epsilon^{AB}
+2 L^\Lambda \bar\psi^A \wedge\psi^B\epsilon_{AB} = 0 \label{Fbi}\\
& \nabla{\cal U}^{A \alpha}&-\quad g \left(F^{\Lambda} -\bar L^\Lambda
\,\bar\psi_A
\wedge\psi_B \epsilon^{AB}-\bar L^\Lambda \,\bar\psi^A\wedge\psi^B
\epsilon_{AB}\right)\,k_\Lambda^u(z) {\cal U}^{A \alpha}_u = 0
\label{Ubi}\\
&  \nabla^2 \zeta _{\alpha}&+\quad {1 \over 4} R^{ab}\,\gamma _{ab}\,\zeta
_{\alpha}
+{{\rm i} \over 2} {\hat K}\,\zeta _{\alpha}\,+
{\hat R}_{\alpha}^{\phantom{\alpha}{\beta}}\zeta _{\beta} = 0
\label{iperibi}\\
&  \nabla^2 \zeta^{\alpha}&+\quad {1 \over 4} R^{ab}\,\gamma
_{ab}\,\zeta^{\alpha}
-{{\rm i} \over 2} {\hat K}\,\zeta^{\alpha}\,+
{\hat R}^{\alpha}_{\phantom{\alpha}{\beta}}\zeta^{\beta} = 0
\label{iperibib}\\
\end{eqnarray}
\par
The covariant derivatives $\nabla$ and $\cal D$ have been defined in eq.s
~\ref{torsdefap}--~\ref{Fcurvap} and include the gauged connections defined
in eq.~\ref{compogauging}.
Furthermore the hat on the scalar manifolds curvatures $\hat K,
\hat R^i_{\phantom{i} j}\,,\,\hat R^{\alpha}_{\phantom{\alpha}\beta}\,,\,
\hat R^A_{\phantom{A}B}$ denotes the gauged
curvatures defined in \ref{compogaugcurv}.
\par
The solution can be
obtained as follows: first of all one requires that the expansion of
the curvatures along the intrinsic $p$--forms basis in superspace
namely: $V^a, V^a\wedge V^b, \psi , \psi \wedge V^b,\psi \wedge \psi$,
is given in terms only of the physical fields (rheonomy). This
insures that no new degree of freedom is introduced in the theory.
\\
Secondly one writes down
such expansion in a form which is compatible with all the symmetries
of the theory, that is: covariance under $U(1)$ K\"ahler and
$SU(2) \otimes Sp(2,m)$, Lorentz transformations and reparametrization of
the scalar manifolds. Besides it is very useful to take into account the
invariance under the following rigid rescalings of the fields (and their
corresponding curvatures):
\begin{equation}
 (\omega^{ab}, A^{\Lambda}, q^u, z^i, z^{i^{\star}}) \rightarrow
 (\omega^{ab}, A^{\Lambda}, q^u, z^i, z^{i^{\star}})
 \end{equation}
 \begin{equation}
 V^a \rightarrow \ell V^a
 \end{equation}
 \begin{equation}
 (\psi^A,\psi_A) \rightarrow \ell^{1 \over 2}(\psi^A,\psi_A)
 \end{equation}
 \begin{equation}
 (\lambda^{iA},\lambda^{i^\star}_A, \zeta^\alpha, \zeta_ \alpha)
 \rightarrow
 \ell^{-{1 \over 2}}(\lambda^{iA},\lambda^{i^\star}_A,
 \zeta^\alpha, \zeta _\alpha)
 \label{resc}
 \end{equation}
 Indeed these rescalings and the corresponding ones for the
 curvatures leave invariant the definitions of the curvatures and the
 Bianchi identities.\\
 Finally we note that we are looking for a solution of the coupled
 system of Bianchi identities of the gravitational sector with those
 of the  matter sectors. The coupling is obtained by setting the
 auxiliary fields of the $N=2$ multiplets to definite expressions in
 the physical fields compatible with all the previously mentioned
 requirements. This fixes completely the ansatz for the curvatures at
 least if we exclude higher derivative interactions.
 \par
 Performing all the steps requires a lot of work. For a more detailed 
explanation
 the interested reader is referred to the standard reference of the geometrical
 approach
 \cite{CaDFb}.
The final parametrizations of the superspace curvatures,
are given by:
\begin{eqnarray}
T^a &=& 0 \label{paramtors}\\
\rho_A &=& \tilde {\rho}_{A\vert ab} V^a\wedge V^b  +
\left ( A_{A}^{\phantom{A} B \vert b}
\eta_{ab}+A_{A}^{\prime \phantom{A}  B \vert b}\gamma_{ab} \right ) \psi _B
\, \wedge \, V^a
\nonumber\\
&\null& + \left [ {\rm i} \, g \,S_{AB}\eta _{ab}+ \epsilon_{AB}( T^-_{ab}\,
+ \,U^+_{ab}) \right ]
\gamma^b\psi^B \wedge V^a
 \label{paramgrav}\\
\rho^A &=& \tilde {\rho}^A_{\vert ab} V^a \wedge V^b+
 \left ( \bar A^{A \vert \phantom{B} b}_{\phantom{A \vert} B} \, \eta_{ab}  +
\bar A^{\prime A \vert \phantom{B} b}_{\phantom{A \vert} B} \,
\gamma _{ab} \right ) \psi^B \wedge V^a \nonumber\\
& \null & + \left [ {\rm i} \, g \,\bar S^{AB} \eta_{ab} \, + \, \epsilon^{AB}
\left ( T_{ab}^{+}+U_{ab}^{-} \right ) \right ] \, \gamma^b \psi _B  \wedge V^a
 \label{paramgravb} \\
R^{ab} &=& \tilde{R}^{ab}_{\phantom{ab}cd} V^c\wedge V^d \, - \, {\rm i}  \,
(\bar\psi_A\theta^{A\vert ab}_c \, + \, \bar\psi^A \theta^{ab}_{A\vert c})
\wedge V^c \nonumber\\
& \null & \, + \, \epsilon^{abcf} \, \bar\psi^A \wedge \gamma _f \psi _B
(A^{\prime B}_{\phantom{B}A\vert c}-\bar A^{\prime \phantom{A}B}_{A \vert c})
\nonumber\\
&\null& + \, {\rm i}\,  \epsilon^{AB}\, \bar\psi _A \wedge
\psi_B (T^{+ab}+U^{-ab})\,  - \, {\rm i} \, \epsilon _{AB} \bar\psi^A \wedge
\psi^B
(T^{-ab}+U^{+ab}) \nonumber\\
& \null & -\,g S_{AB} \, \bar\psi^A \wedge \gamma^{ab} \,\psi^B  \,
- \, g \bar S^{AB} \, \bar\psi _A \wedge \gamma^{ab} \,\psi_B \label{paramriem}
\end{eqnarray}
\begin{eqnarray}
F^\Lambda &=& \tilde{F}^\Lambda_{ab} V^a\wedge V^b+\left({\rm i}
\,f^{\Lambda}_i \,\bar {\lambda}^{iA}
\gamma_a \psi^B \,\epsilon _{AB} +{\rm i} \,{\bar f}^{\Lambda}_{{i}^\star}
\,\bar\lambda^{{i}^\star}_A
\gamma_a \psi _B \,\epsilon^{AB}\right)\wedge V^a \label{gaugparam}\\
\nabla\lambda^{iA} &=& \tilde{\nabla_a \lambda} ^{iA} V^a+ {\rm i}\, {\tilde
Z}^i_a \gamma^a \psi^A+ G^{-i}_{ab} \gamma^{ab} \psi _B \epsilon^{AB}+
\left(Y^{iAB}\,+\,g\,W^{iAB}\right)\psi _B
\label{gauginparam}\\
\nabla\lambda^{{i}^\star}_A &=& \tilde{\nabla_a \lambda}^{{i}^\star}_A V^a+
{\rm i} \,
{\tilde {\bar Z}}^{{i}^\star}_a
\gamma^a\psi _A+G^{+{{i}^\star}}_{ab} \gamma^{ab} \psi^B \epsilon _{AB}
+
\left(Y^{{i}^\star}_{\phantom{{i}^\star}AB}\,+\,g\,W^{{i}^\star
}_{\phantom{{i}^\star}AB}\right)\psi^B
\label{gauginparamb}\\
\nabla z^i &=& \tilde{Z}^i_a V^a+\bar\lambda^{iA}\psi _A \label{scalparam} \\
\nabla \bar z^{{i}^\star} &=& \tilde{Z}^{{i}^\star}_a
V^a+\bar\lambda^{{i}^\star}_A \psi^A \label{scalparamb}
\end{eqnarray}
\begin{eqnarray}
{\cal U}^{A \alpha} &=& \tilde{\cal U}^{A \alpha}_a\,V^a\,+\,
\epsilon^{AB}\,\IC^{\alpha  \beta}\,{\bar \Psi}_B\zeta_{\beta}\,+
\,{\bar \Psi}^A\zeta^{\alpha}\label{iperparam}\\
\nabla\zeta_{\alpha} &=& \tilde{\nabla _a\zeta }_{\alpha} V^a\,+\, {\rm i}\,\,
{\tilde{\cal U}}^{B \beta}_a\,\gamma^a \psi^A
\epsilon _{AB}\,\IC_{\alpha  \beta}
\,+\,g\,N_{\alpha}^A\,\Psi_A \label{iperinparam}\\
\nabla\zeta^{\alpha} &=& \tilde{\nabla _a \zeta}^{\alpha} V^a\,+\, {\rm i}\,\,
{\tilde{\cal U}}^{A \alpha}_a\,\gamma^a \psi_A
\,+\,g\,N^{\alpha}_A\,\Psi^A \label{iperinparamb}
\end{eqnarray}
where:
\begin{eqnarray}
A_{A}^{\phantom{A} \vert a B}
&=&-{{\rm i} \over 4}\, g_{{{k}^\star}\ell}\,
\left(\bar {\lambda}^{{k}^\star}_A \gamma^a \lambda^{\ell B}\,-\,\delta^B_A\,
\bar\lambda^{{k}^\star}_C \gamma^a \lambda^{\ell C}\right)\label{Adefrheo}\\
A_{A}^{\prime \phantom{A} \vert a B}
&=&{{\rm i} \over 4}\,
g_{{{k}^\star}\ell}\,\left(\bar{\lambda}^{{k}^\star}_A \gamma^a
\lambda^{\ell B}-{1\over 2}\, \delta^B_A\, \bar\lambda^{{k}^\star}_C
\gamma^a \lambda^{C\ell}\right) \, +\,  {{\rm i} \over 4}\,\lambda\, \delta
_A^B \,\bar \zeta _{\alpha}
\gamma^a \zeta^{\alpha}\label{A'defrheo}
\end{eqnarray}
\begin{equation}
\,\theta^{ab \vert c}_{A}=2 \gamma^{[a}\rho^{b]c}_A+\gamma^c\rho^{ab}_A;
\quad
\theta^{ab\ A}_c=2 \gamma^{[a}\rho^{b]c\vert A}+\gamma^c
\rho^{ab\vert A}\label{thetadef}
\end{equation}
\begin{eqnarray}
&T^-_{ab} =
\left ({\cal N}-{\bar {\cal N}}\right)_{\Lambda\Sigma} L^{\Sigma}
\left (\tilde{F}_{ab}^{\Lambda -} +{1\over 8} \nabla_i \,f^{\Lambda} _j \,
\bar \lambda^{i A} \gamma_{ab} \, \lambda^{jB} \,\epsilon_{AB}
+{1\over 4} \, \lambda\, \IC^{\alpha  \beta}\,{\bar\zeta}_{\alpha}\gamma
_{ab} \,
\zeta _{\beta}\, L^{\Lambda}
\right )&\label{T-defrheo}\\
&T^+_{ab} =
\left ({\cal N}-{\bar {\cal N}}\right)_{\Lambda\Sigma} {\bar L}^{\Sigma}
\left(\tilde{ F}_{ab}^{\Lambda +} +{1\over 8} \nabla_{{i}^\star} \,\bar
f^\Lambda_{{j}^\star}\,
\bar \lambda^{{i}^\star}_A \gamma _{ab} \, \lambda^{{j}^\star}_B \epsilon^{AB}
+{1 \over 4}\, \lambda\, \IC_{\alpha  \beta}\,{\bar\zeta}^{\alpha}\gamma
_{ab} \,
\zeta ^{\beta}\, {\bar L}^{\Lambda}
\right)&
\label{T+defrheo}
\end{eqnarray}
\begin{eqnarray}
U^-_{ab} &=& {{\rm i} \over 4} \,\lambda\, \IC^{\alpha 
\beta}\,{\bar\zeta}_{\alpha}\gamma _{ab} \,
\zeta _{\beta}\label{U-defrheo}\\
U^+_{ab} &=& {{\rm i} \over 4} \,\lambda\, \IC_{\alpha 
\beta}\,{\bar\zeta}^{\alpha}\gamma _{ab} \,
\zeta^{\beta}\label{U+defrheo}
\end{eqnarray}
\begin{eqnarray}
G^{i-}_{ab} &=& {{\rm i} \over 2}\,g^{i{{j}^\star}} \bar f^\Gamma_{{j}^\star}
\left ( {\cal N}-{\bar {\cal N}}\right)_{\Gamma\Lambda}
\Bigl ( \tilde{F}^{\Lambda -}_{ab} + {1\over 8}
\nabla_{k}  f^{\Lambda}_{\ell} \bar \lambda^{kA}
\gamma_{ab} \, \lambda^{\ell B} \epsilon_{AB}
\nonumber\\
&\null& \qquad \qquad  +
{1\over 4} \,\lambda\, \IC^{\alpha  \beta}\,{\bar\zeta}_{\alpha}\gamma _{ab}
\, \zeta _{\beta}\, L^{\Lambda}
\Bigr )\label{G-defrheo}\\
G^{{{i}^\star}+}_{ab} &=& {{\rm i} \over 2}\,g^{{{i}^\star}j} f^{\Gamma}_j
\left ({\cal N}-{\bar {\cal N}}\right)_{\Gamma\Lambda}
\Bigl ( \tilde{F}^{\Lambda +}_{ab} +{1\over 8}
\nabla _{{k}^\star} \bar f^{\Lambda}_{{\ell}^\star} \bar \lambda^{{k}^\star}_A
\gamma _{ab} \, \lambda^{{\ell}^\star}_B \epsilon^{AB}\nonumber\\
&\null& \qquad \qquad + {1 \over 4}\,\lambda\,
\IC_{\alpha  \beta}\,{\bar\zeta}^{\alpha}\gamma _{ab} \,
\zeta ^{\beta}\, {\bar L}^{\Lambda}
\Bigr )\label{G+defrheo}\\
\end{eqnarray}
\begin{eqnarray}
Y^{iAB} &=& {{\rm i} \over 2}g^{i{{j}^\star}}
C_{{{j}^\star}{{k}^\star}{{\ell}^\star}} \bar\lambda^{{k}^\star}_C
\lambda^{{\ell}^\star}_D
\epsilon^{AC} \epsilon^{BD}\label{3.21arheo}\\
Y^{{i}^\star}_{AB}& =&-{{\rm i} \over 2}g^{{{i}^\star}j} C_{jk\ell}
\bar\lambda^{kC}
\lambda^{\ell D} \epsilon_{AC} \epsilon_{BD}\label{3.21b}
\end{eqnarray}
\begin{eqnarray}
S_{AB}&=&{{\rm i} \over 2} (\sigma_x)_A^{\phantom{A}C} \epsilon_{BC}
{\cal P}^x_{\Lambda}L^\Lambda \nonumber\\
\bar S^{AB}&=&{{\rm i}\over 2}  (\sigma_x)_{C}^{\phantom{C}B} \epsilon^{CA}
{\cal P}^x_{\Lambda}\bar L^\Lambda\label{Sdef}
\end{eqnarray}
\begin{eqnarray}
{N}_{\alpha}^A &=& 2 \,{\cal U}_{\alpha \vert u}^A
\,k^u_{\Lambda}\,\bar L^{\Lambda}\nonumber\\
{N}^{\alpha}_A &=& -\,2 \,{\cal U}^{\alpha}_{A \vert u}
\,k^u_{\Lambda}\, L^{\Lambda}\label{Ndef}
\end{eqnarray}
\begin{eqnarray}
W^{iAB}&=&W^{i[AB]}+W^{i(AB)}\nonumber\\
W^{{i}^\star}_{AB}&=& W^{{i}^\star}_{[AB]}+W^{{i}^\star}_{(AB)}\nonumber\\
\rm {where}:\\
W^{i[AB]}&=&\epsilon^{AB}\,k_{\Lambda}^i \bar L^\Lambda \nonumber\\
W^{{i}^\star}_{[AB]}&=&\epsilon_{AB}\,k_{\Lambda}^{{i}^\star} L^\Lambda
\nonumber\\ W^{i(AB)} &=& {\rm i}(\sigma_x)_{C}^{\phantom{C}B} \epsilon^{CA}
{\cal P}^x_{\Lambda}
g^{i{{j}^\star}} {\bar f}_{{j}^\star}^{\Lambda}\nonumber\\
W^{{i}^\star}_{(AB)}&=& {\rm i}(\sigma_x)_{A}^{\phantom{A}C} \epsilon_{BC}
{\cal P}^x_{\Lambda}
g^{{{i}^\star}j} f _j^{\Lambda}
\label{Wdef}
\end{eqnarray}
\par
As promised the solution for the curvatures is given as
an expansion along the 2--form basis $ \left (V \wedge
V\,,\,V \wedge \psi\,,\,\psi\wedge \psi
\right)$ or the 1--form basis $\left(V\,,\,\psi\right)$
with coefficients given in terms of the physical fields.\\
The "on--shell" auxiliary fields are given in our case by the
composite connections $\hat {\cal Q}\,,\,\hat
{\omega^A_B}$ and by $T^{\mp}_{ab}\,,\,W^{iAB}$ and $S_{AB}$.
\par
It is important to stress that the field strengths
$\tilde{R}^{ab}_{\phantom {ab}cd}$, $\tilde{\rho}_{A \vert{ab}}$,
$\tilde{F}^{\Lambda}_{ab}$ , $\tilde{\cal U}^{A \alpha}_a \equiv {\cal U}^{A
\alpha}_u  \tilde{\nabla_a q}^u$, \break $\tilde{\nabla_a\lambda}^{iA}$,
$\tilde{\nabla_a\zeta}_{\alpha}$
and their hermitian conjugates are not space--time field strengths since
they are components along the bosonic vielbeins $V^a=V^a_{\mu}dx^{\mu}\,+\,
V^a_{\alpha}d \theta^{\alpha}$ where $(V^a_{\mu}\,,\,V^a_{\alpha})$ is
a submatrix of the super--vielbein matrix $E^I \equiv\left(
V^a\,,\,\psi\right)$. The physical field strengths are given by the
expansion of the forms along the $ dx^{\mu}$-differentials and by
restricting the superfields to space--time ($\theta = 0$
component). For example, from the parametrization (27), expanding
along the $dx^{\mu}$--basis one finds:
\begin{equation}
F^\Lambda_{\mu\nu} = \tilde{F}^\Lambda_{ab} V^a_{[\mu} V^b_{\nu]}+{\rm i}
\,f^{\Lambda}_i \,\bar {\lambda}^{iA} \gamma_a \psi^B_{[\mu}
V^a_{\nu]}\,\epsilon _{AB} +{\rm i} \,{\bar f}^{\Lambda}_{{i}^\star}
\,\bar\lambda^{{i}^\star}_A \gamma_a \psi _{B[\mu}
V^a_{\nu]}\,\epsilon^{AB}\label{gaugspt}
\end{equation}
where:
\begin{equation}
F^{\Lambda} = {\cal F}^{\Lambda}\,+\,L^{\Lambda}\bar{\psi}^A \wedge
\psi^B\,\epsilon_{AB}\,
+\,\bar L^{\Lambda}\bar{\psi}_A \wedge \psi_B\,\epsilon^{AB}
\end{equation}
according to equations ~\ref{Fcurv}, ~\ref{gaugparam}.
When all the superfields are restricted to space--time we may treat the
$V^a_{\mu}$ vielbein as the usual 4--dimensional invertible matrix
converting intrinsic indices in coordinate indices and we obtain:
\begin{eqnarray}
\tilde{F}^{\Lambda}_{\mu\nu} &=& {\cal
F}^{\Lambda}_{\mu\nu}\,+\,L^{\Lambda}\bar{\psi}^A_{\mu}\psi^B_{\nu}
\,\epsilon_{AB} \,+\bar L^{\Lambda}\bar{\psi}_{A \mu} \psi_{B
\nu}\epsilon^{AB}\, -{\rm i} \,f^{\Lambda}_i \,\bar {\lambda}^{iA}
\gamma_{[\nu} \psi^B_{\mu]}\,\epsilon _{AB}
\nonumber\\
&&-{\rm i} \,{\bar f}^{\Lambda}_{{i}^\star} \,\bar\lambda^{{i}^\star}_A
\gamma_{[\nu} \psi _{B \mu]} \,\epsilon^{AB}\
\end{eqnarray}
By the same token we also get:
\begin{eqnarray}
\tilde{\nabla_{\mu}\lambda}^{iA} &=& \nabla_{\mu} \lambda ^{iA}\,
-{\rm i}\left( \nabla_{\mu}z^i\,-\bar\lambda^{iB}\psi_{B \vert\nu}\right)
\gamma^{\nu} \psi^A_{\mu}\,
- G^{-i}_{\nu\rho} \gamma^{\nu\rho} \psi _{B \vert\mu} \epsilon^{AB}\nonumber\\
&&- \left(Y^{iAB}\,+\,g\,W^{iAB}\right)\psi _{B \vert\mu}\nonumber\\
\tilde{\nabla_{\mu}\zeta}_{\alpha} &=& \nabla _{\mu}\zeta _{\alpha} \,- {\rm i}
\left({\cal U}^{B \beta}_u \nabla_{\nu} q^u\,-
\epsilon^{BC}\,\IC^{\beta \gamma}\,{\bar \Psi}_{C \vert\nu} \zeta_{\gamma}\,
- {\bar \Psi}^B_{\nu}\zeta^{\beta} \right) \gamma^{\nu} \psi^A_{\mu}
\epsilon _{AB}\,\IC_{\alpha  \beta}
\,-g\,N_{\alpha}^A\,\Psi_{A \vert\mu}\nonumber\\
 \tilde{Z}^i_{\mu} &=&
\nabla_{\mu}z^i\,-\bar\lambda^{iA}\psi_{A \vert\mu}\nonumber\\
 \tilde{\cal U}^{A \alpha}_{\mu} &=& {\cal U}^{A \alpha}_u \nabla_{\mu}q^u\,-
\epsilon^{AB}\,\IC^{\alpha\beta}\,{\bar\Psi}_{B \vert \mu}\zeta_{\beta}\,-
{\bar{\Psi}}^A_{\mu}\zeta^{\alpha}
\end{eqnarray}
We note that in the component approach the "tilded" field strengths defined
in the previous equations are usually referred to as the {\it
supercovariant} field strengths.
\par
The physical fields appearing in the parametrizations are actually
further required to satisfy extra--constraints which are
essentially of two types:
\begin{enumerate}
\item{ The supercovariant field strengths satisfy a set of differential
constraints which are to be identified, when the fields are restricted to
space--time only, with the equations of motion of the theory. Indeed the
analysis of
the Bianchi identities for the fermion fields give such
equations (in the sector containing the 2--form basis
$\bar\psi_A\gamma^a \psi^A$). Further the superspace derivative along the
$\psi_A\,\left(\psi^A\right)$ directions, which amounts to a
supersymmetry transformation, yields the equations of
motion of the bosonic fields. This is not a surprise since the
closure of the Bianchi identities is in fact equivalent to the closure of
the $N=2$ supersymmetry algebra on the physical fields and we know that in
general such closure implies the equations of motion for the
fermion fields. Indeed in our case the usual auxiliary fields of
$N=2$ theory have been determined as suitable expressions in the
physical fields.
\par
Finally we also note that since the expressions for the curvatures imply the
equations of motion it follows that in the ungauged case
$\left(g=0 \right)$ the formulae  ~\ref{trasfgrav}--~\ref{iperintrasf}
are symplectic covariant since the
ungauged theory is on-shell symplectic covariant.}
\item
{ The second type of constraints following from the closure of
Bianchi identities is a set of differential constraints on the upper
part $L^\Lambda,\, \bar L^\Lambda,\, f^\Lambda_i,\,
{\bar f}^\Lambda _{{i}^\star}$ of the
symplectic sections $V$ and   $U_i$ and of the ${\cal TM}^3 \otimes
{\cal L}^2$ sections
$C_{ijk}$ ( together with its complex conjugate $C_{{{i}^\star}{{j}^\star}
{{k}^\star}}$).\\
One finds:
\begin{equation}
\nabla_{{i}^\star} L^\Lambda = \nabla _i {\bar L}^\Lambda =0 \label{3.22}
\end{equation}
\begin{equation}
f^\Lambda_i=\nabla_i L^\Lambda;\quad
{\bar f}^\Lambda_{{i}^\star} = \nabla _{{i}^\star} \bar L^\Lambda\label{3.23}
\end{equation}
\begin{eqnarray}
\nabla_{{\ell}^\star} C_{ijk} &=&
\nabla_\ell C_{{{i}^\star}{{j}^\star}{{k}^\star}}=0 \label{3.24}\\
\nabla_{[\ell} C_{i]jk} &=&
\nabla_{[\bar {\ell}} C_{{{i}^\star}]{{j}^\star}{{k}^\star}}=0
\label{3.25}\\
\nabla_{j} f^\Lambda_{k}&=&{\rm i} g^{i \bar {\ell}}
{\bar f}^\Lambda_{{\ell}^\star} C_{ijk}\label{3.26}
\end{eqnarray}
Using the identities of
Special Geometry (\ref{specialone},~\ref{covholsec},~\ref{ctensor},
~\ref{intriscripen}), $C_{ijk}$ can be written as:
\begin{equation}
C_{ijk} = \left(\cal N\,-\,\bar{\cal N} \right)_{\Lambda\Sigma}f^{\Lambda}_i
\nabla_jf^{\Sigma}_{k}\label{3.27}
\end{equation}
In particular equation \ref{3.26} implies the constraint given in
\ref{specialone}
for the Riemann tensor of the K\"ahler-Hodge manifold while equations
~\ref{3.24}--\ref{3.25} are actually equivalent to the other
equations ~\ref{specialone}, using ~\ref{specialissimo}.
Therefore the constraints \ref{3.22}--\ref{3.26} imply that the
K\"ahler-Hodge manifold we started from is actually a special K\"ahler
manifold.\\
We may also verify that the same equations \ref{3.22}--\ref{3.26} hold
provided we replace $L^{\Lambda}\rightarrow M_{\Lambda}$ and $f^{\Lambda}_i
\rightarrow h_{\Lambda i}$ (together with their c.c.).
Hence we have a set of symplectic covariant constraints, namely:
\begin{eqnarray}
\nabla_i V &=& U_i \nonumber\\
\nabla_j U_j &=& {\rm i}C_{ijk}g^{k \ell^{\star}}U_{\ell^{\star}} \nonumber\\
\nabla_i U_{j^{\star}} &=& g_{ij^{\star}}V \nonumber\\
\nabla_{i^{\star}}V &=& 0 \label{3.27ap}
\end{eqnarray}
which give an alternative definition of Special Geometry in terms of
differential constraints on a symplectic bundle of the K\"ahler-Hodge
manifold. This definition of Special Geometry was in fact first
deduced in \cite{CaDF} from $N=2$ Bianchi identities (i.e. for
the closure of $N=2$ susy algebra). Furthermore there is a close
connection, exploited in ref. \cite{pincopallo},
between the differential constraints ~\ref{3.27ap} and the
Picard--Fuchs equations for the periods of a
3--dimensional Calabi--Yau manifold \cite{anche,pincopallo} .
}
\end{enumerate}
\par
The determination of the superspace curvatures enables us to write
down the $N=2$ SUSY transformation laws. Indeed we recall that from
the superspace point of view a supersymmetry transformation is a Lie
derivative along the tangent vector:
\begin{equation}
\epsilon = \bar\epsilon^A\,\vec D_A\,+\,\bar\epsilon_A\,\vec D^A
\end{equation}
where the basis tangent vectors $\vec D_A\,,\,\vec D^A$ are
dual to the gravitino
1--forms:
\begin{equation}
\vec D_A \left(\psi^B \right) = \vec D^A \left(\psi_B \right) =\bf 1
\end{equation}
where $\bf 1$ is the unit in spinor space.
\par
Denoting by $\mu^I$ and $R^I$ the set of one--forms
$\Bigl ( V^a,\,\psi_A,\,\psi^A,\,A^{\Lambda} \Bigr )$ and of two--forms
$\Bigl ( R^a,\,\rho_a,\,\rho^A,\,F^{\lambda} \Bigr )$ respectively,
one has:
\begin{equation}
\ell \mu^I = \left(i_{\epsilon}d\,+\,di_{\epsilon}\right)\mu^I
\equiv \left(D \epsilon\right)^I\,+\,i_{\epsilon}R^I
\end{equation}
where D is the derivative covariant with respect to the $N=2$
Poincar\'e superalgebra and $i_{\epsilon}$ is the contraction
operator along the tangent vector $\epsilon$.
\par
In our case:
\begin{eqnarray}
\left(D \epsilon\right)^a &=& {\rm
i}\left(\bar\psi_A\gamma^a\epsilon^A\,+\,\bar\psi^A\gamma^a\epsilon_A \right)\\
\left(D \epsilon\right)^{\alpha} &=& \nabla\epsilon^{\alpha}\\
\left(D \epsilon\right)^{\Lambda} &=& 0
\end{eqnarray}
(here $\alpha$ is a spinor index)\\
For the 0--forms which we denote shortly as $\nu^I
\equiv\left(q^u,\,z^i,\,z^{i^{\star}},\,\lambda^{iA},
\,\lambda^{i^{\star}}_A,\,\zeta_{\alpha}
,\,\zeta^{\alpha}\right)$ we have the simpler result:
\begin{equation}
\ell_{\epsilon} = i_{\epsilon}d \nu^I =
i_{\epsilon}\left(\nabla\nu^I\,-\,connection\,terms \right)
\end{equation}
Using the parametrizations given for $R^I$ and $\nabla\nu^I$ and
identifying $\delta_{\epsilon}$ with the restriction of $\ell_{\epsilon}$ to
space--time it is immediate to find the $N=2$ susy laws for all
the fields. The explicit formulae are given in section \ref{omniapanta}.
\section*{Appendix B: Derivation of the space time Lagrangian from
the geometric approach}
\label{appendiceB}
\setcounter{equation}{0}
\addtocounter{section}{1}
In Appendix A we have seen how to reconstruct the $N=2$ susy
transformation laws of the physical fields from the solution of the
Bianchi identities in superspace.\\
In principle, since the Bianchi identities imply the equations of
motion, the Lagrangian could also be completely determined.
However this would be a cumbersome procedure.
\par
In this Appendix we give a short account of the construction of the
Lagrangian on space--time from a geometrical Lagrangian in
superspace.\\
In the geometric (rheonomic) approach the superspace action is a
4--form in superspace integrated on a 4--dimensional (bosonic)
hypersurface ${\cal M}^{  4}$ locally embedded in ${\cal M}^{ {4}\vert  {8}}$:
\begin{equation}
{\cal A} = \int_{{\cal M}^{4} \subset {\cal M}^{4  \vert  8}} \, {\cal L}
\label{campodicalcio}
\end{equation}
Provided we do not introduce the Hodge duality operator in the
construction of $\cal L$ the equations of
motions derived from the generalized variational principle
$\delta \cal A = {\rm{0}}$ are 3--form or 4--form equations independent from the
particular hypersurface $\cal M^{\rm 4}$ on which we integrate.\\
These superspace equations of motion can be analyzed along the $p$--form
basis.
The components of the equations obtained along bosonic vielbeins
give the differential equations for the fields which,
identifying ${\cal M}^{  4}$ with space--time, are the ordinary equations of
motion of
the theory. The components of the same equations along $p$--forms
containing at least one gravitino ("outer components") give instead
algebraic relations which identify the components of the various
"supercurvatures" in superspace.
\par
The Lagrangian must be constructed according to the principles of
rheonomy: the "outer components" computed from the variational
equations must be all expressed in terms of the supercovariant
components (components along the vielbeins basis). Actually if we
have already solved the Bianchi identities this requirement is
equivalent to identify the outer components of the curvatures
obtained from the variational principle with those obtained from the
Bianchi identities.
\par
There are simple rules which can be used in order to write down
the most general Lagrangian compatible with this requirement.\\
The implementation of these rules is described in detail in the
literature to which we refer the interested reader. Actually one
writes down the most general 4--form as a sum of terms with
indeterminate coefficients in such a way that $\cal L$
be a scalar with respect to all the symmetry transformations of the
theory (Lorentz invariance, $SU \left(2\right) \otimes Sp
\left(2m\right)$ and $U\left(1 \right)$ K\"ahler invariance, invariance
under the rescaling \ref{resc}). Varying the action and comparing the outer
equations of motion with the actual solution of the
Bianchi identities one then fixes all the undetermined
coefficients.
\par
Let us perform the steps previously indicated. The most general
Lagrangian has the following form:
\begin{equation}
{\cal L} \,  =  \,{\cal L}_{grav} \,+ \, {\cal L}_{kin} \,
+ \, {\cal L}_{Pauli} \, + \, {\cal L}_{torsion} \, + \,
{\cal L}_{4 ferm} \, + \, {\cal L}_{gauging}
\label{scompo}
\end{equation}
\begin{eqnarray}
{\cal L}_{grav} \,& = & \,\epsilon _{abcd} R^{ab}\wedge V^c
\wedge V^d \, - \, 4 \,
\left (\bar {\Psi} ^A \gamma _a \rho _A \, - \, \bar {\Psi}_A
\gamma _a \rho ^A \right )
\, V^a \nonumber \\
{\cal L}_{kin} \,&= &  \, \beta _1 g_{i {{j}^\star}}\, \left [{Z}^i_a \,
\left (\nabla \bar z^{{j}^\star}\, - \,
\bar {\Psi}^A \lambda ^{{j}^\star}_A  \right ) \, + \,
{\bar Z}^{{j}^\star}_a \,\left (\nabla z^i \, - \, \bar {\Psi}_A
\lambda ^{i A} \right )
\right ]\,
\wedge V^b \wedge V^c \wedge V^d \epsilon ^a_{\phantom {a}{bcd}}
\nonumber \\
&& + \, b_1 \epsilon _{AB} \IC_{\alpha  \beta} \, {\cal U}^{A \alpha}_a
\left ({\cal U}^{B \beta} \, - \, \bar {\Psi}^B \zeta ^{\beta} \, - \,
\epsilon ^{BC} \IC^{\beta \gamma} \, \bar {\Psi}_C \zeta _{\gamma}\right )\,
\wedge V^b \wedge V^c \wedge V^d \epsilon ^a_{\phantom {a}{bcd}}
\nonumber \\
&& - \, {1 \over 4}\, \left (\beta _1 g_{i{{j}^\star}} {Z}^i_l
{\bar Z}^{{{j}^\star}}_m
\, + \, {1\over 2} b_1 \epsilon _{AB} \, \IC_{\alpha  \beta}
{\cal U}^{A \alpha}_l {\cal U}^{B \beta}_m \right )
\eta^{lm} \epsilon _{abcd}\,  V^a \wedge V^b \wedge V^c
\wedge V^d \nonumber \\
&& + \, {\rm i} \,
\beta _2 g_{i{{j}^\star}}\, \left (\bar {\lambda}^{iA}\gamma ^a
\nabla \lambda ^{{j}^\star}_A
\, + \,\bar {\lambda}^{{j}^\star}_A \gamma ^a \nabla \lambda ^{iA}\right )
\wedge V^b \wedge V^c \wedge V^d \epsilon _{abcd} \nonumber \\
&&+ {\rm i} \,
b_2 \, \left (\bar {\zeta}^{\alpha}\gamma ^a \nabla \zeta _{\alpha}
\, + \,\bar {\zeta}_{\alpha}\gamma ^a \nabla \zeta ^{\alpha} \right )
\wedge V^b \wedge V^c \wedge V^d \epsilon _{abcd} \nonumber \\
&& + \,{\rm i} \, \beta _3\,
\left ({\cal N}_{\Lambda \Sigma} {F}^{+ \Lambda}_{ab} \, + \,
\bar {\cal N}_{\Lambda \Sigma} {F}^{- \Lambda}_{ab}\right )
\Bigl [ F^{\Sigma} \, - \, {\rm i} \,
\Bigl ( f^{\Sigma}_i \, \bar {\lambda}^{iA}\gamma _c \Psi ^B
\epsilon _{AB} \nonumber\\
&& \qquad \qquad \qquad \qquad \qquad \qquad \qquad + \,
f^{\Sigma}_{{i}^\star} \,\bar {\lambda}^{{i}^\star}_A \gamma _c \Psi _B
\epsilon ^{AB} \Bigr )
\wedge V^c \Bigr ]
\wedge V^a \wedge V^b \nonumber \\
&& - \,{1 \over {24}}\,\beta _3 \left (
\bar {\cal N}_{\Lambda \Sigma} {F}^{- \Lambda}_{lm}
{F}^{- \Sigma \vert {lm}}
\, - \,
{\cal N}_{\Lambda \Sigma} {F}^{+ \Lambda}_{lm} {F}^{+
\Sigma \vert {lm}} \right )
\epsilon _{abcd}  V^a \wedge V^b \wedge V^c \wedge V^d \nonumber \\
{\cal L}_{Pauli} \,&= &  \, \beta _5 F^{\Lambda} \left (
{\cal N}_{\Lambda \Sigma} L^{\Sigma } \bar {\Psi}^A \Psi ^B \epsilon _{AB}
\, + \, \bar {\cal N}_{\Lambda \Sigma} \bar L^{\Sigma } \bar {\Psi}_A \Psi
_B \epsilon^{AB}
\right ) \nonumber \\
& \null & \,+ \,{\rm i} \, \beta _6 F^{\Lambda} \left (
\bar {\cal N}_{\Lambda \Sigma}
f^{\Sigma} _i \, \bar {\lambda}^{iA} \gamma _a \Psi ^B \epsilon _{AB} \, + \,
{\cal N}_{\Lambda \Sigma}
{\bar f}^{\Sigma}_{{i}^\star}\bar {\lambda}^{{i}^\star}_A \gamma _a \Psi _B
\epsilon ^{AB} \right)
\wedge V^a \nonumber \\
& \null & \,+ \, \beta _7
F^{\Lambda} \left ({\cal N} \, - \, \bar {\cal N} \right )_{\Lambda \Sigma}
\Bigl (\nabla _i f^{\Sigma}_j
\bar {\lambda}^{iA} \gamma _{ab} \lambda ^{jB} \epsilon _{AB} \, \nonumber\\
&& \qquad \qquad \qquad - \,
\nabla _{{i}^\star} f^{\Sigma}_{{{j}^\star}}
{\bar {\lambda}}^{{i}^\star}_A \gamma _{ab} \lambda ^{{{j}^\star}}_B
\epsilon ^{AB}\Bigr )
\wedge V^a \wedge V^b \nonumber \\
& \null & \,+ \,b_5
F^{\Lambda} \left ({\cal N} \, - \, \bar {\cal N} \right )_{\Lambda \Sigma}
\left (
L^{\Sigma} \bar {\zeta}_{\alpha} \gamma _{ab} \zeta _{\beta} C ^{\alpha \beta}
\, - \,\bar L^{\Sigma}
\bar {\zeta}^{\alpha} \gamma _{ab} \zeta ^{\beta} C _{\alpha \beta} \right )
\wedge V^a \wedge V^b \nonumber \\
& \null & \,+ \, \beta _8 g_{i{{j}^\star}} \left (
\bar {\lambda}^{iA} \gamma^{ab} \Psi _A \, \nabla \bar z^{{{j}^\star}} \, + \,
\bar {\lambda}^{{j}^\star}_A \gamma _{ab} \Psi ^A \nabla z^i \right )
\wedge V^c \wedge V^d \epsilon _{abcd} \nonumber \\
& \null & \,+ \,b_3 \, \left (
\bar {\zeta}_{\alpha} \gamma^{ab} \Psi _A \,{\cal U}^{\alpha A} \, + \,
\bar {\zeta}^{\alpha} \gamma^{ab} \Psi ^A \,{\cal U}_{\alpha A}
\right ) \wedge V^c \wedge V^d \epsilon _{abcd} \nonumber \\
{\cal L}_{torsion} \, & = & \, \left (
\beta _4  g_{i{{j}^\star}} \, \bar {\lambda}^{iA}\gamma _b  \lambda
^{{{j}^\star}}_A \,
+ \,b_4 \, \bar {\zeta}^{\alpha}\gamma _b \zeta _{\alpha} \right )
T_a \wedge V^a \wedge V^b \nonumber \\
{\cal L}_{4 ferm} \, & = & \,\alpha _1 \,
 \left ( L^{\Lambda} \bar {\Psi}^A \Psi ^B \epsilon _{AB} \, + \,
  \bar L^{\Lambda} \bar {\Psi}_A \Psi _B \epsilon^{AB} \right) \wedge
  \Bigl  (
{\cal N}_{\Lambda \Sigma} L^{\Sigma } \bar {\Psi}^C \Psi ^D \epsilon _{CD}
\, \nonumber\\
&& \qquad \qquad \qquad \qquad + \,
\bar {\cal N}_{\Lambda \Sigma} \bar L^{\Sigma } \bar {\Psi}_C \Psi _D
\epsilon^{CD}
\Bigr ) \nonumber \\
&& + \, \alpha _2 \, \left (
f^{\Lambda} _i \, \bar {\lambda}^{iA} \gamma _a \Psi ^B \epsilon _{AB} \, + \,
f^{\Lambda}_{{i}^\star} \bar {\lambda}^{{i}^\star}_A \gamma _a \Psi _B
\epsilon ^{AB} \right)
\wedge \Bigl ( \bar {\cal N}_{\Lambda \Sigma}
f^{\Sigma} _j \, \bar {\lambda}^{jC} \gamma _b \Psi ^D \epsilon _{CD} \,
\nonumber\\
&& \qquad \qquad \qquad \qquad \qquad \qquad   + \,
{\cal N}_{\Lambda \Sigma}
f^{\Sigma}_{{{j}^\star}}\bar {\lambda}^{{{j}^\star}}_C \gamma _b \Psi _D
\epsilon ^{CD} \Bigr )\,
\wedge V^a \wedge V^b \nonumber \\
&& + \, \alpha _3 \,
\left ({\cal N} \, - \, \bar {\cal N} \right )_{\Lambda \Sigma}\,
\Bigl (f^{\Lambda} _i \nabla _{{{k}^\star}} f^{\Sigma}_{{{j}^\star}}\,
\bar {\lambda}^{iA} \gamma  _c \Psi ^B \,
\bar {\lambda}^{{{k}^\star}}_C \gamma _{ab} \lambda ^{{{j}^\star}}_D \,
\epsilon _{AB} \epsilon ^{CD} \, \nonumber\\
&& \qquad \qquad \qquad     - \,
\bar f^{\Lambda}_{{i}^\star}\nabla _k f^{\Sigma}_j  \,
\bar {\lambda}^{{i}^\star}_A \gamma _c \Psi _B \,
\bar {\lambda}^{kC} \gamma _{ab} \lambda ^{jD}\,
\epsilon ^{AB} \epsilon _{CD}\Bigr )
\, \wedge V^a \wedge V^b \wedge V^c \nonumber \\
&& + \, a_1 \,
\left ({\cal N} \, - \, \bar {\cal N} \right )_{\Lambda \Sigma} \,
\Bigl  ( \bar f^{\Lambda}_{{i}^\star} L^{\Sigma}
\bar {\lambda}^{{i}^\star}_A \gamma _c \Psi _B \,
\bar {\zeta}_{\alpha} \gamma _{ab} \zeta _{\beta} \,
\epsilon ^{AB} \IC^{\alpha  \beta} \,\nonumber\\
&& \qquad \qquad \qquad \qquad \qquad  - \,
f^{\Lambda} _i \bar L^{\Sigma}
\bar {\lambda}^{iA} \gamma _c \Psi ^B \,
\bar {\zeta}^{\alpha} \gamma _{ab} \zeta ^{\beta} \,
\epsilon _{AB} \IC_{\alpha  \beta} \Bigr )
\, \wedge V^a \wedge V^b \wedge V^c \nonumber \\
&& + \, a_2 \, \left (
\bar {\Psi}^A \Psi ^B \,
\bar {\zeta}^{\alpha} \gamma _{ab} \zeta ^{\beta}
\epsilon _{AB} \IC_{\alpha  \beta} \, + \,
\bar {\Psi}_A \Psi _B \,
\bar {\zeta}_{\alpha} \gamma _{ab} \zeta _{\beta}\,
\epsilon ^{AB} \IC^{\alpha  \beta} \right )
\, \wedge V^a \wedge V^b \nonumber \\
&& + \,\left(
\alpha _4  g_{i{{j}^\star}} \, \bar {\lambda}^{iA}\gamma _b  \lambda
^{{{j}^\star}}_B \,
+ \,a_3 \,\delta ^A_B \bar {\zeta}^{\alpha}\gamma _b \zeta _{\alpha} \right )
\, \bar {\Psi}_A \gamma _a \Psi ^B \wedge V^a \wedge V^b \nonumber \\
&& + \,\alpha _5 \,
\Bigl ( C_{ijk} \bar {\lambda}^{iA} \gamma^a \Psi ^B \,
\bar {\lambda}^{jC} \lambda ^{kD}\,
\epsilon _{AC} \epsilon _{BD} \,\nonumber\\
&& \qquad \qquad - \,
C_{{{i}^\star}{{j}^\star}{{k}^\star}} \bar {\lambda}^{{i}^\star}_A \gamma ^a
\Psi _B \,
\bar {\lambda}^{{i}^\star}_C \lambda ^{{{k}^\star}}_D
\epsilon ^{AC} \epsilon ^{BD}\Bigr )
\, \wedge V^b \wedge V^c \wedge V^ d \,\epsilon _{abcd} \nonumber \\
&& + {1\over 72}\Biggl [\, \gamma _1 \,
\left (R_{i{{j}^\star}l{{k}^\star}}\, + p\,
g_{i{{k}^\star}} \, g_{l{{j}^\star}}\,+ q\,g_{i{{j}^\star}}
\, g_{l{{k}^\star}}\right )
\bar {\lambda}^{iA}\lambda^{lB}
\bar {\lambda}^{{j}^\star}_A\lambda^{{k}^\star}_B  \nonumber \\
&& + \, \gamma _2 \left( \nabla _m \, C_{jkl}
\bar {\lambda}^{jA}\lambda^{mB} \bar {\lambda}^{kC}\lambda^{lD}
\epsilon _{AC} \epsilon _{BD}-h.c.\right) \nonumber \\
&& + \, \gamma _3 \,
\left ({\cal N} \, - \, \bar {\cal N} \right )_{\Lambda \Sigma}
\Bigl (
C_{ijk} \, C_{lmn} g^{k{\bar r}} \, g^{n{\bar s}} f^{\Lambda}_{\bar r}
\, f^{ \Sigma}_{\bar s} \,
\bar {\lambda}^{iA} \gamma _{lm}\lambda^{jB} \bar {\lambda}^{kC}
\gamma^{lm}\lambda^{lD}
\, \epsilon _{AB} \epsilon _{CD}\nonumber\\
&& \qquad\qquad\qquad\qquad\qquad\qquad\qquad\qquad
+h.c.\Bigr ) \nonumber \\
&& + \gamma _4 \, \,g_{i{{j}^\star}} \,
\bar {\zeta}^{\alpha} \gamma _a \zeta _{\alpha}\,
\bar {\lambda}^{iA} \gamma ^a \lambda^{{j}^\star}_A\, \nonumber \\
&& + \, \gamma _5 \,
{\cal R}^{\alpha}_{\phantom{\alpha}\beta ts}
\, {\cal U}^t_{A \gamma}\,{\cal U}^s_{B \delta} \epsilon ^{AB} \, C ^{\delta
\eta}
\bar {\zeta}_{\alpha}\,\zeta _{\eta}\,\bar {\zeta}^{\beta}\,\zeta ^{\gamma}
\nonumber \\
&& + \gamma _6
\left ({\cal N} \, - \, \bar {\cal N} \right )_{\Lambda \Sigma}
\left( L^{\Lambda} \nabla _i f^{\Sigma}_j
\bar {\zeta}_{\alpha} \gamma _{ab} \zeta _{\beta}\,
\bar {\lambda}^{iA} \gamma ^{ab} \lambda^{jB}\,
\,\epsilon _{AB} \,\IC^{\alpha  \beta}+h.c.\right)
\nonumber \\
&& +  \gamma _7
\left ({\cal N} \, - \, \bar {\cal N} \right )_{\Lambda \Sigma}
\Bigl (
L^{\Lambda} \,L^{\Sigma}
\bar {\zeta}_{\alpha} \gamma _{ab} \zeta _{\beta}\,
\bar {\zeta}_{\gamma} \gamma^{ab} \zeta _{\delta}\,
\IC^{\alpha  \beta}\,\IC^{\gamma \delta} \nonumber\\
&& \qquad \qquad\qquad\qquad\qquad\qquad + h.c.\Bigr ) \, \Biggr ]
V^a \wedge V^b \wedge V^c \wedge V^d \, \epsilon _{abcd} \nonumber \\
{\cal L}_{gauging}\,&=&\,-\, {\rm i} \,g\,\delta _1 \left(S_{AB}
\bar {\Psi}^A \gamma
_{ab} \Psi ^B +h.c.\right) V^a \wedge V^b \nonumber\\
&& + {\rm i} \,g\,\delta_2\,g_{i{{j}^\star}}\left(W^{iAB} \bar
{\lambda}^{{j}^\star}_A
\gamma^a \Psi _B +h.c.\right)\wedge V^b \wedge V^c \wedge V^d \,
\epsilon _{abcd}
\nonumber\\
&& + {\rm i} \,g\,\delta_3 \left(N^A_{\alpha} \, \bar {\zeta}^{\alpha}
\gamma^a \Psi _A +h.c.\right)\wedge V^b \wedge V^c \wedge V^d \, \epsilon
_{abcd}
\nonumber\\
&&+ g \Bigl[\delta _4 \nabla _u N^{\alpha}_A \,{\cal U}^u_{B \beta}
\epsilon ^{AB} \IC^{\beta \gamma} \bar {\zeta}_{\alpha} \zeta
_{\gamma} + \delta _5 \nabla _i N^{\alpha}_A \bar {\zeta}_{\alpha}
\lambda^{iA}\nonumber\\
&& + \delta _6\, g_{i \bar{\jmath}} \nabla_{k} W^{\bar{\jmath}}_{AB} \bar
{\lambda}^{iA}\lambda^{kB} +\, h.c.
\Bigr]V^a \wedge V^b \wedge V^c \wedge V^d \epsilon_{abcd}\nonumber\\
&& + \delta _7\, g^2 \, {\rm V}_{potential}
\, V^a\wedge V^b \wedge V^c \wedge V^d \, \epsilon _{abcd}
\end{eqnarray}
where:
\begin{eqnarray}
{\rm V}_{potential}\, &=& \left(g_{i{{j}^\star}} \,
k^i_{\Lambda}\,k^{{j}^\star}_{\Sigma}\,+\,4\,h_{uv}
k^u_{\Lambda}\,k^v_{\Sigma} \right) \bar L^{\Lambda}\,L^{\Sigma}\nonumber\\
&& +\,g^{i{{j}^\star}}\,f^{\Lambda}_i\,f^{\Sigma}_{{j}^\star}\,
{\cal P}^x_{\Lambda}\,{\cal P}^x_{\Sigma}\,
-\,3\,\bar L^{\Lambda}\,L^{\Sigma}\,{\cal P}^x_{\Lambda}\,{\cal P}^x_{\Sigma}
\end{eqnarray}
We note that the kinetic terms of the Lagrangian have been written
in first--order form to avoid the Hodge--operator which would
destroy the independence of the variational equations from the
particular hypersurface of integration. Specifically one introduces
auxiliary 0--forms namely $F^{\pm{\Lambda}}_{ab}$, $Z^i_a\,,\,\bar
Z^{i^{\star}}_a$,
${\cal U}^{A \alpha}_a$ whose
variational equations identify them with
$\tilde{F}^{\pm{\Lambda}}_{ab}$, $\tilde{Z}^i_a\,,\,\tilde{\bar
Z}^{i^{\star}}_a$,
$\tilde{\cal U}^{A \alpha}_{a}$ defined in Appendix A. Of course also the
spin connection $\omega^{ab}$ has to be treated as an independent
field: indeed the term $\cal L_{\mbox {torsion}}$ appearing in the
Lagrangian has been chosen in such a way that the equation of motion
of $\omega^{ab}$ gives $T^a = 0$.
\par
The analysis of the variational equations for the other $p$--forms
containing at least a fermionic vielbein
$\psi^A \left(\psi_A \right)$ then fixes completely all the
coefficients, except the coefficients of terms that are proportional
to $V^aV^bV^cV^d \epsilon_{abcd}$, which, after variation, do not
contain any $\psi^A \left(\psi_A \right)$ and therefore appear in the
space--time equations of motion.\\
These undetermined coefficients, however, can be retrieved by
comparing the space--time equations of motion for the 0--form fermion
fields $\lambda^{iA}\,,\,\lambda^{i^{\star}A}\,,\,\zeta^\alpha\,,\,\zeta_\alpha$
as obtained from the Bianchi identities with those obtained from the
Lagrangian. In this way all the coefficients have been fixed. The
result is:
\begin{eqnarray}
&& \beta _1\,=\,{2 \over 3}\, ;\,\beta _2\,=\,-\,{1 \over 3}\,;
\, \beta _3\,=\,4\,{\rm i}\,;\,\beta _4\,=\,-\,1 \,;\,
\beta _5\,=\,4\,;\, \beta _6\,=\,-\,4\,;\,\beta _7\,=\,{1 \over 2}
\,;\,\beta _8\,=\,-\,1 \,;\nonumber\\
&& b_1\,=\,-\,{4 \over 3}\lambda\, ;\,b_2\,=\,{2 \over 3}\lambda\,;
\,b_3\,=\,2 \lambda\,; \,
b_4\,=\,-\,2 \lambda\,;\,b_5\,=\,\lambda \,;
\nonumber\\
&& \alpha _1\,=\,-\,2\,;\, \alpha _2\,=\,2\,;\,\alpha _3\,=\,{{\rm i} \over
2}\,;\,
\alpha _4\,=\,-\,2\,{\rm i}\,;\,\alpha _5\,=\,-\,{1 \over 9}
\,;\nonumber\\
&& a_1\, =\,-\,{\rm i} \lambda\,;\,a_2 \,=\,-\,{\rm i}
\lambda\,;a_3\,=\,-\,4\,{\rm i} \lambda\,;\nonumber\\ && \gamma _1\,=\,3\,
;\,\gamma _2\,=\,-\,{\rm i}\, ;\,\gamma _3\,=\,{3{\rm i} \over 16}\, ;\,
\gamma _4\,=\,-\,3\,\lambda\, ;\, \gamma _5\,=\,-\,6 \lambda\,
p\,=\,1\,;\,q\,=\,-\,{2 \over 3}\,;\nonumber\\ && \gamma _6\,=\,-\,{3 \over
4}{\rm i}\,\lambda\, ;\,\gamma _7\,=\,-\,{3 \over 4}{\rm i}\,\lambda^2\,;
\nonumber\\
&& \delta _1 \,= \,4\,;\,\delta _2 \,=\, {2 \over 3}\,;
\,\delta _3\, =\, -\,{4 \over 3}\,\lambda \,;\, \delta _4\,=\,-\,{1 \over
12}\,\lambda \,;\,
\delta _5\,=\,-\,{1 \over 3}\,\lambda \,;\,\delta _6\,=\,{1 \over
18}\,;\,\delta _7\,=\,-\,{1 \over 6}
\end{eqnarray}\label{coeff}
In order to obtain the space--time Lagrangian the last step to
perform is the restriction of the 4--form Lagrangian from superspace
to space--time. Namely we restrict all the terms to the
$\theta = 0\,,\,d \theta = 0$ hypersurface ${\cal M}^4$. 
In practice one first goes
to the second order formalism by identifying the auxiliary 0--form
fields as explained before. Then one expands all the forms along the
$dx^{\mu}$ differentials and restricts the superfields to their
lowest ($\theta = 0$) component. Finally the coefficients
of:
\begin{equation}
dx^{\mu}\wedge dx^{\nu}\wedge dx^{\rho}\wedge
dx^{\sigma}\,=\,{\epsilon^{\mu\nu\rho\sigma}\over \sqrt g}\left(
\sqrt g d^4x \right)
\end{equation}
give the Lagrangian density written in chapter $8$.
The overall normalisation of the space--time action has been chosen
such as to be the standard one for the Einstein term.
\section*{Appendix C: Supergravity theory on $ST[2,n]\otimes
HQ[m] $ }
\label{appendiceC}
\setcounter{equation}{0}
\addtocounter{section}{1}
In this appendix, as an illustration of the general method and
also for its interest in applications to tree level effective
lagrangians of heterotic string theory, we consider the
specialization of our formulae to the case where the scalar manifold
of N=2 supergravity is chosen as in eq.~\ref{grancaso}. This choice
is by no means new in the literature, but the interesting point is to
utilize the symplectic gauge where the holomorphic prepotential
$F(X)$ does not exist. This is the gauge chosen by string theory
and also that where partial supersymmetry breaking can be obtained.
\subsection*{The ${\cal ST}[2,n]$ special manifolds and the Calabi
Visentini coordinates}
When we studied the symplectic embeddings of the ${\cal ST}[m,n]$
manifolds, defined by eq.~\ref{stmanif}, a study that lead us to
the general formula in eq.~\ref{maestrino}, we remarked that the
subclass ${\cal ST}[2,n]$ constitutes a family of special K\"ahler
manifolds: actually a quite relevant one. Here we survey the
special geometry of this class.
\par
Besides their applications in the large radius limit of superstring
compactifications, the ${\cal ST}[2,n]$ manifolds are
interesting under another respect. They provide an example
where the holomorphic prepotential  can be non--existing. Furthermore
it is precisely in the symplectic gauge where $F(x)$ does not
exist that the model $n=1$, $m=1$ of eq.~\ref{grancaso} exhibits
partial supersymmetry breaking $N=2 \, \longrightarrow \, N=1$
\cite{FGP1}.
\par
Consider a standard parametrization of the $SO(2,n)/SO(2)\times
SO(n)$ manifold, like for instance that in eq.~\ref{somncoset}.
In the $m=2$ case we can introduce a canonical complex structure
on the manifold by setting:
\begin{equation}
\Phi^\Lambda (X)\,  \equiv \,{\o{1}{\sqrt{2}}} \, \left (
L^\Lambda_{\phantom{\Lambda}0 } + {\rm i} \,
L^\Lambda_{\phantom{\Lambda}1 } \right )\qquad ;\qquad
 \left ( \Lambda=0,1,a \quad a=2,\dots,n+1  \right )
\label{cicciophi}
\end{equation}
The relations satisfied by the upper two rows of the coset
representative (consequence of $L(X)$ being pseudo--orthogonal with
respect to metric $\eta_{\Lambda\Sigma}={\rm diag}(+,+,-,\dots,-)$):
\begin{equation}
L^\Lambda_{\phantom{\Lambda}0 } \, L^\Sigma_{\phantom{\Lambda}0 } \,
\eta_{\Lambda\Sigma} \, =\,  1  \quad ; \quad
L^\Lambda_{\phantom{\Lambda}0 } \, L^\Sigma_{\phantom{\Lambda}1 } \,
\eta_{\Lambda\Sigma} \, =\,  0  \quad ; \quad
L^\Lambda_{\phantom{\Lambda}1 } \, L^\Sigma_{\phantom{\Lambda}1 } \,
\eta_{\Lambda\Sigma} \,= \, 1
\label{pseudodionigi}
\end{equation}
can be summarized into the complex equations:
\begin{equation}
{\bar \Phi}^\Lambda  \,
\Phi^\Sigma
\eta_{\Lambda\Sigma} \, =\,  1  \quad ; \quad
\Phi^\Sigma
\eta_{\Lambda\Sigma} \, =\,  0
 \label{pseudomichele}
\end{equation}
Eq.s ~\ref{pseudomichele} are solved by posing:
\begin{equation}
\Phi^\Lambda \, = \,{\o{X^\Lambda}{\sqrt{{\bar X}^\Lambda \,
X^\Sigma \, \eta_{\Lambda\Sigma}}}}
\label{fungomarcio}
\end{equation}
where $X^\Lambda$ denotes any set of complex parameters, determined
up to an overall multiplicative constant and satisfying the
constraint:
\begin{equation}
 X^\Lambda  \,
X^\Sigma
\eta_{\Lambda\Sigma} \, = \, 0
\label{ciliegia}
\end{equation}
In this way we have proved the identification, as differentiable
manifolds, of the coset space $SO(2,n)/SO(2)\times
SO(n)$ with the vanishing locus of the quadric in
eq.~\ref{ciliegia}. Taking
any holomorphic solution of eq.~\ref{ciliegia}, for instance:
\begin{equation}
 X^{\Lambda}(y)  \, \equiv \,
\left(\begin{array}{c} 1/2\hskip 2pt (1 + y^2) \\
{\rm i}/2\hskip 2pt (1 -
y^2)\\ y^{a}\end{array}\right)
\label{calabivise}
\end{equation}
where $y^a$ is a set of $n$ independent complex coordinates,
inserting it into eq.~\ref{fungomarcio} and comparing with
eq.~\ref{cicciophi} we obtain the relation between whatever
coordinates we had previously used to write the coset representative
$L(X)$ and the complex coordinates $y^a$. In other words
we can regard the matrix $L$ as a function of the $y^a$ that
are named the Calabi Visentini coordinates \cite{calvis}.
\par
Consider in addition the {\it axion--dilaton} field  $S$
that parametrizes the $SU(1,1)/U(1)$ coset according with
eq.~\ref{su11coset}. The special geometry of the manifold
${\cal ST}[2,n]$ is completely specified by writing the
holomorphic symplectic section $\Omega$ as follows
(\cite{CDFVP}):
\par
\begin{equation}
\label{so2nssec}
\Omega (y,S) \, = \, \left (\matrix{
X^{\Lambda}  \cr F_{\Lambda}\cr }\right)
\, = \, \left ( \matrix { X^{\Lambda}(y) \cr
{\cal  S }\, \eta_{\Lambda\Sigma}
X^{\Sigma}(y) \cr } \right  )
\end{equation}
Notice that  with the above choice, it is
not possible to describe $F_\Lambda$ as derivatives of any
prepotential. Yet everything else can be calculated utilizing
the formulae we presented in the text.
The K\"ahler potential is:
\begin{equation} \label{so2nkpotskn}
\cK \, =\, \cK_1(S)+\cK_2(y)\,
 =\, -\mbox{log} {\rm i} (\bar S - S) -\log X^T\eta X
\end{equation}
The K\"ahler metric  is  block diagonal:
\begin{equation}
\label{so2n5}
g_{ij^\star} \, = \,
\left(\begin{array}{cc}g_{S\bar S} & {\bf
0}\\ {\bf 0} & g_{a \bar b}\end{array}\right)
\qquad \quad
\left\{\begin{array}{l} g_{S\bar S} = \partial_S \partial_{\bar S} \cK_1 =
{-1\over (\bar S - S)^2}\\ g_{a\bar b}(y)=
\partial_{a}\partial_{\bar b} \cK_2 \end{array}\right.
\end{equation}
as expected.
The anomalous magnetic moments-Yukawa
couplings $C_{ijk}$ ($i=S, a$) have a very simple expression
in the chosen coordinates:
\begin{equation}
\label{so2n6bis} C_{Sab} = -{\rm exp}[{\cal K}] \,
\delta_{ab},
\end{equation}
all the other components being zero.
\par
Using the definition of the {\it period matrix} given
in eq.~\ref{intriscripen} we obtain
\begin{equation} \label{so2n10} \cN_{\Lambda\Sigma} = (S -
\bar S) {{X_{\Lambda} {\bar X}_{\Sigma} + {\bar X}_{\Lambda} X_{\Sigma}
}\over {{\bar X}^T\eta X}} + \bar S\eta_{\Lambda\Sigma}.
\label{discepolo}
\end{equation}
In order to see that eq.~\ref{discepolo} just coincides
with eq.~\ref{maestrino} it suffices to note that as a
consequence of its definition~\ref{cicciophi} and of the
pseudo--orthogonality of the coset representative $L(X)$,
the vector $\Phi^\Lambda$ satisfies the following
identity:
\begin{equation}
\Phi^\Lambda \, {\bar \Phi}^\Sigma + \Phi^\Sigma \, {\bar
\Phi}^\Lambda  \, = \,
  {\o{1}{2}} \,  L^{\Lambda}_{\phantom{\Lambda}\Gamma} \,
L^{\Sigma}_{\phantom{\Sigma}\Delta} \left ( \delta^{\Gamma\Delta} +
\eta^{\Gamma\Delta} \, \right )
\label{meraviglia}
\end{equation}
Inserting eq.~\ref{meraviglia} into eq.~\ref{discepolo},
formula ~\ref{maestrino} is retrieved.
\par
This completes the proof that the choice~\ref{so2nssec} of
the special geometry holomorphic section corresponds to the
symplectic embedding~\ref{ortoletto} and ~\ref{ortolettodue}
of the coset manifold ${\cal ST}[2,n]$. In this symplectic
gauge the symplectic transformations of the isometry group
are the simplest possible ones and  the entire group $SO(2,n)$
is represented by means of {\it classical} transformations
that do not mix electric fields with magnetic fields.
The disadvantage of this basis, if any,
is that there is no holomorphic prepotential. To find
an $F(X)$ it suffices to make a symplectic rotation to
a different basis.
\par
 If we set:
\begin{eqnarray}
X^1 = {{1}\over{2}} (1 + y^2) &=& -{{1}\over{2}} (1 - \eta_{ij} t^{i}
 t^{j})\nonumber\\
X^2 = i {{1}\over{2}} (1 - y^2) &=& t^2\nonumber\\
X^a = y^a &=& t^{2+a} \quad
{a=1,\dots,n-1}\nonumber\\
X^{a=n} = y^{n} &=& {{1}\over{2}} (1+\eta_{ij} t^{i} t^{j})
\end{eqnarray}
where
\begin{equation}
\eta_{ij} = {\rm diag} \left ( + , -, \dots , -\right ) ~
i,j=2,\dots , n +1
\end{equation}
Then we can show that $\exists \, {\cal C}\, \in Sp(2 n+2,\IR)$ such that:
\begin{equation}
 {\cal C} \,
\left (
       \matrix {
                 X^\Lambda \cr
                 S\eta_{\Lambda\Sigma} \, X^\Lambda \cr
                }
\right ) \, =\,  \exp[\varphi(t)]\,
\left (
\matrix{
         1 \cr
         S \cr
           t^{i} \cr
          2 \, {\cal F} - t^{i}
         {{\partial}\over{\partial t^{i}}}{\cal F}
          - S{{\partial}\over S}{\cal F} \cr
           S{{\partial}\over S}{\cal F}\cr
       {{\partial}\over{\partial t^{i}}}{\cal F}\cr
      }
\right )
\end{equation}
with
\begin{eqnarray}
{\cal F}(S,t) &=&{{1}\over{2}} \,
S\,  \eta_{ij} t^{i} t^{j}
= {{1}\over{2}} \, d_{IJK} t^{I} t^{J} t^{K}
\nonumber\\
t^{1} &=&S\nonumber\\
d_{IJK}&=&\cases{
d_{1jk}=\eta_{ij}\cr
0 ~~\mbox{otherwise}\cr}
\label{vecchiume}
\end{eqnarray}
and
\begin{equation}
W_{IJK} = d_{IJK} = {{\partial^3{\cal F}(S, t^{i})}
\over{\partial t^{I}\partial
t^{J} \partial t^{K} } }
\end{equation}
This means that in the new basis
the symplectic holomorphic section ${\cal C}\Omega$ can be derived
from the following cubic prepotential:
\begin{equation}
F(X) \, =\,{\o{1}{3!}}\, {\o{ d_{IJK} \, X^I \, X^J \, X^K}{X_0}}
\label{duplocubetto}
\end{equation}
For instance in the case $n=1$  the matrix which does such a job
is:
\begin{equation}
{\cal C}=\left (\matrix{ 1 & 0 & -1 & 0 & 0 & 0 \cr 0 & 0 & 0 & 1
& 0 & 1 \cr 0 & -1
   & 0 & 0 & 0 & 0 \cr 0 & 0 & 0 & {1\over 2} & 0 & -{1\over 2} \cr
  -{1\over 2} & 0 & -{1\over 2} & 0 & 0 & 0 \cr 0 & 0 & 0 & 0 & -1
   & 0 \cr  }\right )
\end{equation}
\subsection*{Comments on the ${\cal ST}[2,2]$ case: S duality and
R symmetry}
To conclude let us focus  on the case ${\cal ST}[2,2]$.
This manifold has two coordinates that we can either call $S$ and
$t$, in the parametrization of eq.~\ref{vecchiume} or  $S$
and $y$ in the Calabi Visentini basis. The relation between $t$ and
$y$ simplifies enormously in this case:
 \begin{equation}
t ~=~i {{y+1}\over{y-1}}
\end{equation}
It is then a matter of choice to regard the holomorphic section
in whatever basis as a function of $y$ or of $t$, in addition to
$S$. Independently from this choice the manifold
${\cal ST}[2,2]$ emerges as {\it moduli space (at tree--level)}
in a locally N=2
supersymmetric gauge theory of a rank one gauge group, namely
$SU(2)$. The two fields spanning the manifold have very different
interpretations. The field $y$ is the scalar partner of the
gauge field that remains massless after Higgs mechanism. Its vacuum
expectation value is the modulus of the gauge theory. It is the
same field that occurs also in a globally supersymmetric theory.
On the other hand the field $S$ is the dilaton--axion.
It plays the role of generalized coupling constant and generalized
$theta$--angle.
There are two $SL(2,\IR)$ groups embedded in $SP(6,\IR)$, they
act as standard fractional linear transformations on the
{\it dilaton--axion} $S$ and on the special coordinate $t$
for the gauge modulus.
Using the Calabi--Visentini section of eq.~\ref{so2nssec} and the
embedding eq.s~\ref{ortoletto} and ~\ref{ortolettodue},
we have that
\par
{\bf S--duality} $ S \longrightarrow \, - 1/S$ is generated
by the symplectic matrix:
\begin{equation}
S_{duality}~=~\left ( \matrix{ 0 & 0 & 0 & 1 & 0 & 0 \cr 0 & 0 & 0 & 0 & 1 &
0 \cr 0 & 0 &
  0 & 0 & 0 & -1 \cr -1 & 0 & 0 & 0 & 0 & 0 \cr 0 & -1 & 0 & 0 & 0 &
  0 \cr 0 & 0 & 1 & 0 & 0 & 0 \cr  }\right )
\label{sduality}
\end{equation}
while {\bf T--duality} $ t \longrightarrow \, - 1/t$ is generated
by the symplectic matrix:
\begin{equation}
R_{symmetry}~=~\left (\matrix{ -1 & 0 & 0 & 0 & 0 & 0 \cr 0 & -1 & 0 & 0 & 0
& 0 \cr 0 & 0
   & 1 & 0 & 0 & 0 \cr 0 & 0 & 0 & -1 & 0 & 0 \cr 0 & 0 & 0 & 0 & -1
   & 0 \cr 0 & 0 & 0 & 0 & 0 & 1 \cr  } \right )
   \label{tduality}
\end{equation}
If we think of the $t$--field as the {\it modulus} of some compact
internal manifold then T--duality is just the transformation from
small to large compactification radius. Looking at the same
transformation in terms of the $y$ variable its meaning
becomes more clear. It is
$R$--symmetry $ y \longrightarrow - y$, an exact global
symmetry of the {\it microscopic} lagrangian. The fact that
the matrix generating T--duality or R--symmetry is block--diagonal
agrees with the fact that this is a perturbative symmetry, holding
at each order in perturbation theory and never exchanging electric
with magnetic states. Very different is the nature of
S--duality. Since it inverts the coupling constant it is by definition
non--perturbative. It exchanges strong and weak coupling regimes and
because of that it is supposed to exchange elementary states
with soliton states. For this reason it must mix electric with magnetic
field strengths and it is off--diagonal. These symmetries exist in the
microscopic theory which is derived by {\it gauging}
the abelian theories possessing continuous duality symmetries (in
this case the two $SL(2,\IR)$ groups). After gauging the continuous
duality symmetries will be broken. The question is will the integer
valued symplectic generators of S--duality and R--symmetry survive
given that they respect the Dirac quantization condition? The answer
is yes, but in the effective quantum theory they will be
represented by new integer valued elements of $Sp(6,\ZZ)$ not
derivable from the classical embedding.  Since the special
geometry in the effective theory is corrected by the instanton
contributions and has a new complicated transcendental structure,
the duality generators must change basis to adapt themselves to the new
situation and be integer valued in the new non--perturbative geometry.
Alternatively one can turn matters around. If we know the new quantum
symplectic embedding of the discrete duality group we have
essentially determined the non perturbative geometry.
It is this point of view that has proven very
fruitful in the very recent literature.
\subsection{Momentum maps of $HQ[m]$ and mass matrices}
\vskip 0.2cm
As we are just going to see the quaternionic manifold $HQ[m]$ is the
closest quaternionic analogue of a flat HyperK\"ahler manifold and
the relevant formulae for the metric and the momentum maps are almost
identical, {\it mutatis mutandis}, with the equations surveyed in
subsection 9.3, when we discussed the renormalizable microscopic N=2
super Yang--Mills lagrangian.
\par
To describe the coset manifold $SO(4,m)/SO(4)\times SO(m)$ we use a
family of
coset representatives $L(q) \in SO(4,m)$. A typical choice is the
$(4+m)\times(4+m)$ matrix:
\begin{equation}
 L(q) \, = \, \left ( \matrix{\sqrt{\bfone + q \, q^T} & q \cr
 q^T & \sqrt{\bfone + q^T \, q }\cr }\right )
\end{equation}
function of an independent $4\times m$ matrix $q$. By definition of
the group $SO(4,m)$ we have:
\begin{equation}
L^T \, \eta \, L \, = \, \eta \qquad ; \qquad \eta \, = \,
\mbox{diag} \, \left ( +,+,+,+,-,\dots,- \right )
\end{equation}
We can regard the index range in the fundamental representation of
$SO(4,m)$ as split in the following way:
\begin{equation}
L\, =\, L^I_{\phantom{I} J}    \qquad \qquad I,J=\cases { a,b=0,1,2,3 \cr
 t,s=1,2,\dots \, m \cr }
\end{equation}
and introducing the left invariant one--form:
\begin{equation}
L^{-1}\, dL \,\equiv \, \Theta
\end{equation}
we can split it into the vielbein and the connections on the coset
manifold:
\begin{equation}
\label{so4m3}
\Theta = \left(\begin{array}{cc}\theta^{ab} & E^{at}\\ (E^T)^{ta} &
\Delta^{st}
\end{array}\right) \hskip 1cm \left\{\begin{array}{ll}
\theta^{ab} & \mbox{SO(4) connection} \\
E^{at} & \mbox{Vielbein on the coset}\\
\Delta^{st} & \mbox{SO(m) connection.} \end{array}\right.
\end{equation}
From the very definition of $\Theta$ one immediately
obtains the Maurer-Cartan equations:
\begin{equation}
\label{so4m4}
\left\{\begin{array}{ll}
\d  E^{at} +\theta^{ab}\wedge E^{bt} - \Delta^{ts}\wedge E^{as} = 0 &
\mbox{Torsion equation}\\
\d  \theta^{ab} + \theta^{ac}\wedge \theta^{cb} =- E^{as}\wedge E^{bs}  &
\mbox{SO(4) curvature} \\
\d  \Delta^{ts} -\Delta^{tr}\wedge \Delta^{rs} = E^{at}\wedge E^{as} = 0 &
\mbox{SO(m) curvature}
\end{array}\right.
\end{equation}
Notice that the vielbein $E^{at}=E_u^{at} dq^u$
 carries a vector index
$a=0,1,2,3$ of SO(4) and an index $t$ in the vector representation
of SO(m) just as it does the coordinate ${\bf q}$ of the flat HyperK\"ahler
manifold discussed in eq.~\ref{qfieldo}. Accordingly the quaternionic
generalization of eq.~\ref{qfieldo} is obtained by setting:
\begin{eqnarray}
 {\bf l} & \equiv & L^{a \vert t}\nonumber\\
 {\bf E} & \equiv & E^{a \vert t}
 \label{qfieldoap}
\end{eqnarray}
 The quaternionic metric and the corresponding triplet of
HyperK\"ahler 2--forms are given by:
\begin{eqnarray}
ds^2 &\equiv & h_{uv} \, dq^u \,  dq^v \, = \, {\bf E}^T \, \left (
\bfone_{4\times 4} \otimes \bfone_{m\times m} \right ) \,  {\bf E}
\nonumber\\
K^x & = &  {\bf E}^T \, \wedge \, \left ( J^{+ \vert x}
\otimes \bfone_{m \times m} \right ) \,  {\bf E} \ ,
\label{spiattoap}
\end{eqnarray}
which is the quaternionic counterpart of eq.~\ref{spiatto}
Alternatively in the above formula one can use the triplet of
antiself dual t'Hooft matrices to define the HyperK\"ahler structure.
Using the identities ~\ref{ide3}
 and rearranging the $4m$ vielbein $E^{a\vert t}$ into an
$m$-vector of quaternions:
\begin{equation}
{\bf QE} \, = \, \left ( \matrix {QE^1 \, = \, E^{a \vert 1} \, e_a \cr
QE^2 \, = \, E^{a \vert 2} \, e_a \cr
\dots \cr
QE^t \, = \, E^{a \vert t} \, e_a \cr
\dots \cr } \right )
\label{quatvectorap}
\end{equation}
which is the quaternionic counterpart of eq.~\ref{quatvector},
eq.s~\ref{spiattoap} can be rewritten in a form completely
analogous to eq.s~\ref{riscrivo}:
\begin{eqnarray}
ds^2 & = &  {\o{1}{2}} \hskip 3pt \mbox{tr}  \, \left (
  {\bf QE}^\dagger  \, \bfone_{m\times m}   {\bf QE} \right
)\nonumber\\
K & = &   \,{\o{1}{2}} \,  {\bf QE}^T \, \wedge \,
\bfone_{m\times m}  {\bar {\bf QE}}  \, = \,
{\o{1}{2}} \, K^x \, e_x^T
\label{riscrivoap}
\end{eqnarray}
Just as in the flat Hyperk\"ahler case
the action of the gauge group ${\cal G}$ on the hypermultiplets is
assumed to be linear and be generated by a set of $4m \times 4m $
matrices $T_I$:
\begin{equation}
\delta_I {\bf l} \, = \, T_I \, {\bf l} \, \quad\longrightarrow \quad
k^u_I \, = \, \left ( T_I \right )^{u}_{\phantom{u}v} \, q^v
\label{linearazioneap}
\end{equation}
In order for this action to be an  isometry of the Euclidean diagonal
metric ~\ref{spiatto} it is necessary and sufficient that the
matrices $T_I$ belong to the  linearly realized part
of the isometry algebra $SO(4,m)$, namely  $SO(4)\times SO(m)$.
namely:
\begin{equation}
  T_I \, \in \, SO(4)\times SO(m) \, \subset \, SO(4,m)
\end{equation}
The action of ${\cal G}$ however is not only required to be isometrical
but also to be triholomorphic. This means:
\begin{equation}
\ell_{I} \, K^x \, \equiv \, {\bf i}_I \, d K^x \, + \, d\, {\bf i}_I \,
K^x \, =  \nabla W^x_I
\end{equation}
where $W_x$ is the infinitesimal parameter of some $SU(2)$ transformation
A straightforward calculation
shows that the triholomorphicity condition is that the generators $T_I$
should commute with the tensor product  of the 't Hooft matrices with
the unit matrix in $m$--dimensions. When this last condition is
verified we can write the momentum maps as:
\begin{equation}
{\cal P}^x_I \, = \, {\bf l}^T \,
J^{+\vert x}\otimes \bfone_{m \times m} \, T_I \,{\bf l}
\label{unmomentoap}
\end{equation}
 Using these ingredients the mass matrices and the scalar potential
 can be written down without any further difficulty.
The quaternionic vielbein is given in full analogy to eq.s~\ref{scordo1},
\ref{scordo2}, by
\begin{equation}
{\cal U}^{A \alpha} \, \equiv \, {\cal U}^{A \alpha}_{b \vert s} \,
dq^{b \vert s} \, = \,    = \, E^{a \vert t} \,
( e_a )^{A}_{\phantom{A}B}
\label{scordo1ap}
\end{equation}
and, as before, we identify the symplectic index $\alpha$
running on $2m$ values
with the pair of indices $B,t$ ( $B=1,2$; $t=1,\dots,m$).
\section*{Appendix D: Normalizations and conventions}

\par
{\it Minkowski metric}:
\begin{equation}
\eta_{ab}\equiv \left(1,-1,-1,-1\right)
\end{equation}
 \par
{\it Definition of the Riemann tensor}:
\begin{equation}
R^{\mu}_{\phantom{\mu} \nu}\,=\, d \Gamma^{\mu}_{\phantom{\mu} \nu} +
\Gamma^{\mu}_{\phantom{\mu} \rho} \wedge \Gamma^{\rho}_{\phantom{\rho} \nu}
\equiv \,-\,{1 \over 2} R^{\mu}_{\phantom{\mu} \nu\rho \sigma}dx^\rho \wedge
dx^\sigma
\end{equation}

{\it Decomposition of tensors in self--dual and antiself--dual parts
($\epsilon _{0123}\, \equiv\,1$)}:
\begin{equation}
T^{\mp}_{\mu \nu}\,=\,{1 \over 2} \left(T_{\mu \nu} \,{\mp} \,{{\rm i} \over 2}
\epsilon _{\mu \nu \rho \sigma} T^{\rho \sigma} \right)
\end{equation}

{\it Clifford Algebra}:
\begin{eqnarray}
 \left\{\gamma_a,\gamma_b \right\}\, &=& \,2\,\eta_{ab} \nonumber\\
   \left[\gamma_a,\gamma_b \right]\,&=&\,2\,\gamma_{ab} \nonumber\\
  \gamma_5 \,& \equiv & -\, {\rm i}\, \gamma_0 \gamma_1 \gamma_2 \gamma_3
\nonumber\\
  \gamma _0^{\dagger}\,&=&\,\gamma _0; \qquad \qquad
 \gamma _0 \gamma _i^{\dagger}  \gamma _0 \,=\, \gamma _i  \qquad
 (i=1,2,3);\qquad \qquad  \gamma _5^{\dagger}\,=\,\gamma _5 \nonumber\\
  \epsilon _{abcd} \gamma^{cd}\,&=&\,2\,{\rm i}\, \gamma_{ab} \gamma_5
 \label{clifdef}
\end{eqnarray}

{\it Decomposition of fermions in chiral and antichiral parts}:\\
the indices of the spinors also fix their chirality according to the
following conventions:
\begin{equation}
 \gamma _5
\left(\begin{array}{c}  \lambda ^{iA} \\
 \zeta _{\alpha} \\ \psi _A \end{array}\right)\, =\,
\left(\begin{array}{c}  \lambda ^{iA} \\
 \zeta _{\alpha} \\ \psi _A \end{array}\right)\ ,
 \label{gammachi}
\end{equation}
 \begin{equation}
 \gamma _5
\left(\begin{array}{c}  \lambda ^{i^\star}_A \\
 \zeta^{\alpha} \\ \psi^A \end{array}\right)
\, =\,-\,
 \left(\begin{array}{c}  \lambda ^{i^\star}_A \\
 \zeta^{\alpha} \\ \psi^A \end{array}\right)
\label{gammachi-}
\end{equation}

{\it Majorana conventions}:\\
For any fermion $ \phi$ :
\begin{equation}
{\bar \phi} \, \equiv \, \phi ^{\dagger}\gamma _0\,=\,\phi^T C
\end{equation}

{\it Fierz rearrangements}\\
Let us denote by a lower or upper dot right and left chirality
respectively. Then:
{\it for 0--form spinors $\chi, \xi$}:
\begin{eqnarray}
&& {\chi}_{\bullet} \bar{\xi}_{\bullet}\, = \,-\,{1 \over 2} \,
 \bar{\xi}_{\bullet} {\chi}_{\bullet}  \, +\,
 {1 \over 8}\, \gamma_{ab} \bar{\xi}_{\bullet} \gamma^{ab}{\chi}_{\bullet}
 \nonumber\\
 && {\chi}_{\bullet} \bar{\xi}^{\bullet}\, = \,-\,{1 \over 2} \,
\gamma _{a} \bar{\xi}^{\bullet} \gamma^{a}{\chi}_{\bullet}
\label{fierzchi}
\end{eqnarray}
 {\it for 1--form spinors $\psi _A, \psi^B$}:
\begin{eqnarray}
&& {\psi}_{A} \bar{\psi}_{B}\, = \, {1 \over 2} \,
 \bar{\psi}_{B} {\psi}_{A} \,  -  \,
 {1 \over 8} \,\gamma _{ab} \bar{\psi}_{B}
 \gamma^{ab}{\psi}_{A}\nonumber\\
 && {\psi}_{A} \bar{\psi}^{B}\, = \, {1 \over 2} \,
\gamma _{a} \bar{\psi}^{B} \gamma^{a}{\psi}_{A}\label{fierzpsi}
\end{eqnarray}

{\it Charge conjugation matrix properties}:
\begin{equation}
C^2 \,=\,-1 ; \qquad C^T \, =\, - \,C ; \qquad \left(C \gamma^a \right)^T \,=\,
 C \gamma^a   ; \qquad  \left(C \gamma^{ab} \right)^T \,=\,
 C \gamma^{ab}    \label{cmatr}
\end{equation}

{\it Hermiticity of currents}\\
  for 0--form spinors:
\begin{eqnarray}
&& \left({\bar \chi}_ \bullet \xi _ \bullet \right)^\dagger \, =\,{\bar
\xi}^\bullet \chi^\bullet
\,=\, {\bar \chi}^\bullet \xi^\bullet\\
&& \left({\bar \chi}_ \bullet \gamma^a \xi^\bullet \right)^\dagger \, = \,
{\bar \xi} _ \bullet \gamma^a \chi^\bullet
\,=\,-\, {\bar \chi}^\bullet \gamma^a \xi _ \bullet\\
&& \left({\bar \chi}_ \bullet \gamma^{ab} \xi _ \bullet \right)^\dagger \, =
\,- \,
{\bar \xi}^\bullet \gamma^{ab} \chi^\bullet
\,=\,  {\bar \chi}^\bullet \gamma^{ab} \xi^\bullet \label{hermi0}
\end{eqnarray}
 for 1--form spinors:
 \begin{eqnarray}
 && \left({\bar \psi}_ A \psi _ B \right)^\dagger\, =\,-\,  {\bar \psi}^B
\psi^A\,=\, {\bar \psi}^A \psi^B\\ && \left({\bar \psi}^A \gamma^a \psi _ B
\right)^\dagger\, =\,-\,  {\bar \psi}^B \gamma^a \psi_ A\, =\,-\, {\bar
\psi}_ A \gamma^a  \psi^B\\ && \left({\bar \psi}^A \gamma^{ab} \psi^B
\right)^\dagger\, =\, {\bar \psi}_B \gamma^{ab} \psi_ A\, =\, {\bar \psi}_ A
\gamma^{ab}  \psi_B \label{hermi}
\end{eqnarray}
\par
{\it Conventions on K\"ahler geometry}:
The hermitean metric is locally given by:
\begin{equation}
g_{i {j^\star}}= \partial_i \partial_{j^\star} {\cal K}
\label{popov}
\end{equation}
where the real function  ${\cal K}={\cal K}^\star = {\cal K}( z, z^*)$
is named the {\it K\"ahler potential}.
It is defined up to the real part of a holomorphic function $f(z)$.
Indeed one sees that
\begin{equation}
{\cal K}^\prime (z, z^{i^\star} )={\cal K} (z, z^{i^\star} ) +
{\rm Re}f(z)
\label{041}
\end{equation}
gives rise to the same metric $g_{i {j^\star}}$ as ${\cal K}$.
The transformation in eq.~\ref{041} is named a
{\it K\"ahler transformation}.
\par
To fix our notations we write the formulae for the Levi--Civita
connection 1--form  and Riemann curvature 2--form on a K\"ahler manifold:
\begin{equation}
\begin{array}{ccccccc}
\Gamma^i_j &=&  \Gamma^i_{kj} d z^k   & ; &
\Gamma^i_{kj}   & = &  \,  g^{i\ell^*}(\partial_j g_{k\ell^*})  \cr
\Gamma^{i^*}_{j^*}  &=&  \Gamma^{i^*}_{k^*j^*} d \bar
z^{k^*}  &; &
\Gamma^{i^*}_{k^*j^*}  & =&   g^{i^*\ell}(\partial_{j^*} g_{k^*\ell})
  \cr
 {\cal R}^i_j   &=&  {\cal R}^i_{jk^*\ell} d \bar{z}^{k^*} \wedge d
z^\ell  &; &
{\cal R}^i_{jk^*\ell}   & =&   \partial_{k^*} \Gamma^i_{j\ell}  \cr
 {\cal R}^{i^*}_{j^*} &=&  {\cal R}^{i^*}_{j^*k \ell^*}d z^k \wedge d\bar
z^{\ell^*}   &; &
{\cal R}^{i^*}_{j^* k \ell^*}  & =&    \partial_k \Gamma^{i^*}_{j^*\ell^*} \cr
\end{array}
\label{curvelievi}
\end{equation}
\par
{\it $SU(2)$ and $Sp(2n)$ metrics}:
\begin{equation}
\epsilon^{AB}\,\epsilon _{BC} \,=\,-\, \delta^A _C ; \qquad  \qquad
\epsilon^{AB}\,=\,-\,   \epsilon^{BA} \label{epsi}
\end{equation}
 \begin{equation}
\IC^{\alpha\beta}\,\IC _{\beta\gamma} \,=\,-\, \delta^\alpha _ \gamma ;
\qquad  \qquad
\IC^{\alpha\beta}\,=\,-\,   \IC^{\beta\alpha};  \label{cpsi}
\end{equation}
For any $SU(2)$ vector $P_A$ we have:
\begin{equation}
\epsilon _{AB}\, P^B \,=\,P_A ; \qquad  \qquad  \epsilon^{AB}\, P_B \,=\,-\,P^A
\end{equation}
 and equivalently for $Sp(2n)$ vectors  $P_ \alpha$:
 \begin{equation}
\IC_{\alpha\beta}\, P^\beta \,=\,P_ \alpha ; \qquad  \qquad
\IC^{\alpha\beta}\, P_ \beta \,=\,-\,P^\alpha
\end{equation}

{\it Reality condition for $SU(2)$ valued matrices $H^{AB}$}:
\begin{equation}
\bar{\left(H^{AB}\right)} \,=\,\epsilon ^{AC}\,\epsilon^{BD} H^{\star}_{CD}
\end{equation}
\begin{table}
\begin{center}
\caption{\sl Homogeneous Symmetric Special Manifolds}
\label{homospectable}
\begin{tabular}{|c||c||c||c|}
\hline
n  & $G/H$  & $Sp(2n+2)$  & symp rep of G
\\
\hline
~~&~~&~~& ~~\\
$1$ & $\o{SU(1,1)}{U(1)}$ & $Sp(4)$ &
${\underline {\bf 4}}$ \\
~~&~~&~~& ~~\\
\hline
~~&~~&~~& ~~\\
~~&~~&~~& ~~\\
$n$ & $\o{SU(1,n)}{SU(n)\times U(1)}$ & $Sp(2n+2)$ &
${\underline {\bf n+1}}\oplus{\underline {\bf n+1}}$ \\
~~&~~&~~& ~~\\
\hline
~~&~~&~~& ~~\\
$n+1$ & $\o{SU(1,1)}{U(1)}\otimes \o{SO(2,n)}{SO(2)\times SO(n)}$
 & $Sp(2n+4)$ & $
{\underline {\bf 2}}\otimes \left ({\underline {\bf n+2}}
\oplus{\underline {\bf n+2}}\right )  $ \\
~~&~~&~~& ~~\\
\hline
~~&~~&~~& ~~\\
$6$ & $\o{Sp(6,\IR)}{SU(3)\times U(1)}$ & $Sp(14)$ & $
{\underline {\bf 14}}$ \\
~~&~~&~~& ~~\\
\hline
\hline
~~&~~&~~& ~~\\
$9$ & $\o{SU(3,3)}{S(U(3)\times U(3))}$& $Sp(20)$ & $
{\underline {\bf 20}}$ \\
~~&~~&~~& ~~\\
\hline
~~&~~&~~& ~~\\
$15$ & $\o{SO^\star(12)}{SU(6)\times U(1)}$& $Sp(32)$ & $
{\underline {\bf 32}}$ \\
~~&~~&~~& ~~\\
\hline
~~&~~&~~& ~~\\
$27$ & $\o{E_{7(-6)}}{E_6\times SO(2)}$& $Sp(56)$ & $
{\underline {\bf 56}}$ \\
~~&~~&~~& ~~\\
\hline
\end{tabular}
\end{center}
\end{table}
\vfill
\eject
\begin{table}
\begin{center}
\caption{\sl Homogeneous symmetric quaternionic manifolds}
\label{quatotable}
\begin{tabular}{|c||c|}
\hline
m  & $G/H$
\\
\hline
~~&~~\\
$m$ & $\o{Sp(2m+2)}{Sp(2)\times SP(2m)}$ \\
~~&~~\\
\hline
~~&~~\\
$m$ & $\o{SU(m,2)}{SU(m)\times SU(2)\times U(1)}$  \\
~~&~~\\
\hline
~~&~~\\
$m$ & $\o{SO(4,m)}{SO(4)\times SO(m)}$  \\
~~&~~\\
\hline
\hline
~~&~~\\
$2$ & $\o{G_2}{S0(4)}$ \\
~~&~~\\
\hline
~~&~~\\
$7$ & $\o{F_4}{Sp(6)\times Sp(2)}$\\
~~&~~~\\
\hline
~~&~~\\
$10$ & $\o{E_6}{SU(6)\times U(1)}$\\
~~&~~\\
\hline
~~&~~\\
$16$ & $\o{E_7}{S0(12)\times SU(2)}$\\
~~&~~\\
\hline
~~&~~\\
$28$ & $\o{E_8}{E_7\times SU(2)}$\\
~~&~~\\
\hline
\end{tabular}
\end{center}
\end{table}
\vfill
\eject
\section*{Acknowledgements}
A.C. and S. F. would like to thank the Institute for Theoretical Physics
at U. C. Santa Barbara for its kind hospitality and where part of this work
was completed.

\end{document}